\documentclass[aps,prx,twocolumn
,onecolumnshowpacs,showkeys,superscriptaddress, 11pt tightenlines]{revtex4-2}

\usepackage{amsmath}
\usepackage[dvipdfmx]{graphicx}

\usepackage{enumitem}   
\usepackage{tikz-cd}

\usepackage{chngcntr}

\DeclareMathOperator{\sgn}{sgn}

\newcommand{\hl}[1]{#1}
\usepackage{here}

\usepackage{bm}

\begin{document}

\title{
Conditions for Darwinian evolution in compartmentalized autocatalytic reaction networks}

\author{Yoshiya J. Matsubara}
\email{yoshiyam@uchicago.edu}
\affiliation{Department of Physics, University of Chicago, USA}
\author{Sandeep Ameta}
\affiliation{Department of Biology, Trivedi School of Biosciences, Ashoka University, Sonipat, India}
\author{Shashi Thutupalli}
\affiliation{Simons Centre for the Study of Living Machines, National Centre for Biological Sciences, Tata Institute of Fundamental Research, Bengaluru, India}
\affiliation{International Centre for Theoretical Sciences, Tata Institute of Fundamental Research, Bengaluru, India}
\author{Philippe Nghe}
\affiliation{Laboratoire de Biologie Structurale de la Cellule, BIOC, CNRS, Ecole polytechnique, Institut Polytechnique de Paris, 91120, Palaiseau, France.}
\author{Sandeep Krishna}
\affiliation{Simons Centre for the Study of Living Machines, National Centre for Biological Sciences, Tata Institute of Fundamental Research, Bengaluru, India}

\date{}

\begin{abstract}
\hl{The first forms of life likely consisted of protocells endowed with metabolism, growth, and division. Such systems may have evolved due to variation and heredity in their chemical composition, even before the advent of genetics. However, whether compositional heredity is robust enough to sustain evolution by natural selection remains unknown, especially given that early compartmentalization cycles were likely imperfect, potentially disrupting stable inheritance across generations. Here, we show that multistable autocatalytic reaction networks can maintain heritable compositional states across a broad class of growth-division regimes, including continuous, serial, symmetric division, and multi-fragmentation cycles. We further identify parameter domains that preserve stable inheritance in the presence of stochastic variation, such that selection can operate efficiently. We finally demonstrate rudimentary forms of evolution by natural selection in populations of protocells with two heritable states, which we illustrate in an experimentally feasible setting. Our findings establish conditions for natural selection in compartmentalized autocatalytic systems and set the stage for understanding the minimal requirements for open-ended evolution.}
\end{abstract}


\maketitle

\section{Introduction}

A critical transition for the origin of life on Earth was the emergence of a ``Darwinian population" of self-reproducing individuals evolving under natural selection \cite{benner2010defining}. In one scenario -- ``replication-first" or ``genes-first" -- a template replicator, such as an RNA ribozyme that can copy itself, was the earliest self-replicating entity. Eigen and others \cite{eigen1971selforganization, eigen1977principle} have demonstrated the conditions under which a population of protocells containing such a replicator can implement the three key properties of a Darwinian population, namely heredity, variation, and differential reproduction \cite{lewontin1970units, godfrey2007conditions}. However, amid the lack of spontaneity in the emergence of even simpler error-prone template replicators \cite{dyson1982model, dyson1999origins}, an alternate approach, the ``metabolism-first" scenario, was proposed, in which the first protocells contained autocatalytic chemical systems (ACSs) that could collectively self-reproduce despite lacking an individually self-replicating molecule. In this scenario, the chemical composition of the encapsulated chemical system acts as the phenotype of a protocell. 
 
Using ACSs, various schemes \cite{segre1998graded, segre2000compositional} have been proposed to demonstrate how such compositional information can be stably inherited as protocells grow and divide, despite stochasticity and other sources of variation. Earlier studies questioned compositional heredity in ACSs because their chemical compositions may be either too unstable to persist across growth–division cycles or too stable to generate variation on which selection can act~\cite{vasas2010lack, vasas2012evolution, vasas2015primordial}. Vasas et al. proposed a scenario where open-ended evolution may occur through the emergence and the competition of multiple viable autocatalytic `cores'~\cite{vasas2012evolution}. 
\hl{However, the conditions under which ACSs can simultaneously sustain heredity, variation, and selection across growth–division cycles remain unclear.}

In this paper, we \hl{show that} autocatalytic chemical systems, enclosed in growing and dividing compartments, can exhibit heredity of their chemical composition, {and identify the precise parameter regimes under which this occurs}. 
{We analyze the stability of compositional growth states under a generalized class of growth–division dynamics, encompassing a wide range of compartmentalization cycles (e.g., division in two as in biological cells as well as division into multiple protocells). 
We then examine a broad class of bistable autocatalytic systems that exhibit two growth states and determine the conditions under which these states remain stable during growth and division. 
We show that such dynamics can be bounded by two limiting cases: chemostat-like continuous dilution (continuous stirred-tank reactor, CSTR) and discrete serial dilution protocols.} 
We then introduce variation (via stochasticity) and selection (via differential reproductive rates of the two states) to identify parameter regimes in which the stability of multiple compositional states can be propagated across cycles of growth and division. 
Our work shows that, in the absence of template replicators, the emergence of a Darwinian population on the prebiotic Earth via autocatalytic chemical systems is indeed feasible.

 \subsection{Defining heredity in autocatalytic systems}

\begin{figure}
    \centering
    \includegraphics[width=7.8cm]{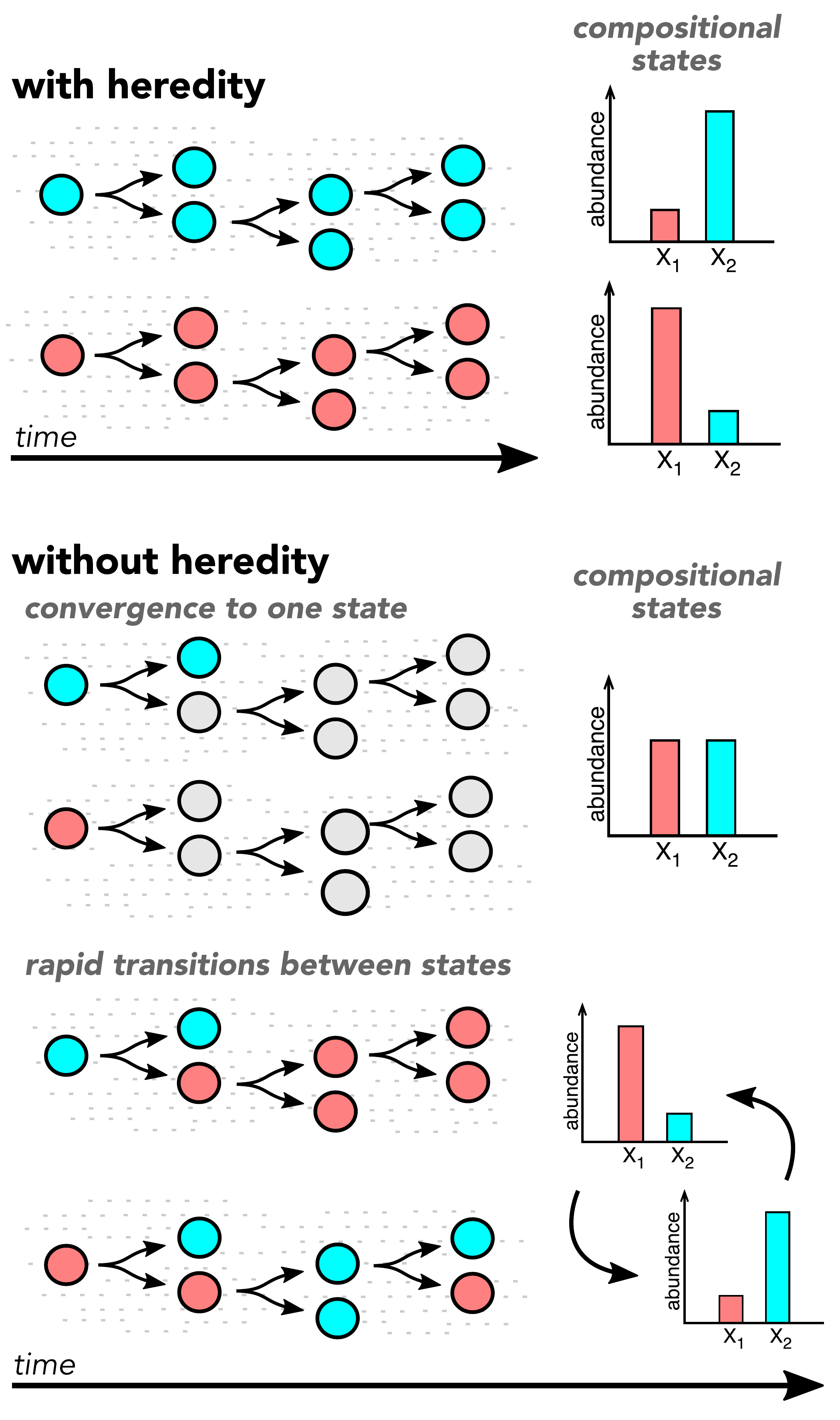}
    \caption{Schematic representation of compartmentalized autocatalytic systems with and without heredity. Red, blue, and grey circles represent compartments or protocells with different chemical compositions under the same environment: $X_1$-dominant, $X_2$-dominant, and uniform compositions, respectively. \hl{Grey bars in the background represent the food molecules.} (\emph{top}) reproduction of compartments with heredity: daughters have the same composition as their parents.
    Heredity can break down in two ways: (\emph{middle}) either the chemical composition converges to the same single (gray) state irrespective of the initial state. (\emph{bottom})
    spontaneous transitions between the two growth states (red and blue) are so rapid that all information about the initial state is lost very quickly.}
    \label{fig:heredity}
\end{figure}

\hl{We posit} that the \emph{minimum} requirements for a chemical system to exhibit heredity 
are the following (see more rigorous definitions in Appendix Sec.~\ref{sec:appdx-definition}):
\begin{enumerate}[label=(\roman*)]
\item The existence of two distinct growth states for the same substrate environment (food set),
\item The stability of these states under the growth and division of the enclosing compartment.
\end{enumerate}
We define a growth state as one where, with a continuous supply of food molecules and the absence of any dilution, the concentrations of the chemicals comprising the system would grow without bound (often exponentially) as the system consumes food molecules, but where the chemical composition
-- the relative concentrations -- 
reach a stationary state (Fig.\,\ref{fig:heredity}, \emph{top}). 
The first requirement provides distinct compositional phenotypes on which differential reproduction can act, whereas the second ensures that these phenotypes persist across generations. 

A minimum of two growth states is necessary for heredity to combine with both variation and differential reproduction. By this criterion, models of autocatalytic systems with just a single stable state \cite{west2017origin, nunes2022limits}, or with an inactive and active state \cite{schlogl1972chemical, dyson1982model, giri2012origin, matsubara2016optimal, higgs2021reaction}, or which exhibit different active states only when the nutrient or environmental conditions are changed \cite{colomer2020selection, ameta2021darwinian, peng2022hierarchical}, are not considered here. This \hl{scenario} is also depicted in Fig.\,\ref{fig:heredity} (middle panel), where, irrespective of the initial compositional state, after one or a few growth and division cycles, the composition converges to one state and all information from previous generations is lost. 
\footnote{Furthermore, we also do not examine the conditions for ``open-ended evolution'', where new growth states (new autocatalytic cores) continually arise over time. For us, a system with even two growth states is sufficient to form a (simple) Darwinian population, in analogy with the evolution of a gene with two alleles.} 
We thus focus on ACSs that exhibit such `bistability', and will study heredity (or lack of it) of the two compositional states when these ACSs are enclosed in compartments that grow and divide. \hl{While most of our analysis focuses on the simplest bistable ACS shown in Fig.}\,~\ref{fig:growth-division}\hl{A, the framework applies more generally to a broader class of ACS network motifs (see  Fig.}~\ref{fig:SImodel}\hl{ and Fig.~}\ref{fig:SI2steps}\hl{).}

Robust inheritance of chemical composition \hl{requires} the stability of two growth states \hl{across successive growth-division cycles}. In our case, \hl{this corresponds to the dynamical stability of two growth states, such that} systems do not spontaneously transition from one state to the other \hl{during growth and division} (Fig.\,\ref{fig:heredity}). \hl{In addition, these states must remain stable} against random perturbations, e.g., due to thermal noise~\cite{van1992stochastic, gardiner1985handbook} \hl{or stochastic fluctuations, mixing between compartments, uncoordinated production or growth, stochastic partitioning upon division}. A small probability of transitions to different states due to noise can be subsumed under phenotypic variation (indeed, this may be the only available source of variation). Still, too much will destroy the heredity of states. This scenario is depicted in Fig.\,\ref{fig:heredity} (\emph{bottom panel}), where two growth states exist, but the transitions between them occur \hl{on timescales comparable to or shorter than the} growth and division cycles; therefore, the information about past generations is lost. Thus, robust heredity requires sufficient stability of the chemical composition both across growth and division cycles, and \hl{against stochastic perturbations} and other sources of variation.

\subsection{Mapping general growth and division scenarios to the serial dilution protocol}
\label{sec:growth-division}
\begin{figure*}
    \centering
    \includegraphics[width=16cm]{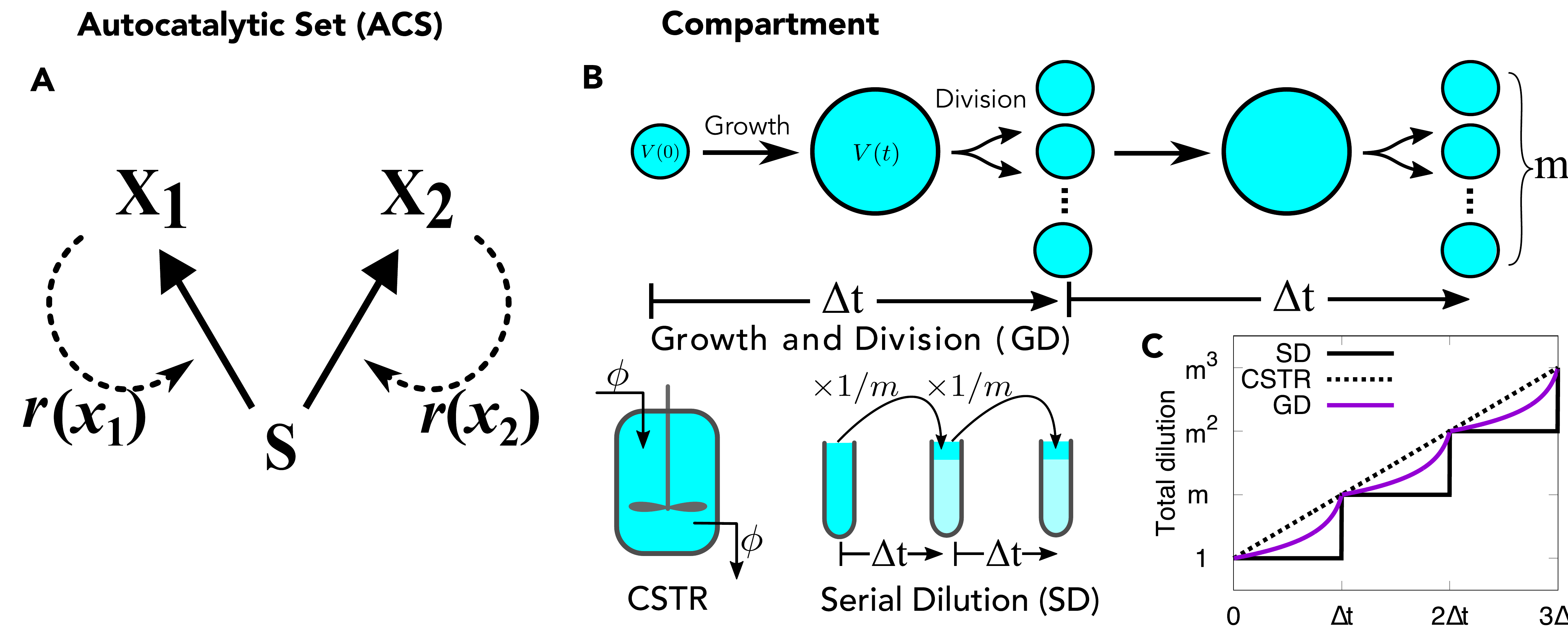}
    \caption{Schematics of autocatalytic sets and dilution protocols composing a self-reproducing chemical system. (A) Schematics of the competing autocatalytic entities, $\rm{X}_1$ and $\rm{X}_2$; they are converted from a substrate S (solid arrows) catalyzed by itself (dashed arrows). 
    (B) Schematics of dilution protocols. Growth and Division (GD) cycles of a compartment with volume $V(t)$: After a period $\varDelta t$, during which the compartment grows by a factor $m=e^{\bar{\phi} \varDelta t}$, it divides into $m$ equal-sized compartments each with the initial volume. Continuous Stirred-Tank Reactor (CSTR) protocol: a substrate is supplied, and compositions are diluted with the same constant rate $\bar \phi$. Serial Dilution (SD) protocol: for each interval $\varDelta t$, compositions are diluted with the factor $m$. (C) The time course of the cumulative dilution experienced by the compartment as a function of time, $\exp(\int_0^t \phi(t') dt')$. 
    The black solid and dotted lines correspond to the SD and CSTR protocols, respectively. The blue line corresponds to a case where the compartment grows at rate $\frac{dV}{dt} \propto V^\alpha$ ($\alpha=4$) and splits into $m$ equal-sized daughters when it divides (see Methods and Models).}
       \label{fig:growth-division}
\end{figure*}

We \hl{show that} growth and division (GD) dynamics of compartments \hl{can impose constraints on the inheritance of chemical composition and may even disrupt heredity, depending on their dynamics, even when the underlying reaction systems support multiple compositional growth states}. \hl{Division events need not correspond to binary fission, but may involve fractionation into many subcompartments, as observed in coacervates, vesicles}~\cite{zhu2009coupled}\hl{, and other protocell models in prebiotic environments}~\cite{chakraborty2026novo}\hl{.}
As compartments grow and divide, the chemical concentrations within them change due to the chemical reactions occurring, but also get diluted due to increases in the compartment volume. If the compartment volume grows between divisions as $V(t)$, then the dilution rate is $\phi(t)\equiv \frac{dV}{dt}(t) / V(t)$. An influx of substrates from outside the compartment may also occur, increasing the concentration of those components. Thus, the chemical rate equations must include terms for such influx and dilution. Initially, we consider three assumptions on such a GD process, some of which will be relaxed in later sections: 
\begin{enumerate}
    \item Compartments divide into $m$ equal-sized smaller ones periodically, at time intervals of $\varDelta t$ (see Fig.\,\ref{fig:growth-division}B). 
    \item We assume a well-mixed condition inside a compartment. This implies that the \emph{concentrations} do not change at divisions because the chemical components are partitioned proportionally to the volumes of the daughter compartments.
\item For simplicity, we assume the influx rate of substrates $\sigma(t)$ is proportional to the dilution rate $\phi(t)$ (this is not an important assumption; if $\sigma(t)$ is an arbitrary function with periodicity $\varDelta t$ our results do not qualitatively change).
\end{enumerate}

Typical dilution protocols that are used in laboratories, such as serial dilution (SD) or the continuous stirred-tank reactor (CSTR), are special cases of the compartment GD scenario:
If $\phi(t)$ (and $\sigma(t)$) have sharp spikes at times $t = n \varDelta t \quad (n = 1,2,3, \dots)$, it corresponds to SD, in which after each time interval $\varDelta t$ the chemical compositions are diluted by $m = e^{\varDelta t \bar{\phi}}$ fold, and $s^{tot} (1-\frac{1}{m})$ substrate is added at the beginning of the next cycle 
(here $\bar{\phi}$ and $\bar{\sigma}$ are the average dilution and influx rates over one division cycle; see Methods and Models for details). 
Similarly, a CSTR corresponds to a GD process where influx and dilution rates are constants: $\phi(t) = \bar{\phi}$ and $\sigma(t) = \bar{\sigma}$. It also corresponds to an SD protocol with an infinitesimal short interval of cycles, $\varDelta t \rightarrow 0$ (see Methods and Models for details).

General GD protocols interpolate between the impulsive SD and constant  CSTR dilution protocols (see Fig.\,\ref{fig:growth-division}B). We first investigate the reaction dynamics of competing autocatalytic reaction sets (ACSs) under SD. Later, we will return to general GD protocols and show that their bistable parameter regions are bounded by the corresponding  SD and  CSTR limits.

\section{Results}
\subsection{\hl{Heredity under serial dilution requires a concentration-dependent growth rate}}
\label{sec:acs-model}

First, to test whether an autocatalytic chemical reaction system can exhibit bistability under the SD protocol, we consider a simple class of autocatalytic reaction systems consisting of two identical (but distinguishable) autocatalytic \hl{species with symmetric autocatalytic kinetics}, ${\rm{X}}_1$ and ${\rm{X}}_2$, which consume the same substrate S (see Fig.\,\ref{fig:growth-division}A). This system can exhibit two distinct growth states.
The rate equations for this class of systems are:
\begin{equation}
    \frac{dx_i}{dt} = s r(x_i) x_i,
    \label{eq:model}
\end{equation}
where $i=1,2$, and $s$ is the concentration of the substrate S.
$r(x_i)$ is the reproduction rate of ${\rm{X}}_i$
~\footnote{Here, reproduction rate functions are kept to be symmetric between ${\rm{X}}_1$ and ${\rm{X}}_2$ for the sake of simplicity. However, the asymmetric cases are also feasible with this approach (see Appendix Sec.\,\ref{sec:asymmetric}).}. We assume that $r(x)$ is a differentiable and non-negative function for $x \geq 0$, but otherwise leave its form unrestricted; experimentally motivated examples are considered in  Sec.\,\ref{sec:azoarcus}. We study this reaction system under the SD protocol with cycle interval $\varDelta t$ and the dilution factor $m$ ($=e^{\bar{\phi} \varDelta t}$) (see Methods and Models). Because substrate replenishment compensates for dilution, the total concentration of the components $s^{tot} = s + x_1 + x_2$ remains constant at $s^{tot} = \frac{\bar{\sigma}}{\bar{\phi}}$. 

\begin{figure*}
    \centering
    \includegraphics[width=16.cm]{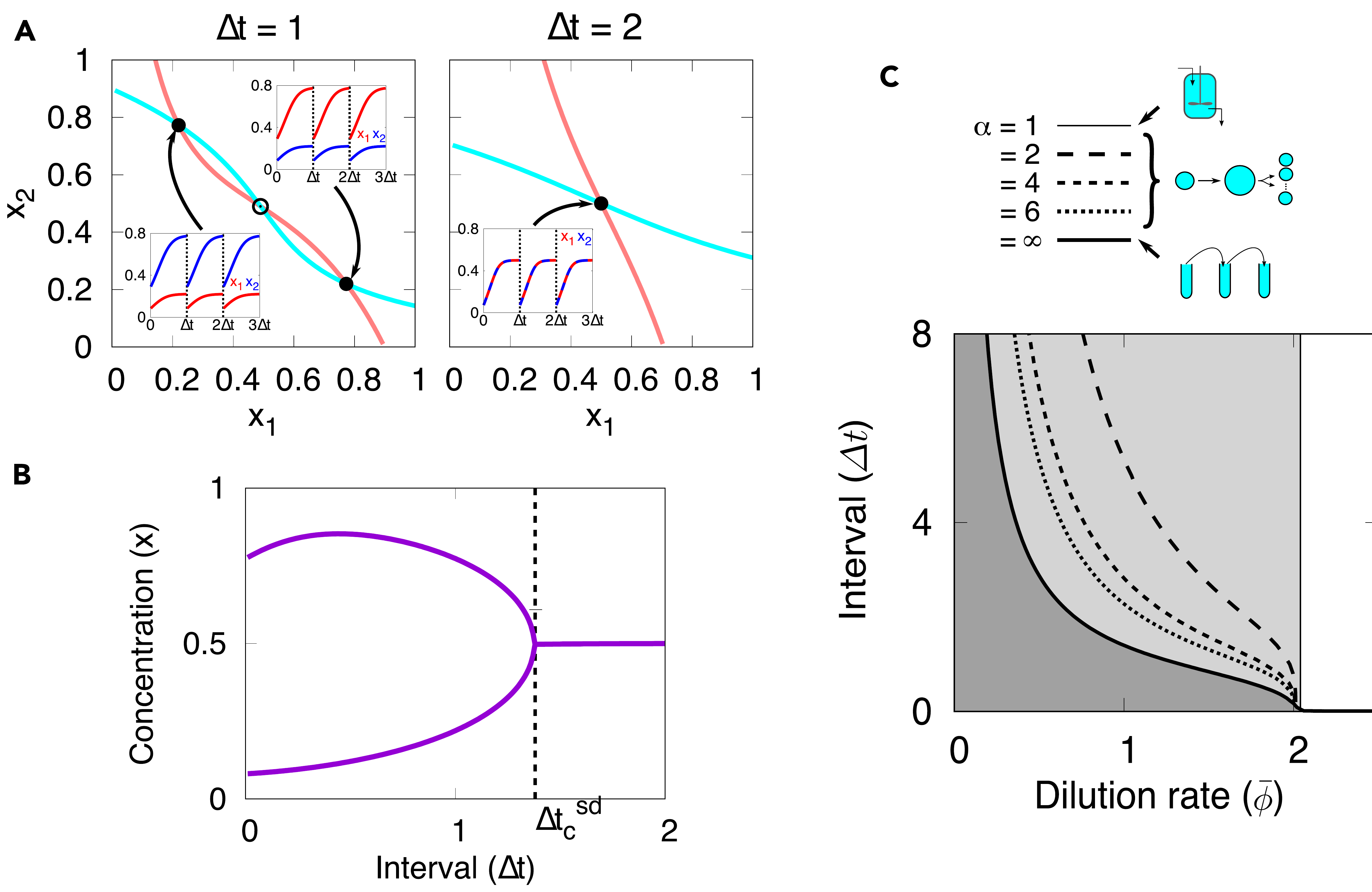}
    \caption{Constraints for composition heredity under dilution protocols. 
    (A)
    The red and blue curves represent the nullclines for the composition of the entities $\rm{X}_1$ and $\rm{X}_2$ at just before the dilution, $x_1(-0)$ and $x_2(-0)$, respectively, in a case with $r(x)x = \epsilon + \kappa x^2$. The cross points represent the stable/unstable fixed points of concentrations of $\rm{X}_1$ and $\rm{X}_2$ at just before the dilution, $x_1^*(-0)$ and $x_2^*(-0)$. See Methods and Models for details of the drawing of the nullclines.
    We set $\varDelta t = 1$, $2$, $\kappa = 8$, $\epsilon = 0.5$, and $\bar{\phi} = 1$. 
    (B) The bifurcation diagram with a varying interval of each dilution cycle $\varDelta t$ in a case with $r(x)x = \epsilon + \kappa x^2$. 
    The pitchfork bifurcation occurs at $\varDelta t = \varDelta t_c^{sd}$ when the upper and lower fixed points merge into one fixed point. 
    (C) Phase diagram for the system with/without heredity using the dilution rate $\bar\phi$ and dilution interval $\varDelta t$ as parameters, in a case with $r(x)x = \epsilon + \kappa x^2$. The solid line represents the boundary between with and without bistability under SD.
     The dashed lines represent that under the GD with different growth laws (the exponents $\alpha$ of the growth law for compartment; $\alpha = 2,4,6$).
     The thin vertical line represents that under GD with $\alpha = 1$, corresponding to the CSTR protocol.
    We set the parameters as $\kappa = 8, \epsilon=0.5$. 
    }
    \label{fig:nullclines}
\end{figure*}

\hl{We observe} that, on the long timescale, after sufficiently many SD cycles, the trajectory of the chemical composition reaches a stationary periodic orbit (Fig.\,\ref{fig:nullclines}A). If the system does \emph{not} exhibit inheritance, then it must settle into the same stable trajectory for every initial condition. Since we assumed two ACSs are completely symmetric, this trajectory must be one in which the concentrations are equal (i.e., $x_1=x_2$). In contrast, if the compositional state is inherited, then across different initial conditions the system must exhibit bistability, i.e., two stable trajectories. Again, due to symmetry, in each of these two trajectories, one of the components, ${\rm{X}}_1$ or ${\rm{X}}_2$, must be dominant. 
Therefore, a sufficient condition for bistability under the SD protocol~\footnote{Note that this is only a sufficient condition since the system could have the stable symmetrical state and $\rm X_1$- and $\rm X_2$-dominant states at the same time.  
} can be obtained by showing the \emph{instability} of the symmetrical ($x_1=x_2$) trajectory 
\footnote{Note that the dynamics of $x_1$ and $x_2$ in the Poincar\'e section, as described by Eq.\,\ref{eq:model} with the SD protocol, are bounded and do not exhibit oscillations for any choice of $r(x)$ (see Appendix Sec.\,\ref{sec:appdx-noosci}). The presence of a fixed point at which the dynamics are unstable in one direction (i.e., a saddle point) thus guarantees the existence of multiple stable fixed points (multistability).}.

Introducing the notation $\chi = x_1 + x_2$ and $\delta = x_1 - x_2$, for trajectories close to the symmetrical one\hl{, we can assume that} $\delta \ll \chi$. One can then derive the following relation (see details of the derivation in Appendix Sec.\,\ref{sec:derivation}):
\begin{equation}
    \frac{\delta(t)}{\chi(t)} = \frac{r(\frac{\chi(t)}{2})}{r(\frac{\chi(0)}{2})} \frac{\delta(0)}{\chi(0)}.
\end{equation}
If $\delta(t)/\chi(t)$ at the end of a cycle, $\delta(\varDelta t)/\chi(\varDelta t)$, is larger than that at the beginning, $\delta(0)/\chi(0)$, the trajectory is unstable, otherwise it is stable. Therefore, the sufficient condition for bistability under the SD protocol is
\begin{equation}
    r\big(\frac{ \chi(\varDelta t)}{2}\big) > r\big(\frac{\chi(0)}{2}\big).
    \label{eq:condition}
\end{equation}
That is, the stability of the compositional trajectory is determined by whether the reproduction rate at the end of a cycle $r\big(\frac{ \chi(\varDelta t)}{2}\big)$ is larger than that at the beginning $r\big(\frac{ \chi(0)}{2}\big)$ or not \footnote{Note that if $r(x)x$ is monotonic, this is also a necessary condition. However, in general, this is only a sufficient and not necessary condition. For example, there are cases where both symmetric and asymmetric trajectories are stable if $r(x)x$ is non-monotonic. For example, $r(x)x = \epsilon + \kappa x (x^2 - \frac{3}{2} (\alpha + \beta) + 3 \alpha \beta)$.}.

For example, if $r(x)x$ is linear, (e.g., $r(x)x = {\epsilon + \kappa x}$, as is the case for the competitive ACSs discussed later in Sec.\,\ref{sec:azoarcus}) only the growth state with $\delta = 0$ (i.e., the symmetrical trajectory) is always stable. Thus, for the system to show bistability and heredity, $r(x)x$ must be a nonlinear function of $x$.

\subsection{Heredity of compositional state requires serial dilution interval to be below a critical threshold}

We next determine how bistability depends on the SD cycle interval $\varDelta t$ and the dilution factor $m$ ($=e^{\bar{\phi} \varDelta t})$. As an illustrative nonlinear reproduction law satisfying Eq.\,\ref{eq:condition}, we use \begin{equation}
r(x)x = \epsilon + \kappa x^2.
\end{equation} This function can arise from a ``spontaneous" or ``background" chemical reaction at rate $\epsilon$ combined with catalyzed reaction with the efficiency $\kappa$, for example through a dimeric catalyst~\cite{wagner2020programming}, or multi-step reactions such as those in the modified \emph{Azoarcus} system (discussed in Sec.\,\ref{sec:azoarcus}.).  However, the bounds derived below apply more generally when $r(x)$ is convex,  $\frac{d^2r}{dx^2}>0$.

Fig.\,\ref{fig:nullclines}B shows the bifurcation diagram of the concentrations just before a dilution in the stationary trajectory, as the cycle interval $\varDelta t$ in the SD protocol is varied while keeping the dilution rate $\bar{\phi}$ fixed (note that the dilution factor $m$($=e^{\bar{\phi} \varDelta t}$) is \emph{not} fixed).
The bifurcation occurs at $\varDelta t = \varDelta t_c^{sd}$. If $\varDelta t$ is more than this critical value, the system is no longer bistable, i.e., it does not exhibit heredity of the compositional state. 
Similarly, if we fix the interval $\varDelta t$ and vary the dilution rate $\bar{\phi}$, the same bifurcation at which the bistability disappears is observed at $\bar{\phi} = {\phi}_c^{sd}$ (Fig.\,\ref{fig:si_nullclines}A). The phase diagram of the parameters in the protocols, $\varDelta t$, and $\bar{\phi}$ is drawn in Fig.\,\ref{fig:nullclines}C.

The critical value $\varDelta t_c^{sd}$ depends on the reproduction rate function $r(x)$ and its kinetic parameters (Fig.\,\ref{fig:si_nullclines}B). 
Using Eq.\,\ref{eq:condition}, the critical $\varDelta t$ at which the system loses bistability, $\varDelta t_c^{sd}$, in a case with $r(x)x = \epsilon + \kappa x^2$ is derived as 
\begin{math}
    \varDelta t_c^{sd} \sim \frac{1}{\bar{\phi}} \log \left( \frac{\kappa}{4\epsilon} (s^{tot})^2 \right)
\end{math}
(Fig.\,\ref{fig:si_nullclines}C).
Intuitively, this form can be realized as the condition that the background reaction dominates the catalyzed reaction at the start of each cycle (just after each dilution), i.e., $\epsilon > \kappa (x^*)^2$, and $x^*$ is roughly $x^* \sim \frac{1}{2}\frac{s^{tot}}{m}$ if all of the substrate $\rm{S}$ added at the beginning of a cycle is converted to the $\rm{X}_1$ or $\rm{X}_2$ by the end of the cycle. 

The critical value for $\bar \phi^{sd}$ can be determined in a similar way. Especially, in the CSTR limit (i.e., $\varDelta t \rightarrow 0$), the condition Eq.\,\ref{eq:condition} becomes $\frac{dr}{dx}(\frac{\chi^*}{2}) > 0$, where $\chi^*$ is such that $\chi^* r(\frac{\chi^*}{2}) - \bar\phi =0$ (see Appendix Sec.\,\ref{sec:appendix-cstr}). Then, the critical dilution rate under CSTR, $\phi_c^{cstr}$ is the value of $\bar \phi$ at which this condition is violated. In a case with $r(x)x = \epsilon + \kappa x^2$, $\phi_c^{cstr}$ is derived as $\phi_c^{cstr} = 2 s^{tot} \sqrt{\epsilon \kappa} - 4 \epsilon$.

\subsection{Critical interval in a general growth and division process is bounded by that in the serial dilution protocol}
The ACSs can also exhibit bistability under general GD protocols with the dilution rate $\phi(t) = {\frac{dV}{dt}}/{V}$, cycle interval $\varDelta t$ and long-term dilution rate $\bar{\phi}$, similar to the SD protocol. Our result shows that the parameter region exhibiting bistability for the general GD protocol is bounded by that of SD and CSTR: if $r(x)$ is a convex function of $x$, i.e., $\frac{d^2r}{dx^2}>0$, all of the parameter regions ($\varDelta t$ and $\bar{\phi}$) where there is bistability under SD are included within the bistable region under the general dilution protocols, which in turn is included within the region exhibiting bistability under the CSTR protocol (see Appendix Sec.\,\ref{sec:proof} for the proof).

For example, consider the compartment growth dynamics obeying,
\begin{math}
    \frac{dV}{dt} = \bar \phi_\alpha V^\alpha, 
\end{math}
where $\alpha$ is the order of the growth, and $\bar \phi_\alpha$ is a constant depending on $\alpha$. Here, to compare sensibly across the different protocols, the growth rate of the compartment volume on long timescales in each case is assumed to be the same, i.e., SD with the dilution factor $m=e^{\bar{\phi}\varDelta t}$ or the CSTR with the dilution rate $\bar{\phi}$ (see Methods and Models for details).
As Fig.\,\ref{fig:nullclines}C shows, the critical interval $\varDelta t_c$ that is the upper limit for a system with bistability in the general case is bounded from below by the critical $\varDelta t^{sd}_c$ for the SD protocol:
\begin{math}
    \varDelta t_c^{sd} \leq \varDelta t_c  < \varDelta t_c^{cstr},
\end{math}
under the fixed $\bar{\phi}$, where $\varDelta t_c^{cstr}$ is infinite or otherwise zero ($\varDelta t_c^{cstr} = 0$ means that there is no bistability under any $\varDelta t$ values). 
On the other hand, if $\varDelta t$ is fixed, 
\begin{math}
    \phi_c^{sd} \leq \phi_c \leq \phi_c^{cstr},
\end{math}
where  $\phi_c^{sd}$, $\phi_c^{cstr}$ and $\phi_c$ are the thresholds for $\phi$ under SD, CSTR and general protocols, respectively. 

The above results are for systems in which all chemical reactions are irreversible; however, reversible reactions are more chemically realistic and allow convergence to thermal equilibrium in the absence of dilution protocols. Interestingly, we find that, unlike the irreversible reaction case, the region of bistability is bounded in the reversible case for the parameter $\bar\phi$. That is, for the SD protocol with fixed $\varDelta t$, there is both an upper and a lower critical $\bar\phi$ (see Fig.\,\ref{fig:rev-cdt} in Appendix). Importantly, even with reversible reactions, we found that the parameter space for the general GD protocol is inclusive for SD protocols, as observed in the case of irreversible reactions.\\

\subsection{Robustness of heredity to the introduction of variation and differential reproduction}
\subsubsection{Combining heredity with variation}\label{sec:variation}

\begin{figure}
    \centering
\includegraphics[width=7.8cm]{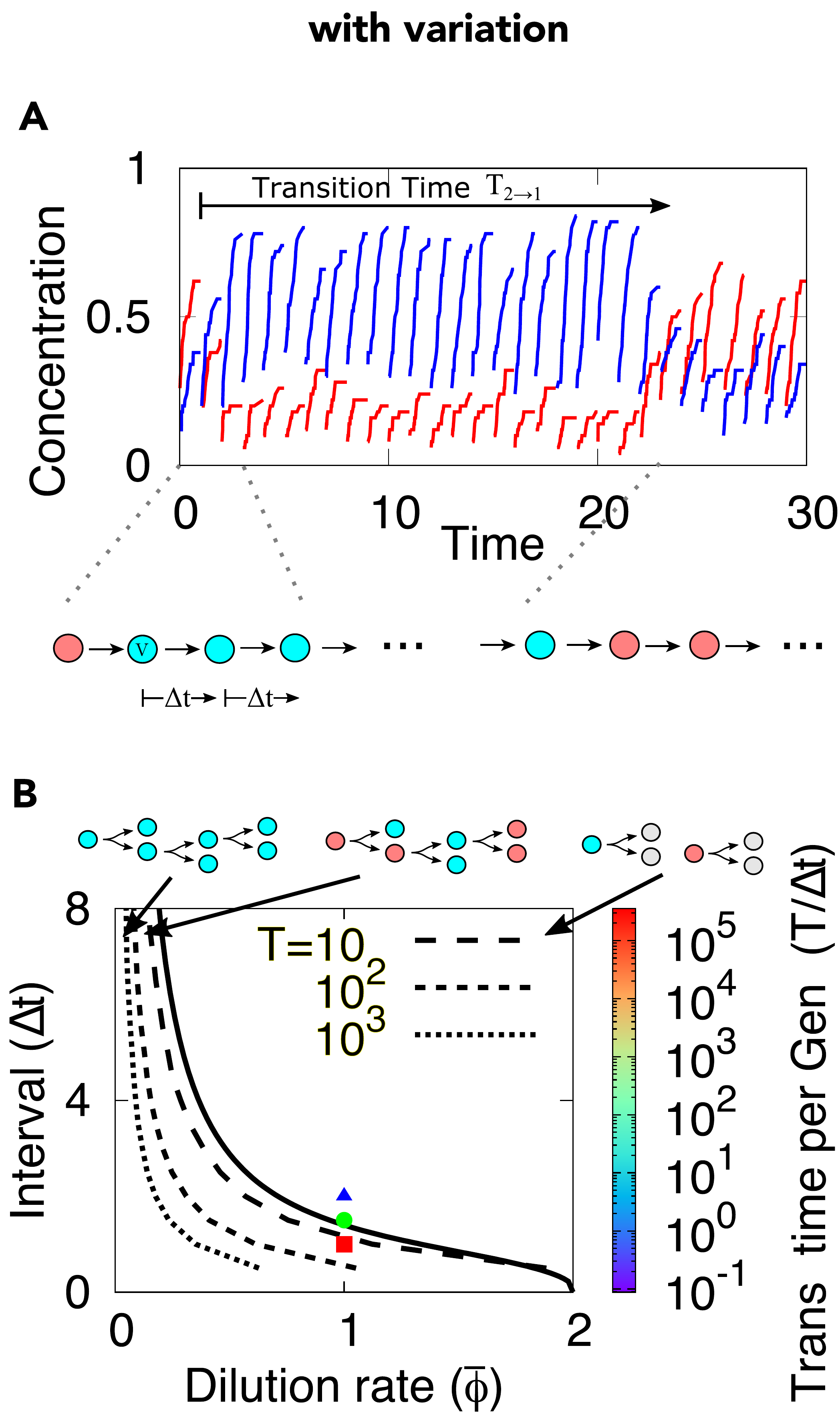}
    \caption{
    Effect of variation on compositional heredity. 
    (A) Time course of the concentrations of $\rm X_1$ and $\rm X_2$, below which the schematics of the SD protocol are depicted. Over the time course, transitions occur; we define the time between two transitions as $T_{2 \to 1}$. We set the parameters as $\bar \phi = 1$, $\varDelta t = 1$, $\kappa$ = 8, $\epsilon$ = 0.5, and $V=50$. (B) The color intensity represents the transition time divided by the interval ($T/\varDelta t$). The solid curve is the boundary between with and without bistability in a deterministic case (the same as shown in Fig.\,\ref{fig:nullclines}C). The dashed lines are the contours of $T/\varDelta t= 10$, $10^2$, and $10^3$.
    We set the parameters as $\kappa$ = 8, $\epsilon$ = 0.5, and $V=50$.
    }
    \label{fig:stochastic-reaction}
\end{figure}

In the deterministic systems we have examined so far, heredity comes without any variation in chemical composition. However, a number of sources of variation in reaction networks may exist. For instance, if the number of molecules is small (e.g., the reaction dynamics occur inside sufficiently small compartments), stochastic fluctuations are not negligible. Close to the deterministic bifurcation transitions, these fluctuations cause random transitions between states (Fig.\,\ref{fig:stochastic-reaction}A). If these transitions are rare, then they are a source of variation that does not destroy heredity; however, if they occur rapidly enough, they will destroy the information to be inherited (see Fig.\,\ref{fig:heredity}, \emph{bottom panel}). Thus, it is not surprising that we find that the parameter space where the system exhibits heredity is narrower than the deterministic case (Fig.\,\ref{fig:stochastic-reaction}B). But the critical point is that this regime does not shrink to zero; i.e., heredity can combine with variation through stochastic transitions between the two growth states. 

Moreover, our results on the existence of a critical threshold in the dilution interval or the dilution rate, and on the bistable parameter region for general GD cycles bounded by those of SD and CSTR are robust to the addition of stochastic noise in the chemical reaction system. Thus, although transitions between states are inevitable due to the presence of noise, below the critical thresholds previously computed for the deterministic system, the transition time rises very rapidly, as shown in Fig.\,\ref{fig:si_stochastic-reaction}. The transition time from state 1 to 2 is defined as the average number of growth and division cycles the system remains in state 1 before it transitions to state 2 (a similar transition time can be defined for the reverse transition). When the transition timescale is of order unity, i.e., identical to the growth-division timescale, the information about the current state is rapidly lost, and one can say that heredity does not exist (Fig.~1, \emph{bottom panel}). Conversely, if the transition times are much larger than unity, the heredity is robust to such stochasticity.

\subsubsection{Combining heredity with differential reproduction}

\begin{figure}
    \centering
    \includegraphics[width=8.0cm]{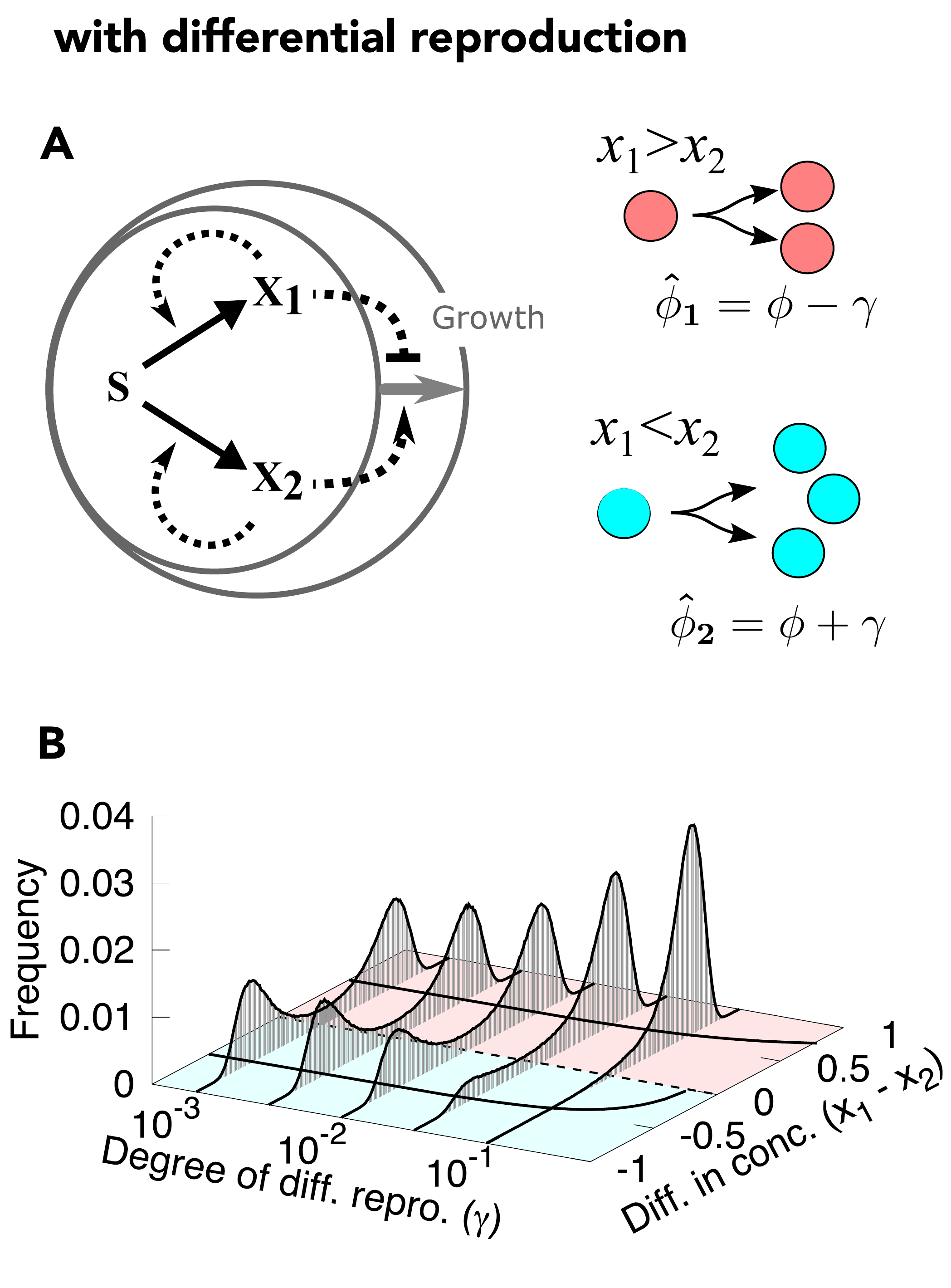}
    \caption{ 
    Effect of differential reproduction on compositional heredity.
    (A) \hl{Schematic of competing autocatalytic entities encapsulated by a growing compartment. The grey arrow represents the growth of the compartment, which is promoted (dotted arrow) or inhibited (bar-headed dotted arrow) by entities. The entities have symmetric catalytic strength
    , but affect the compartment growth asymmetrically (see Eq.\,\ref{eq:compo_dep_dilution}).} Schematic of competing autocatalytic entities under the serial dilution protocol with `differential reproduction’, i.e., the dilution factor depends on the state of the system, $x_1 > x_2$ or $x_1 < x_2$.
(B) Bifurcation diagram with varying the differential reproduction $\gamma$. The solid lines represent the steady states of $x_1 - x_2$ in the deterministic case. The figure also displays the probability density profile of $x_1-x_2$ at each $\gamma$ value, with finite system size case (stochastic case) $V=200$. We set $\varDelta t = 1$, $\bar \phi = 1$, $\epsilon = 0.5$, $\kappa = 8$. 
    }
    \label{fig:differential_fitness}
\end{figure}

The third property necessary for a Darwinian population is differential reproduction, upon which selection can act. In protocellular systems, the growth and division of compartments can depend on their internal chemical composition in a context-dependent manner. For example, autocatalytic molecules or peripheral species produced by them may influence compartment growth by synthesizing compartment precursors \cite{zwicker2017growth}, stabilizing its structure \cite{Jha2025vescileOuteq}, or modulating osmotic pressure \cite{Heng2024formose}. Alternatively, in laboratory settings, differential reproduction can be implemented in a controlled manner by imposing composition-dependent dilution rates in CSTR or SD.
We therefore investigated variations of our models in which dilution rates depend on chemical composition and the system exhibits different growth rates in the two growth states by considering the simple case that the system dilutes more slowly ($\hat \phi_1$) or faster ($\hat \phi_2$) if its state is 1 ($x_1>x_2$) or 2 ($x_1<x_2$) (see Fig.~\ref{fig:differential_fitness}A). Here, the dilution rate for state 1 or 2 is $\hat \phi_{1} = \bar \phi - \gamma$ or $\hat \phi_{2} = \bar \phi + \gamma$, where $\gamma$ is the degree of differential reproduction. In the case of SD, the system experiences dilution with either factor $m_1$ or $m_2$ ($m_i = e^{\hat \phi_i \varDelta t}$), depending on the state at the end of the cycle. Here, too much differential reproduction, i.e., high $\gamma$, leads to the disappearance of the faster state 2.

Moreover, the combination with the stochastic fluctuations further makes the maintenance of the heredity of the system difficult. As shown in Fig.\,\ref{fig:differential_fitness}B, the system remains more often at the state with the slower growth rate at the steady state probability distribution if the difference in the reproduction rate is larger. This is because the transition between states caused by stochastic fluctuation is far more likely from the state with a faster growth rate to the slower growth rate (see Fig.\,\ref{fig:stochastic-reaction}). This effect is even more substantial if the system size is larger (Fig.\,\ref{fig:si_differential-reproduction}). 
These results hold even when the growth and division protocol is used (see Appendix Sec.\,\ref{sec:robust_differential_rate} for details). By considering scenarios where $\phi$ depends symmetrically (Fig.\,\ref{fig:si_differential_fitness}A) or asymmetrically (Fig.\,\ref{fig:si_differential_fitness}C) on $x_1$ and $x_2$, we also examine the case where catalytic rates are asymmetric in addition. In both cases, the chemical composition exhibits bistability when the protocol interval is below the threshold (Fig.\,\ref{fig:si_differential_fitness}B and \ref{fig:si_differential_fitness}D), and the growth rates (reproduction rates) of the compartment in these states differ.

\subsection{Building a Darwinian population of autocatalytic protocells}

\begin{figure*}
    \centering
    \includegraphics[width=17.8cm]{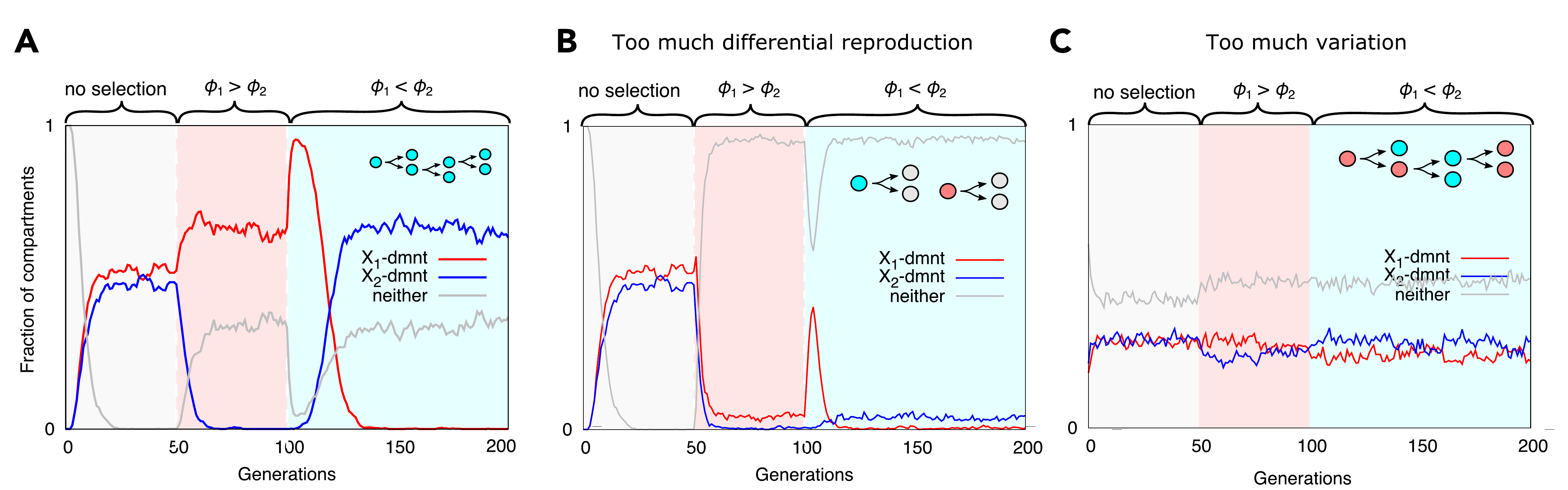}
\caption{
The population dynamics of compartmentalized autocatalytic sets with heredity, variation, and differential reproduction. 
(A) Fractions of $\rm X_1$-dominant, $\rm X_2$-dominant, and neither compartments through generations. Initially (from 0th to 50th generations), we set no selection (i.e., no differential reproduction), while from the 50th to 100th, or from 100th generations, we set $\rm X_1$- or $\rm X_2$-dominant to reproduce faster than the other, respectively. (B) If the differential reproduction is too large, significant selection is not observed.  (C) When the variation is too large, there is no significant selection, compared with a smaller variation case. We classify a compartment as $\rm X_1$ or $\rm X_2$-dominant, if $x_1-x_2 > 0.5$ or $< -0.5$, respectively. We set $\varDelta t = 1.1$, $\epsilon = 1/3$, $\kappa = 8$, the volume of each compartment $V=1000$, and the population size $N = 1000$, $\phi_i = \log(m_i)/ \varDelta t$, where $(m_1, m_2) = (3,3)$ $ (t <50)$, $(m_1, m_2) = (4,3)$ $(50 \leq t < 100)$ and $(m_1, m_2) = (3,4)$ $(100 \leq t)$. 
Under excessive differential reproduction, we use $(m_1, m_2) = (5,3)$ $(50 \leq t < 100)$ and $(m_1, m_2) = (3,5)$ $(100 \leq t)$. Under excessive variation, we set $V=10$.
}
  \label{fig:moran}
\end{figure*}

To examine whether an autocatalytic chemical system could form a Darwinian population of growing and dividing protocells, we next combine all three elements: heredity, variation, and differential reproduction. We consider  $N$-parallel lineages undergoing GD cycles; each has a volume $V$ containing the autocatalytic system described by Eq.\,\ref{eq:model} (see Fig.\,\ref{fig:si_WF_schematics} for the schematics and the details for Methods and Models). One could imagine a laboratory implementation of such a Darwinian population using a very large parallel realization of the SD cycles. 
Initially, all cells were given random chemical compositions. The population undergoes a Wright-Fisher-like process \cite{ewens2004mathematical}: after the interval $\varDelta t$, all test cells are divided into $m_1$ or $m_2$ cells, depending on the composition and the environmental conditions (i.e., the selection pressure). After the division, $N$ cells were randomly chosen to maintain the population size of $N$. On shorter timescales, the chemical reactions in each cell occur stochastically as in section \ref{sec:variation}.
The system was subjected to three regimes of selection pressure (see Fig.\,\ref{fig:moran}A):\\
1. Initially, no selection pressure was imposed; neither state is favored. The dilution factor $m$ for each cell is given by $m = \exp (\varDelta t \phi)$. The serial dilution cycle was run under these conditions until the population stabilized.\\
2. We then impose selection favoring state 1 by setting $\phi_1 > \phi_2$ (i.e., $m_1 > m_2$).\\
3. After the population stabilizes, we reverse the selection pressure so that state 2 is favored by setting $\phi_1 < \phi_2$ (i.e., $m_1 < m_2$).\\
In all cases, the fraction of protocells as a function of the number of cycles was plotted for state 1, 2, and neither of them, denoted as $f_1$, $f_2$ and $f_0$; defined arbitrarily as those where the $x_1 - x_2 > 0.5$ (and vice versa). In a Darwinian population, we expect to see: (i) in the case of no selection, $f_1\approx f_2\approx 0.5$, and there will be very few cells which are in neither state; (ii) when selection favors state 1, $f_1$ should rise rapidly and $f_2$ fall; (iii) when selection favors state 2, $f_2$ should recover to a high value, while $f_1$ should fall. This is indeed what we see, as shown in Fig.\,\ref{fig:moran}A.  The result in the parallelized serial dilution process is the same as the process of a Moran-like~\cite{ewens2004mathematical} population of $N$ growing protocells (Fig.\,\ref{fig:moran}A), which is discussed in Appendix Sec.\,\ref{sec:appdx-darwinian}. 

Conversely, if variation or differential reproduction destabilizes heredity, we should observe different behavior. As shown in Fig.\,\ref{fig:moran}C, where stochasticity is significant due to $V$ being sufficiently small, large fluctuations in $f_1$ and $f_2$ in all selection regimes are caused by the rapid transitions between the two growth states (for schematic, see Fig.\,\ref{fig:heredity}, \emph{bottom}). Thus, there is no evolution under selection, and the population cannot be called a Darwinian one. In contrast, Fig.\,\ref{fig:moran}B shows a different scenario, where the stochasticity is not too large, but the selection pressure being larger (due to too much differential fitness) leads to the system no longer being bistable but only having one stable state (because the dilution rate exceeds the critical value $\phi_c$). Thus, again, the population is not Darwinian, and there is no evolution under selection (for a schematic, see Fig.\,\ref{fig:heredity}, \emph{middle}).

\subsection{Application to experimental autocatalytic systems}
\label{sec:azoarcus}
Next, we assessed whether the identified parameter space for dilution interval and differential growth can be implemented in an established experimental system. Although several chemical systems can form ACSs \cite{ameta2021self}, they often exhibit poor differential growth, limited variation, and a scarcity of selection experiments. However, simulation with one of the RNA-based systems indicates that our general criteria for compositional heredity can be implemented in an experimental setting. We used RNA networks based on engineered \emph{Azoarcus} ribozymes \cite{vaidya2012spontaneous,yeates2016dynamics,ameta2021darwinian} (see Appendix Sec.\,\ref{sec:azo_result} for details). In these systems, ribozymes assemble from fragments and compete for a shared limiting substrate, allowing their chemical composition to be mapped onto the competing-ACS framework introduced earlier. The \emph{Azoarcus} system is a particularly intriguing experimental system for studying heredity because it can go beyond our present theoretical analysis in two ways: \\
\begin{enumerate}[label=(\roman*)]
\item It can be engineered to exhibit up to 48 compositional states \cite{vaidya2012spontaneous}. For example, by choosing different bases in recognition sites, such a system can encode more than one bit of information.
\footnote{Moreover, the \emph{Azoarcus} ribozyme can form cross-catalytic networks~\cite{vaidya2012spontaneous, ameta2021darwinian} (e.g., choosing bases M and N to be CC and GG, or AA and UU); in such cases, each network is a unit of self-reproduction (called an `autocatalytic core') and could compete with other units.}. \\
\item Each compositional state can be composed of a large number of chemical species. The \emph{Azoarcus} ribozyme can catalyze not only the formation of itself but also the production of diverse RNA sequences~\cite{jeancolas2021rna}. Here, self-reproducing ribozymes (ACSs) correspond to the `autocatalytic cores,' and the sequences produced by the ribozyme correspond to their `peripheries'~\cite{jain2002crashes, kaneko2005recursive, vasas2012evolution}. Species in the periphery can nevertheless play important roles, for instance, in the differential reproduction of the compositional states.
\end{enumerate}

For the parametrization here, the standard engineered \emph{Azoarcus} system, in which two self-catalyzing ribozymes compete for a common fragment, exhibits an effective reproduction law of the form $x\,r(x)=\epsilon+\kappa x$, arising from background assembly combined with linear self-catalysis. As predicted by our general analysis (Sec.~\ref{sec:acs-model}), such linear growth does not support bistability under SD; regardless of the initial conditions, trajectories converge to the symmetric compositional state. This absence of compositional heredity is consistent with experimental observations \cite{ameta2021darwinian}. Bistability can be generated by coupling ribozyme assembly to additional catabolic and anabolic processing steps, as realized experimentally in metabolically coupled \emph{Azoarcus} networks \cite{arsene2018coupled}. This coupling introduces higher-order nonlinearities in the effective growth law $x\,r(x)$, satisfying the criterion identified in Sec.~\ref{sec:acs-model} for the existence of multiple exponential growth states. 
Applying our analysis to a parameterization of the modified Azoarcus system, we predict that compositional heredity should be observable for serial-dilution intervals of 50–125 min and dilution factors of 2.5–11 per cycle (Figs.~\ref{fig:azoarcus}, \ref{fig:az_dt-m}). This range provides a concrete, experimentally testable regime for observing the inheritance of distinct ribozyme compositions.

Although the \emph{Azoarcus}-based RNA system is one of the few experimental systems capable of generating diverse multi-species ACS reaction networks \cite{ameta2021darwinian,ameta2021self}, the diversity is still limited. With the current fragmented system, up to 48 different catalytic variants \cite{vaidya2012spontaneous} can be used to generate thousands of reaction networks \cite{ameta2021darwinian}. However, further variation in the system can be by mutating the recognition elements (IGS-tag~\cite{vaidya2012spontaneous}) or by encapsulating them in small protocells (sub-femtoliter droplets) in a microfluidic set-up that introduces stochasticity in encapsulation~\cite{ameta2021darwinian}.

\section{Discussion}
In this work, we studied mathematical models of a very general class of chemical reaction systems in which two ACSs compete for a shared resource. When enclosed within growing and dividing compartments, such a system serves as a simple example that exhibits heredity in its compositional state and remains stable with respect to the growth and division of the compartments. This, along with differential reproduction rates and compositional-state variation, is a key property for a chemical system to form a Darwinian population.

We show that the bistable region for general GD protocols is bounded by the corresponding SD and CSTR limits. In particular, the SD boundary provides a conservative sufficient condition for heredity under more general GD dynamics. SD experiments therefore provide a practical means of testing whether an autocatalytic chemical system can preserve distinct compositional states across GD cycles.
Crucially, we also found that the inheritance of compositional information is robust to the introduction of both variation, in the form of noise, and mechanisms of differential reproduction. With all three elements—heredity, variation, and differential reproduction—working together, our study suggests a plausible parameter space for building a Darwinian population of growing and dividing protocells containing such autocatalytic systems. 
Not surprisingly, the introduction of noise does reduce the parameter regime under which heredity occurs, so an experimentalist aiming to build such a system must be careful to control noise or use large enough volumes that transitions from one compositional state to another do not occur too often, nor too rarely. Interestingly, \hl{too} strong selection pressure can sometimes destroy bistability and thereby eliminate the Darwinian population. In modern cells with template replication, increasing selection pressure does not convert a multistable system into a monostable one; however, autocatalytic chemical reaction networks are more susceptible to this.

Furthermore, extending our present theoretical analysis to multistable systems that can encode more than 1 bit of information, and to more complex autocatalytic systems consisting of cores and peripheries, is feasible. The \emph{Azoarcus} system can guide such theoretical extensions, but we expect our core results to remain the same for such more complex reaction systems, provided the current experimental system can enhance differential fitness among the variants.

The conditions for bistability and multistability in autocatalytic systems have been previously discussed in the context of self-reproduction and the origins of life. For example, Giri et al.~\cite{giri2012origin} found a class of ACSs that exhibit bistability under CSTR conditions, with one state growing and the other a non-growing state~\cite{matsubara2016optimal}. Remarkably, their network model also requires two steps to exhibit bistability (under mass-action kinetics), as does the \emph{Azoarcus} system coupled to metabolic reactions. Note that bistability in their model requires a high catalytic efficiency of self-catalysts (e.g., on the order of $10^4$). \footnote{This type of bistability, low and high catalyst concentration states, also appears in our model if the background reaction rate is small enough (nearly zero). However, such bistability is easily destroyed under the SD protocol unless the interval is very short}.
It is noteworthy that our model exhibits bistability even with relatively low catalyst efficiency (or, equivalently, a high rate of background reactions), which is more plausible in a prebiotic scenario. Note that some previous studies have shown that heredity of composition can arise even in the absence of bistability due to differences in reproduction rates and competition between compartments~\cite{segre2000compositional, kaneko2002kinetic, doulcier2020eco}. 
However, such mechanisms require fine-tuning of protocol parameters, such as compartment size, and it is not clear whether the compositional states can be sustained long enough for selection pressures to act on them~\cite{vasas2010lack}. In contrast, we demonstrated that compositional information in our models is robustly inherited within a single lineage of compartments across a wide range of kinetic constants and protocol parameters. 

As a compartment, lipid vesicles can be considered a protocell model, which has been well established for various functional studies~\cite{chen2004emergence, szostak2001synthesizing}. Furthermore, recent studies also envisage liquid-liquid phase-separated droplets (`coacervates'~\cite{oparin1953origin}) as a suitable compartment, as they have been shown to support various functions~\cite{drobot2018compartmentalised, ameta2023multispecies, chakraborty2025temperature} and are amenable to growth and division protocols~\cite{zwicker2017growth, taylor2017autonomous, matsuo2021proliferating, ianeselli2022non}. Even though coacervates are permeable, the inheritance of compositional information could be robust against unwanted reactions~\cite{ameta2023multispecies, singh2024constrained}. In addition, autocatalytic molecules can influence the formation and stability of LLPS droplets~\cite{soria2025primitive}, providing a direct physical link between chemical composition and compartment growth, division, and differential reproduction.

Still in the given parameter space, it is challenging to demonstrate `open-ended' evolution of ACS-containing protocells. As indicated by Vasas et al. \cite{vasas2012evolution}, this would likely require a chemical network comprising multiple autocatalytic cores that can arise stochastically over time via rare reactions, competing with existing cores. Our results would provide bounds on such a chemical network for maintaining heredity but do not provide additional information about which might be capable of such open-ended evolution.

\section{Methods and Models}
\subsection{Models and simulation details}
Deterministic chemical reaction dynamics were simulated by numerically integrating the ordinary differential equations and associated discrete maps using an adaptive Dormand–Prince Runge–Kutta method~\cite{press2007numerical} implemented in C++. Stochastic reaction dynamics were simulated using the exact Gillespie algorithm, as described below.

\paragraph{The serial dilution protocol:} The concentration vector of entities $\bm{x} = (x_1(t), x_2(t), ..., x_M(t) )$ under the serial dilution (SD) protocol evolves according to a rate equation,
\begin{equation} \label{eq:method_rate}
    \frac{d \bm{x}}{dt} = \bm{f}(\bm{x}),
\end{equation}
where $\bm{x}(t)$ is a time-dependent compositional vector (e.g., in a case with a model in Sec.\,\ref{sec:azo_result}, $\bm{x}(t)= \{s(t), x'_1(t), x'_2(t), x_1(t), x_2(t) \}$).
At each interval $t = n\varDelta t$ ($n = 1, 2, \dots$), all entities are diluted (and the substrate S is added), that is, the composition changes according to the discrete mapping:
\begin{equation} \label{eq:method_map}
    \bm{x}( n \varDelta t + 0) = \frac{1}{m} \bm{x}(n \varDelta t - 0) + s^{tot} (1-\frac{1}{m}) \hat{\bm{1}}_s, 
\end{equation}
where 
$\hat{\bm{1}}_s$ is a unit vector for the substrate S (e.g., in a case with a model in Sec.\,\ref{sec:azo_result}, $\hat{\bm{1}}_s = \{1,0,0,0,0\}$), $n = 0, 1, \dots$, and 
$n\varDelta t - 0$ and $n\varDelta t + 0$ represent the time right before and after the dilution at $t = n \varDelta t$ ($n = 0, 1, \dots$). 
If the reaction dynamics in Eq.\,\ref{eq:method_rate} do not change the total sum of composition, $x^{tot} = \sum_i x_i$, the repeats of the dilution cycle, i.e., the mapping in Eq.\,\ref{eq:method_map} results in the steady state with $x^{tot} = s^{tot}$. In the main text, we fix $x^{tot}$ as $s^{tot}$, and the concentration of free substrate is $s = s^{tot} - \sum x_i$ in the rate equations.

If we set $\varDelta t$ small enough, i.e., the system is diluted repeatedly at quite a short interval, and set $m$ and $s^{tot}$ as $m=e^{\bar\phi\varDelta t}$ and $s^{tot}=\frac{\bar\sigma}{\bar\phi}$, the dynamics and the steady-state of the species are the same as that in the continuous stirred-tank reactor (CSTR) with a dilution rate $\bar\phi$ and a substrate supply rate $\bar\sigma$~\cite{blokhuis2018reaction}.

\paragraph{The dilution by the growth of the compartment:} 
Formally, the rate equation for the chemical composition $\bm{x}$ under the general dilution protocols is expressed as,
\begin{equation}
    \frac{d \bm{x}}{dt} = \sigma(t) \hat{\bm{1}}_s + \bm{f}(\bm{x}) - \phi(t) \bm{x},
\end{equation}
where $\bm{f}(\bm{x})$ is an arbitrary reaction dynamics, and $\sigma(t)$ and 
$\phi(t)$ are the time-dependent supply rate of S and the 
dilution rate. If $\sigma(t) = s^{tot} \phi(t)$ and the dynamics $\bm{f}(\bm{x})$ conserves the total concentration $x^{tot}$, then $x^{tot}$ is constant at the steady state.

In the case of the compartment growth scenario, $\phi(t)$ is determined as $\phi(t)={\frac{dV}{dt}}/{V}$, where $V$ is the volume of the compartment. For example, we consider the power-law model of compartment growth. 
\begin{equation}
    \frac{dV}{dt} = \bar \phi_\alpha V^\alpha, 
    \label{eq:growth}
\end{equation}
where $\alpha$ is the order of the growth, and $\bar\phi_\alpha$ is a constant depending on $\alpha$. If $\alpha=1$, the growth is exponential. For example, if we consider the volume growth is proportional to the surface area, i.e., $\frac{dV}{dt}=r_\alpha S$, and if the vesicle is spherical, the surface area $S$ is  $S=V^{\frac{2}{3}}$, then $\alpha$ = 2/3~\cite{shirt2015emergent,ruiz2019dynamics}. The volume $V$ is solved as $V(t) = \left(\bar\phi_\alpha (1-\alpha) t + V_0^{1-\alpha}\right)^{\frac{1}{1-\alpha}}$. Here, we assume the growth speed of the compartment's volume in the long time scale is the same as the exponential growth with the rate $\bar{\phi}$ (i.e., CSTR with dilution rate $\bar{\phi}$); that is, $V(\varDelta t) = V_0 e^{\bar{\phi}\varDelta t}$. Then, $\bar \phi_\alpha$ should be $\bar \phi_\alpha = \frac{V_0^{1-\alpha}}{ (1-\alpha)\varDelta t} ( e^{\bar{\phi}(1-\alpha) \varDelta t} -1)$. Therefore, the dilution rate is
\begin{equation} 
    \phi(t) = \frac{e^{\bar{\phi}(1-\alpha) \varDelta t}-1} {(1-\alpha)\varDelta t \left(1+\frac{t}{\varDelta t} (e^{\bar{\phi}(1-\alpha) \varDelta t} -1) \right)}. 
    \label{eq:phi_t}
\end{equation}
Note that $\phi(t)$ approaches $\phi(t) = \bar{\phi}$ as $\alpha \rightarrow 1$ (i.e., the same as the condition under the CSTR). On the other hand, $\phi(t)$ approaches $\phi(t) = \infty$ if $t=n\varDelta t \quad (n=1,2, \dots)$ or $\phi(t)=0$ otherwise as $\alpha\rightarrow \infty$ (i.e., the serial dilution condition).

\paragraph{Nullclines under SD protocol:} 
The map between the chemical composition at the beginning (end) of one cycle to that at the beginning (end) of the next cycle, $P: \bm{x}( n\varDelta t + 0) \mapsto \bm{x}\big( (n+1) \varDelta t + 0\big),  n=0,1,..$, is interpreted as the Poincar\'e map obtained using the Poincar\'e section: $t = n \varDelta t$. The trajectory $\bm{x}(t)$ is stable if and only if the corresponding fixed point in the Poincar\'e map is  stable. Then, the stability of the trajectories can be determined from the intersections of the nullclines of the discrete map (see Fig.\,\ref{fig:nullclines}B for a depiction of these nullclines). 
The nullcline for $\rm X_1$, $x_1 = f(x_2)$, is obtained by fixing the concentration of $\rm X_2$ at the beginning of every cycle, $x_2(n\varDelta t+0) = x_2$, and calculating the stationary concentration of $\rm X_1$ at the beginning of cycles $x_1 = x_1^*(+0)$ by repeating the map enough times. 

\paragraph{Chemical reaction dynamics with stochasticity:}
When the compartment volume is small, stochastic fluctuations in chemical reaction dynamics become non-negligible \cite{gardiner1985handbook}. We therefore describe the autocatalytic reaction system shown in Fig.~\ref{fig:growth-division}A using a stochastic formulation based on discrete molecule numbers.

Let $n_i$ ($i=1,2$) denote the number of molecules of species $\mathrm{X}_i$ in a compartment of volume $V$, with concentrations $x_i=n_i/V$. The probability $P(\mathbf{n},t)$, with $\mathbf{n}=(n_1,n_2)$, obeys the chemical master equation
\begin{equation}
\frac{d}{dt}P(\mathbf{n},t)
= \sum_{i=1}^2 \Big[
P(\mathbf{n}-\mathbf{e}_i,t)\,\tau_i(\mathbf{n}-\mathbf{e}_i)
- P(\mathbf{n},t)\,\tau_i(\mathbf{n})
\Big],
\end{equation}
where $\mathbf{e}_i$ is the unit vector in the $i$-th direction and
$\tau_i(\mathbf{n})=s\,r(x_i)\,n_i$. Here we assume $sV = s^{tot}V - n_1 - n_2$, i.e., the total concentration $x^{tot}$ is fixed.

The stochastic dynamics were simulated using the exact Gillespie algorithm~\cite{gillespie1977exact}.
For the SD protocol, the volume $V$ is held fixed during each growth phase.
At dilution events separated by a fixed interval $\varDelta t$, each molecule is retained independently with probability $m^{-1}=e^{-\bar{\phi}\varDelta t}$, and molecule numbers immediately after dilution are sampled from a binomial distribution $ n_i(\varDelta t+0)\sim \text{Binomial}(n_i(\varDelta t-0),m^{-1})$.
If a reaction event was scheduled to occur later than the next dilution time $\varDelta t$, the reaction was discarded and dilution was applied at $t=\varDelta t$.

\paragraph{Wright--Fisher--like population dynamics:}
We consider a population of $N$ compartments undergoing synchronized SD cycles.
Each compartment evolves internally according to the stochastic chemical dynamics described above.

At the first generation, each compartment $j\in\{1,\dots,N\}$ is initialized with molecule numbers
$(n_1^{(j)}(0),n_2^{(j)}(0))$ drawn independently from a binomial distribution on $n_i^{(j)}(0)\sim \text{Binomial} (V/2, m^{-1})$.
During each generation, intracellular reaction dynamics in each
compartment evolve for a fixed duration $\Delta t$ according to the
stochastic reaction process, yielding molecule numbers
$(n_1^{(j)}(\Delta t-0),n_2^{(j)}(\Delta t-0))$ immediately before
division.

At division, each parental compartment $j$ is assigned a division
factor $m_j\in\{m_1,m_2\}$ according to its compositional state, determined by the sign of
$n_1^{(j)}-n_2^{(j)}$, at
$\Delta t-0$ and produces $m_j$ daughter compartments. The molecules
of each species are partitioned uniformly among these daughters.
Consequently, the molecule number of species $i$ in daughter $\ell$ of
parent $j$ is marginally distributed as
\begin{equation}
n_{i,\ell}^{(j)}(\Delta t+0)
\sim
\mathrm{Binomial}\!\left(
n_i^{(j)}(\Delta t-0),\,m_j^{-1}
\right),
\ell=1,\ldots,m_j.
\end{equation}

After division, all daughter compartments are combined to form the
post-division population. The next generation is formed by sampling
$N$ compartments uniformly without replacement at random from this population, thereby
maintaining a constant population size. This resampling step defines a
Wright--Fisher--like population process. The sampled compartments
provide the initial conditions for the next generation, and the
procedure is repeated.
\subsection{Experimental autocatalytic system based on \emph{Azoarcus} ribozyme}

We apply our framework to an experimentally realized ACS based on the \emph{Azoarcus} ribozyme~\cite{vaidya2012spontaneous, yeates2016dynamics, ameta2021darwinian}. We consider simplified models of this system, including competition between two distinct \emph{Azoarcus} ribozymes for shared resources.

\paragraph{Autocatalytic reaction scheme:}
The \emph{Azoarcus} ribozyme $\mathbf{WXYZ}$ is assembled from two fragments via
\begin{equation}
{}_{\mathrm{M}}\mathbf{WXY}_{\mathrm{N}} + \mathbf{Z}
\;\longrightarrow\;
{}_{\mathrm{M}}\mathbf{WXY}_{\mathrm{N}}\mathbf{Z},
\end{equation}
where ${}_{\mathrm{M}}\mathbf{WXY}_{\mathrm{N}}$ and $\mathbf{Z}$ denote RNA fragments, and $\mathrm{M},\mathrm{N}\in\{A,U,C,G\}$ specify the internal guide sequence and tag bases~\cite{yeates2016dynamics}. The reaction is catalyzed specifically by ribozymes with complementary $\mathrm{M}$ and $\mathrm{N}$ bases, while weak background reactions arise from non-covalent complexes and nonspecific catalysis~\cite{yeates2016dynamics}.

\paragraph{Competing ribozymes without metabolic coupling:}
We consider two self-catalyzing ribozymes $\mathrm{X}_1$ and $\mathrm{X}_2$ that compete for a common substrate $\mathrm{Z}$, assumed to be limiting, while the $\mathbf{WXY}$ fragments are abundant. Assuming symmetric kinetic parameters, the concentrations $x_1$ and $x_2$ obey the general model Eq.\,\ref{eq:model} with a linear reproduction term
\begin{equation}
r(x)x = \epsilon + \kappa x,
\end{equation}
where $\epsilon$ represents background reaction rates and $\kappa$ denotes the catalytic efficiency of the ribozymes.

\paragraph{Metabolically coupled Azoarcus system;}
To incorporate higher-order autocatalysis, we consider a modified \emph{Azoarcus} system coupled to additional catabolic and anabolic reactions~\cite{arsene2018coupled},
\begin{equation}
\begin{aligned}
{}_{\mathrm{M}}\mathbf{WXY}_{\mathrm{N}}\text{-mod} &\rightarrow {}_{\mathrm{M}}\mathbf{WXY}_{\mathrm{N}} + \text{-mod},\\
{}_{\mathrm{M}}\mathbf{WXY}_{\mathrm{N}} + \mathbf{Z} &\rightarrow {}_{\mathrm{M}}\mathbf{WXY}_{\mathrm{N}}\mathbf{Z},
\end{aligned}
\end{equation}
where $\text{-mod}$ denotes an additional sequence that must be processed before ribozyme assembly. We identify $\mathbf{Z}$ as substrate $\mathrm{S}$, $\mathbf{WXYZ}$ as ribozymes $\mathrm{X}_i$, and $\mathbf{WXY}$ as intermediates $\mathrm{X}'_i$ ($i=1,2$).

The concentrations of the chemical species evolve according to
\begin{equation}
\begin{aligned}
\frac{d x'_i}{dt} &= \left( \epsilon + \kappa x_i \right)
\left[ 1 - (s+b) x'_i + b x_i \right],\\
\frac{d x_i}{dt} &= \left( \epsilon + \kappa x_i \right)
\left( s x'_i - b x_i \right),
\end{aligned}
\end{equation}
where $\epsilon$ is the spontaneous reaction rate, $\kappa$ is the catalytic efficiency, $b\ll1$ is the backward reaction rate, and $s$ denotes the concentration of the shared substrate. The total concentration $s^{\mathrm{tot}} = s + x_1 + x_2$ is conserved.

\begin{acknowledgements}
We thank Sanjay Jain, Angad Yuvraj, and Nayan Chakraborty for the fruitful discussions. We thank Martin Falk for his comments on the final draft.
We acknowledge support from the Indo-French Centre for the Promotion of Advanced Research under project no. 5904-3, the Department of Atomic Energy (India) under project no.\,RTI4006, the Simons Foundation (Grant No. 287975), EU Horizon 2020 Grant ERC AbioEvo (101002075), France 2030 PEPR Origins ANR-22-EXOR-0013, and computational facilities at NCBS.
\end{acknowledgements}

\bibliography{ACSreviewBib_final,ref}

\clearpage
\widetext
\setcounter{equation}{0}
\def\theequation{S\arabic{equation}}
\setcounter{figure}{0}
\def\thefigure{S\arabic{figure}}
\appendix

\counterwithout{equation}{section}
\setcounter{equation}{0}
\renewcommand{\theequation}{S\arabic{equation}}

\section{On the definition of chemical Darwinian population}
\label{sec:appdx-definition}

According to the so-called NASA definition, life is defined as
a self-sustaining chemical system capable of undergoing Darwinian evolution.
Here we describe the definitions for \emph{chemical system} and \emph{Darwinian evolution} used throughout the present paper.

\paragraph{(self-sustaining) chemical system} There have been a number of proposals for the nature of the earliest self-reproducing entities, ranging from RNA~\cite{Woese1967, Crick1968} to clay~\cite{cairns1966origin} to X. Even within broadly supported frameworks like the RNA world~\cite{gilbert1986origin}, there are many possibilities -- some have advocated for collective autocatalytic sets~\cite{kauffman1986autocatalytic, jain1998autocatalytic, blokhuis2020universal, ameta2021self} while others search for the simplest RNA ribozyme that can copy itself~\cite{adamski2020self}. There are also multiple proposals for a compartment that separates a self-reproducing individual from its environment and other individuals~\cite{mizuuchi2021primitive}, ranging from lipid membranes~\cite{chen2004emergence} to coacervates~\cite{zwicker2017growth} to spatial separation on surfaces~\cite{boerlijst1991spiral, szabo2002silico}, or within hydrodynamic flows~\cite{krieger2020turbulent}.
Regardless of the details, one could describe the emergent population as consisting of autocatalytic chemical entities confined within compartments that grow and divide to produce new offspring individuals. We use ``autocatalytic chemical entities" and ``compartments" very broadly to encompass all the possibilities described above. 

\paragraph{Darwinian evolution} For such a system to be considered a population evolving under natural selection~\cite{charlat2021natural}, it must have certain properties. Following Godfrey-Smith~\cite{godfrey2007conditions}, who analyzes and builds on formulations by Lewontin~\cite{lewontin1970units}, Endler~\cite{endler2020natural}, Ridley~\cite{ridley2003evolution}, and others, we expect the individuals to exhibit: 
\begin{enumerate}
    \item Phenotypic variation
    \item Differential reproduction
    \item Inheritance of phenotypic traits (``heredity")
\end{enumerate}

\paragraph{Darwinian evolution in the context of (dynamical) chemical systems}
One of the simplest dynamical (chemical) systems that have the properties enumerated by Godfrey-Smith consists of a bistable chemical system, which exhibits two growth states with different chemical compositions and, in general, different growth rates. We define a growth state as one where the concentrations of the chemicals comprising the system grow without bound (often exponentially) as the system consumes food molecules, but where the chemical composition -- the relative concentrations of the chemical components -- reaches a steady state. Identifying the chemical composition to be the individual's phenotype accounts for phenotypic variation, and assuming the chemical composition affects the growth and division process accounts for differential reproduction. 
{This could occur via a variety of mechanisms.  
For example, osmotic pressure due to the difference in the composition between the inside and outside of the compartment may induce its growth~\cite{chen2004emergence}.
The ACS could also produce the precursors of the compartment (e.g., lipid molecules)~\cite{luisi1985enzymes, luisi1989self}. Another possibility is that the ACS energetically drives the growth, and shape instability triggers the division of compartments~\cite{zwicker2017growth}.}

In the language of dynamical systems, the third property of inheritance translates to the stability of the two growth states (hence our term `bistable chemical system') (see Fig.\,\ref{fig:heredity} in the main text). Here, by stability, we mean that the system does not spontaneously transition from one growth state to the other \emph{when it divides into two offspring entities}. 
{In dynamical systems, stability often refers to stability against stochasticity, e.g., due to thermal noise. It is true that for an autocatalytic chemical system to exhibit an inheritance of its phenotype (its chemical composition), it must also exhibit a certain amount of stability against noise. Some small enough probability of transitions to different states due to noise can be subsumed under phenotypic variation (indeed, this may be the only source of variation available), but too much will destroy the property of the heredity of states. Later we provide some results from stochastic simulations, but we largely assume that if the system, in the absence of noise, is stable upon division, then it satisfies the third property of inheritance of phenotypic traits.}

\paragraph{Multistability in chemical systems} has been extensively studied in the context of epigenetic memory in gene regulatory networks~\cite{gardner2000construction}, signaling pathways~\cite{ferrell2001bistability}, metabolic networks~\cite{peng2020ecological}, chiral symmetry breaking~\cite{frank1953spontaneous, laurent2021emergence}, or enzymatic cascades~\cite{maity2019chemically, wagner2020programming, schlogl1972chemical, giri2012origin, matsubara2016optimal}, etc. Generally speaking, positive feedback in the network structure is necessary though not sufficient for multistability~\cite{wilhelm2009smallest}, and more detailed conditions have also been suggested in specific contexts~\cite{craciun2006understanding}.
{The necessary conditions for bistable chemical reactions are: (i) positive feedback (e.g., autocatalysis), (ii) filtering noise, and (iii) preventing explosion (e.g., conservation law of the components)~\cite{craciun2006understanding}. In addition, `nonlinearity (or `ultrasensitivity') is required in the positive feedback~\cite{ferrell2001bistability}. Later, we discuss the minimum autocatalytic chemical system that satisfies the above condition}. However, these conditions for bi/multistability have been investigated mainly in chemostat or continuously-stirred-tank-reactor (CSTR) scenarios where there is a constant influx and outflux~\cite{blokhuis2018reaction}. It has not systematically been investigated under the conditions where these reaction systems are enclosed within a compartment that dynamically grows and divides.

The inheritance of the compositional information has been previously debated in a number of models of prebiotic autocatalytic networks~\cite{segre2000compositional,vasas2010lack, vasas2012evolution, hordijk2012structure, guttenberg2015transferable}. 
One suggestion has been that the network must contain multiple `viable autocatalytic cores'~\cite{vasas2012evolution, lancet2018systems} in order to exhibit heredity. 
However, it is unclear whether such mechanisms could be stable enough against stochastic noise or environmental fluctuations, let alone to the growth and division dynamics of compartments, which would be needed for the inheritance of information across generations. Interestingly, such autocatalytic cores are also one of the necessary conditions for multistability as discussed. We assumed two such cores (called `entities' more generally) competing for the same substrate in the model.

\section{Inheritable variety in general autocatalytic systems}
\label{sec:appdx-model}

\subsection{More rigorous arguments for the criteria Eq.3}
\label{sec:appdx-noosci}

We provide a more rigorous discussion of whether the criteria Eq.3 is sufficient for the bistability, in both the cases with CSTR and SD.

 First, we discuss the case with CSTR:
 \begin{enumerate}
     \item The system is bounded, i.e., $0 \leq x_1 + x_2 \leq s^0$. This follows because $\frac{dx_i}{dt}$ is negative when $x_1 + x_2 > s^0$ and positive when $x_1 + x_2 < 0$, as $r(x) \geq 0$.
     \item The system cannot exhibit oscillations and heteroclinic cycles: here, we follow the arguments in \cite{pigolotti2007oscillation}. First, the nullclines for $\frac{dx_1}{dt} = 0$ and $\frac{dx_2}{dt} = 0$ are single-valued functions of $x_1$ and $x_2$, respectively. (Note that this is the case for even reproduction rate functions with asymmetric catalytic strength or reversible reaction (Fig.\,\ref{fig:SImodel}B and C), while not for with inhibition (Fig.\,\ref{fig:SImodel}D).) These nullclines divide the $x_1-x_2$ plane into regions designated by the signs of $\frac{dx_1}{dt}$ and $\frac{dx_2}{dt}$, 
denoted as $(\sgn \frac{d x_1}{dt}, \sgn \frac{d x_2}{dt}) \equiv (+,+), (+,-), (-,+)$ and $(-,-)$.
Then, all the possible transitions between the areas are represented as 
\begin{equation} \label{eq:symbolic_dynamics}
    \begin{tikzcd}
    (-,+)   & \arrow[l, ""] (-,-) \arrow[d, ""] \\
    (+,+) \arrow[u, ""] \arrow[r, ""] & (+,-)
    \end{tikzcd}
\end{equation}
For example, the transition $(+,+) \rightarrow (-,+)$, i.e., from the area with $x_2 < f(x_1)$ to $x_2 > f(x_1)$ is possible, but its reverse direction is impossible.
 Consequently, the system ultimately reaches the $(-,+)$ or $(+,-)$ region and cannot display oscillatory dynamics.
 \end{enumerate}
Given that the system is bounded, and the absence of oscillations, a saddle fixed point (i.e., stable in one direction and unstable in another) is sufficient for the existence of multiple stable fixed points.

Next, we discuss Eq.\,1 in the main text under SD. 
Here, we assume the map from $\bm{x}(+0)$ to $\bm{x}(\varDelta t+0)$: $P: \bm{x}(+0) \mapsto \bm{x}(\varDelta t+0)(=\bm{x}( \varDelta t - 0 )m^{-1})$, where $+0$ refers to the beginning of a cycle, i.e., the time just after the dilution, and $\varDelta t -0$ refers to the end of a cycle, just before the dilution. 

\begin{enumerate}
    \item First of all, this map satisfies the monotonicity: 
we consider two trajectories $(x_1(t), x_2(t))$ and $(x'_1(t), x'_2(t))$. If $x_1(+0) < x_1'(+0)$ and $x_2(+0) \geq x'_2(+0)$ (or $x_1(+0) \leq x_1'(+0)$ and $x_2(+0) > x'_2(+0)$),
then $x_1(\tau) < x_1'(\tau)$ and $x_2(\tau) > x'_2(\tau)$ for all $\tau > 0$. This is because that $\frac{dx_1}{dt} < \frac{d x'_1}{dt}$ if $x_1 = x'_1$ and $x_2 > x'_2$, and $\frac{dx_2}{dt} > \frac{d x'_2}{dt}$ if $x_2 = x'_2$ and $x_1 < x'_1$.
\item This follows that the `nullcline' in the Poincar\'e section (see the definition for Methods and Models) for $x_1(+0) = x_1(\varDelta+0)$ is a single-valued function of $x_1$, the same as in the case of CSTR. That is, assuming ($x^*_1(t)$, $x^*_2(t)$) is a trajectory where ($x^*_1(+0)$, $x^*_2(+0)$) is a point on the nullcline (i.e., $x^*_1(+0) = x^*_1(\varDelta t+0)$), we consider the other trajectory ($x_1(t)$, $x_2(t)$) where $x_1(t) = x^*_1(t)$; if $x_2(+0) > x^*_2(+0)$ then $x_1(\varDelta t+0) < x^*_1(\varDelta t+0)(=x^*_1(+0))$, thus $x_1(\varDelta t+0) - x_1(+0) <0$, if otherwise $x_1(\varDelta t+0) - x_1(+0) > 0$.
\item Then, also similar to the case of CSTR, these nullclines divide the $x_1-x_2$ plane into regions designated by the signs of $x_1(\varDelta t +0) - x_1(+ 0)$ and $x_2(\varDelta t +0) - x_2(+ 0)$. Here also, the transition that crosses the nullcline for $x_1$ to the direction such as from $(+,-)$ to $(+,+)$ or $(-,-)$ is not allowed, even though the dynamics under the map $P$ in the Poincar\'e section is discrete in general. This is confirmed as follows: we consider a point ($x_1(+0)$, $x_2(+0)$) in the region designated by $(+,-)$. This point is transferred into ($x_1(\varDelta t+0)$, $x_2(\varDelta t+0)$) by the map $P$, where $x_1(+0) < x_1(\varDelta t+0)$ and $x_2(+0) > x_2(\varDelta t+0)$. Here, we assume that this point is in the region designated by $(-,-)$ (i.e., above the nullcline for $x_1$), then, further, we consider a point ($x^*_1(+0)$, $x^*_2(+0)$) on the nullcline for $x_1$, where $x_1(\varDelta t+0) = x^*_1(+0)$ and $x_2(\varDelta t+0) > x^*_2(+0)$. Since $x_1(+0) < x^*_1(+0)$ and $x_2(+0) > x^*_2(+0)$, it should be that $x_1(\varDelta t+0) < x^*_1(\varDelta t+0) = x^*_1(+0)$. However, this contradicts with $x_1(\varDelta t+0) = x^*_1(+0)$. Thus, the point ($x_1(\varDelta t+0)$, $x_2(\varDelta t+0)$) cannot be in $(-,-)$.
Therefore, the same as in the case of CSTR, the transition between the area is only allowed for (\ref{eq:symbolic_dynamics}), so oscillatory dynamics are not allowed.
\end{enumerate}

A trajectory of the dynamical system is stable if and only if a fixed point in the Poincar\'e section is stable.
This holds even under the existence of periodic force (i.e., time-dependent $\phi(t)$ in our case), since if we assume $\phi(t)$ as the third variable other than $x_1$ and $x_2$. (In a case with serial dilution, $\phi(t)$ has a singular point (i.e., discontinuous), but it holds if we approximate the delta function $\phi(t)$ by a continuous function.

\subsection{The sufficient condition under the serial dilution protocol} \label{sec:derivation}

We consider the competing autocatalytic entities, $\rm{X}_1$ and $\rm{X}_2$, under the SD protocol. The rate equation for the concentration for the entities, $x_1$ and $x_2$, during one cycle until the dilution, are Eq.\,1 in the main text,
\begin{equation}
    \frac{d{x}_i}{dt} = s(\{x_j\},t) r (x_i) x_i, 
\end{equation}
where $s(\{x_j\},t)$ is the concentration of the substrate S that is consumed in the replication reactions and $r(x_i)$ is the reproduction rate of $x_i$, respectively. Here we assume $s$ is symmetrical under exchange of $x_i$s, i.e., $s(x_1, x_2) = s(x_2, x_1)$ and $s$, $x_i$s satisfy some conservation law (e.g., $s+x_1+x_2 = s^{tot}$). Also, we assume $s$ and $r$ are differentiable and non-negative functions for $x_1 \geq 0$ and $x_2 \geq 0$. 

Here, we define
$\chi = x_1 + x_2$ and  $\delta = x_1 - x_2$, respectively. Then, the time derivative of them are 
\begin{equation}
    \frac{d \chi}{dt} = s r (\frac{\chi}{2}) \chi + \mathcal{O}(\delta^2),  \quad \frac{d \delta}{dt}  =  s \left (r (\frac{\chi}{2}) +\frac{1}{2} \chi  \frac{dr}{dx} (\frac{\chi}{2}) \right) \delta + \mathcal{O}(\delta^2),
\end{equation}
where we used the expansion $r \left( \frac{1}{2} (\chi \pm \delta) \right) = r (\frac{\chi}{2}) \pm \frac{dr}{dx} (\frac{\chi}{2}) \frac{\delta}{2} + \mathcal{O}\left( (\frac{\delta}{2})^2\right)$, and assumed $\delta$ is small compared with $\chi$, i.e., the concentrations of two catalysts are nearly equal, $x_1 \sim x_2$.
Then, 
\begin{equation}
    \frac{d}{dt} \bigl(\frac{\delta}{\chi}\Bigr) = \frac{1}{\chi^2} (\frac{d\delta}{dt}\chi - \frac{d\chi}{dt}\delta) = s \left( \frac{1}{2}  \frac{dr}{dx} \Bigl(\frac{\chi}{2}\Bigr) \chi \right) \frac{\delta}{\chi}.
\end{equation}
The integration of $\frac{d}{dt} (\frac{\delta}{\chi})/(\frac{\delta}{\chi})$ lead to
\begin{equation}
    \begin{aligned}
        \log \left| \frac{\delta(t)}{\chi(t)} \right| &=  \int_0^t  \frac{1}{2} s  \frac{dr}{dx} \Bigl(\frac{\chi}{2}\Bigr) \chi dt &+& \log \left| \frac{\delta(0)}{\chi(0)} \right| \\
        &=  \int_{\frac{\chi(0)}{2}}^{\frac{\chi(t)}{2}}    \frac{\frac{dr}{dx}(x)}{r(x)}  dx &+& \log \left| \frac{\delta(0)}{\chi(0)} \right|,
    \end{aligned}
\end{equation}
where we used $\frac{dt}{d\chi} = 1/ (  s r (\frac{\chi}{2}) \chi )$.
Therefore, 
\begin{equation}
    \frac{\delta(t)}{\chi(t)} =  \frac{ r(\frac{ \chi(t)}{2}) }{ r(\frac{\chi(0)}{2})} \frac{\delta(0)}{\chi(0)},
\end{equation}
which is Eq.\,2 in the main text.

Now, we consider the serial dilution protocol, i.e., the amounts of $x_1$ and $x_2$ are multiplied by $m^{-1}$ at the end of a cycle $t=\varDelta t$. In the stationary trajectory, $\chi(t)$ should satisfy the condition $\chi(\varDelta t) = m \chi(0)$. If the condition 
\begin{equation}
    \frac{\delta(\varDelta t)}{\chi(\varDelta t)} > \frac{\delta(0)}{\chi(0)}
\end{equation}
is met, 
the difference between $x_1$ and $x_2$, $\delta/\chi$ is magnified during a cycle. Therefore, the stationary trajectory with the equal concentration of $\rm{X}_1$ and $\rm{X}_2$ (i.e., $\delta = 0$) is unstable; if otherwise, the stationary trajectory is stable. 

Thus, surprisingly, whether the trajectory is stable or not is determined by only whether the replication rate at the end of a cycle $r(\frac{ \chi(\varDelta t)}{2})$ is larger than that at the beginning $r(\frac{ \chi(0)}{2})$ or not. Roughly, the replication rates at the beginning and the end are interpreted as the background and catalyzed reaction rates. 
For example, if $r(x) = \frac{\epsilon + \kappa x}{x}$ (the system based on the \textit{Azoarcus} ribozyme) the stationary trajectory with $\delta = 0$ is always stable. If $r(x) = \frac{(\epsilon + \kappa x)^2}{x}$, $\varDelta t_c$ is calculated as $\varDelta t_c = \log \left|  \frac{\epsilon + \kappa s^{tot}}{\epsilon + \kappa s^{tot} m^{-1}} \right| ^2 \sim 2 \log \left( 1 + \frac{\kappa s^{tot}}{\epsilon} \right)$.

\subsection{The sufficient condition for the heredity under the CSTR}
\label{sec:appendix-cstr}

Next, we consider the competing autocatalytic systems under the CSTR condition, where the constant dilution rate is $\bar \phi$. Similar as in the previous section, the rate equations for $x_1$ and $x_2$ are
\begin{equation}
    \frac{dx_i}{dt} = s(\{x_j\},t) r(x_i) x_i - \bar\phi x_i,
\end{equation}
where the notations are the same as in the previous section.

Here, the time derivative of $\chi$ and $\delta$ ($\chi = x_1 + x_2$ and $\delta = x_1 - x_2$) are
\begin{equation}
    \frac{d \chi}{dt} = s r \Bigl(\frac{\chi}{2}\Bigr) \chi - \bar \phi \chi + \mathcal{O}(\delta^2), \quad \frac{d \delta}{dt} = s \left( r \Bigl(\frac{\chi}{2}\Bigr) + \frac{1}{2} \chi \frac{dr}{dx} \Bigl(\frac{\chi}{2}\Bigr) \right) \delta - \bar \phi \delta + \mathcal{O}(\delta^2),
\end{equation}
where we used $r \left( \frac{1}{2} (\chi \pm \delta) \right) = r (\frac{\chi}{2}) \pm \frac{dr}{dx} (\frac{\chi}{2}) \frac{\delta}{2} + O\left( (\frac{\delta}{2})^2\right)$, and assumed $\delta$ is small.
The same calculation in the previous section leads to 
\begin{equation}
    \frac{d}{dt} \Bigl( \frac{\delta}{\chi} \Bigr) = \frac{1}{\chi^2} (\frac{d\delta}{dt}\chi - \frac{d\chi}{dt}\delta) = s \left( \frac{1}{2} \frac{dr}{dx} \Bigl(\frac{\chi}{2}\Bigr) \chi \right) \frac{\delta}{\chi} + \mathcal{O}( (\delta/\chi)^2 ).
\end{equation}
Thus, given the steady state concentration $\chi^*$ such that $s^* r(\frac{\chi^*}{2}) = \bar\phi$,  the state with $\delta = 0$ is  stable if $ \frac{dr}{dx}(\frac{\chi^*}{2})$ is negative. For example, if $r(x) = \frac{\epsilon+\kappa x}{x}$ (the \emph{Azoarcus} based system with only one step), $\frac{dr}{dx}(x) = -\epsilon/x^2$ is negative for all $x$, therefore the state with $\delta = 0$ is always stable. While in a case with $r(x) = \frac{\epsilon+\kappa x^2}{x}$, $\frac{dr}{dx}(x) = -\epsilon/x^2 + \kappa$. Therefore, if $\bar\phi > \phi_c^{cstr}$, where $\phi_c^{cstr} = 2s^{tot}\sqrt{\epsilon \kappa} - 4 \epsilon$, the state with $\delta = 0$ is stable.

Note that for the stability of the steady state $\chi^*$, the condition $-r (\frac{\chi^*}{2}) + \frac{1}{2} ( s^{tot} - \chi^*) \frac{dr}{dx}(\frac{\chi^*}{2}) < 0$ should be satisfied. The $\chi$-direction is always stable if $\delta$-direction is stable, i.e., $\frac{dr}{dx}(\frac{\chi^*}{2}) < 0$.

\subsection{Under the general dilution protocol (compartment growth and division)} \label{sec:proof}
Lastly, we discuss the heredity of the system under the general dilution scenarios:
\begin{equation}
    \frac{dx_i}{dt} = s(\{x_j\}, t) r(x_i) x_i - \phi(t) x_i,
\end{equation}
where $\phi(t)$ is the time-dependent dilution rate due to the growth of the compartment volume $V$, $\phi(t) = \frac{dV}{dt}/V$.
Here we restrict $\phi(t)$ to periodic functions $\phi(t+\varDelta t)=\phi(t)$ such that $\int_0^{\varDelta t} \phi(t) dt = \bar{\phi} \varDelta t$. Also, $\sigma(t) = s^{tot} \phi(t)$, thus the total mass of the substrate is kept as a constant, $s+\sum_i x_i = s^{tot}$.
We further assume that $r(x)$ is a convex function, i.e., $\frac{d^2 r}{d x^2}(x) > 0$. 

Under the above setup, the critical interval $\varDelta t_c$, which divides the region with and without heredity, is bounded by both that under the serial dilution and the CSTR:
\begin{enumerate}
    \item The period $\varDelta t_c$ for the protocol under any $\phi(t)$ is bounded from the above by that under the CSTR $\varDelta t_c^{cstr}$,  $\varDelta t_c \le \varDelta t_c^{cstr}$. Here, $\varDelta t_c^{cstr}$ is either $\infty$ or $0$.
    \item $\varDelta t_c$ is bounded from the bottom by that under the serial dilution $\varDelta t_c^{sd}$, $\varDelta t_c^{sd} \le \varDelta t_c$.
\end{enumerate}
To prove this, we use the dilution rate function $\phi(t)$
\begin{equation}
    \phi(t) \equiv 
    \begin{cases}
        \bar{\phi}_1 \quad (0 \le t \le \tau), \\
        \bar{\phi}_2 \quad (\tau \le t \le \varDelta t). \\
    \end{cases}
\end{equation}

\begin{figure}[H]
    \centering
    \includegraphics[width=14cm]{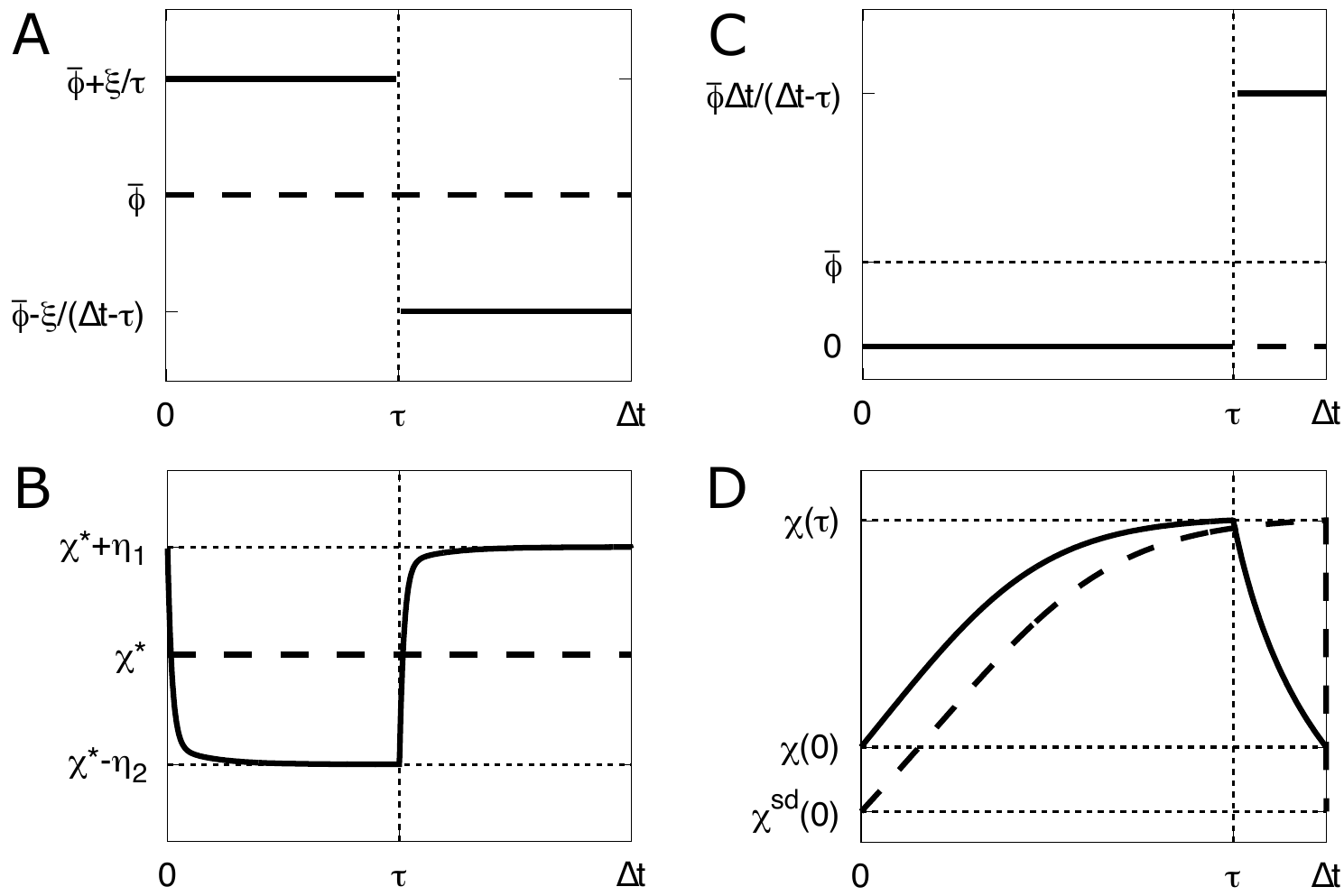}
    \caption{Schematics for the proof in Sec.\,\ref{sec:proof}. The left figures represent the time-dependent dilution rate in one dilution cycle, $\phi(t)$, and the right figures represent $\chi(t)$. The top figures are for the perturbation from the CSTR condition (The dashed lines represent the corresponding case under the CSTR condition). The bottom figures are for the perturbation from the SD condition (The dashed lines represent the corresponding SD). }
    \label{fig:for-proof}
\end{figure}

\paragraph{The perturbation from the CSTR condition:}
Firstly, we consider the perturbation from the CSTR protocol fixing $\int_0^{\varDelta t} \phi(t) = \bar{\phi}\varDelta t$, i.e., $\bar{\phi}_1 = \bar{\phi} + \frac{\xi}{\tau}$ and $\bar{\phi}_2 = \bar{\phi} - \frac{\xi}{\varDelta t - \tau}$ (see Fig.\,\ref{fig:for-proof}). Here, we define the steady-state concentration under the CSTR with the constant dilution rate $\bar \phi$,  $\chi^*$, is such that $(s^{tot}-\chi^*) r(\chi^*) \chi^* - \bar{\phi} = 0$. 
Here we consider a small perturbation for $\bar{\phi}$, $\bar{\phi} \rightarrow \bar{\phi}+\xi_i$, where $\xi_1=\frac{\xi}{\tau}$ and $\xi_2=- \frac{\xi}{\varDelta t - \tau}$, and the deviation for $\chi^*$ due to it, $\chi^* \rightarrow \chi^* + \eta_i$, which should satisfies
\begin{equation}
\begin{aligned}
         (s^{tot}-\chi^*-\eta_i) r\Bigl(\frac{\chi^*+\eta_i}{2}\Bigr) - (\bar{\phi} + \xi_i) = 0,\\
        (s^{tot}-\chi^*-\eta_i) \left( r\Bigl(\frac{\chi^*}{2}\Bigr) + \frac{\eta_i}{2} \frac{dr}{dx}\Bigl(\frac{\chi^*}{2}\Bigr) + \frac{\eta_i^2}{8} \frac{d^2r}{dx^2}\Bigl(\frac{\chi^*}{2}\Bigr)  + \mathcal{O}(\eta_i^3)\right) - (\bar{\phi} + \xi_i) = 0,\\
        \eta_i \left( - r\Bigl(\frac{\chi^*}{2}\Bigr) + (s^{tot}-\chi^*) \frac{1}{2} \frac{dr}{dx}\Bigl(\frac{\chi^*}{2}\Bigr) \right) +  \eta_i^2 \left( -\frac{1}{2} \frac{dr}{dx}\Bigl(\frac{\chi^*}{2}\Bigr) + (s^{tot}-\chi^*) \frac{1}{8} \frac{d^2r}{dx^2}\Bigl(\frac{\chi^*}{2}\Bigr) \right)  + \mathcal{O}(\eta_i^3) -  \xi_i = 0,\\
\end{aligned}
\end{equation}
where we used the expansion $r\bigl(\frac{\chi^*+\eta_i}{2}\bigr) = r\bigl(\frac{\chi^*}{2}\bigr) + \frac{\eta_i}{2} \frac{dr}{dx}\bigl(\frac{\chi^*}{2}\bigr) + \frac{\eta_i^2}{8} \frac{d^2r}{dx^2}\bigl(\frac{\chi^*}{2}\bigr)  + \mathcal{O}(\eta_i^3)$, and $(s^{tot}-\chi^*) r(\chi^*) - \bar{\phi} = 0$.
Then, 
$\eta_i$ is determined as 
\begin{equation}
    \eta_i = -\tilde \xi_i - \frac{1}{2} \mathcal{R} \tilde \xi_i^2 + \mathcal{O}(\xi^3),
\end{equation}
where we define 
\begin{equation}
    \begin{aligned}
        \tilde \xi_i = \frac{\xi_i}{r(\frac{\chi^*}{2}) - \frac{1}{2}(s^{tot}-\chi^*)\frac{dr}{dx}(\frac{\chi^*}{2})}, \quad
        \mathcal{R} =  \frac{  \frac{dr}{dx}\bigl(\frac{\chi^*}{2}\bigr) -  \frac{1}{4}  (s^{tot}-\chi^*)\frac{d^2r}{dx^2}\bigl(\frac{\chi^*}{2}\bigr)}{r\bigl(\frac{\chi^*}{2}\bigr) - \frac{1}{2} (s^{tot}-\chi^*) \frac{dr}{dx}\bigl( \frac{\chi^*}{2}\bigr)}.
    \end{aligned}
\end{equation}
Note that from the condition for the stability of the steady state $\chi^*$, $\frac{d}{dx} \left( (s^{tot} - x) r(\frac{x}{2}) - \bar\phi\right)|_{x=\chi^*} < 0$, $\frac{1}{2}(s^{tot}-\chi^*)\frac{dr}{dx}(\frac{\chi^*}{2}) - r(\frac{\chi^*}{2}) < 0$.

The same as the previous sections, the integration of $\frac{d}{dt} (\frac{\delta}{\chi})/(\frac{\delta}{\chi})$ from $t=0$ to $=\varDelta t$ leads to
\begin{equation}
    \begin{aligned}
                \log \left| \frac{\delta(\varDelta t)}{\chi(\varDelta t)} \right| - \log \left| \frac{\delta(0)}{\chi(0)} \right| =  \frac{1}{2} \int_0^{\varDelta t} (s^{tot} - \chi) \chi \frac{dr}{dx} \Bigl(\frac{\chi}{2}\Bigr) dt , \\
                \approx \tau (s^{tot} - \chi^* - \eta_1) \frac{\chi^*+\eta_1}{2} \frac{dr}{dx} \Bigl(\frac{\chi^*+\eta_1}{2}\Bigr) + (\varDelta t - \tau) (s^{tot} - \chi^* - \eta_2) \frac{\chi^*+\eta_2}{2} \frac{dr}{dx} \Bigl(\frac{\chi^*+\eta_2}{2}\Bigr),
    \end{aligned}
\end{equation}
where 
we assumed that  when $\phi$ is changed (at $t=0$ and $t=\tau$), the relaxation time to the steady concentration is much smaller than $\tau$ and $\varDelta t - \tau$ (see Fig.\,\ref{fig:for-proof}), whose contribution is of the order of $\xi^3$. Then, substituting Eq.\,S13 into the above, 
\begin{equation}
\begin{aligned}
     =& \varDelta t (s^{tot} - \chi^*) \frac{\chi^*}{2} \frac{dr}{dx} \Bigl(\frac{\chi^*}{2}\Bigr) \\
&+  (\tau \eta_1 + (\varDelta t - \tau) \eta_2) \left( ( \frac{1}{2} s^{tot} - \chi^*)  \frac{dr}{dx} \bigl(\frac{\chi^*}{2}\bigr) + (s^{tot} - \chi^*) \frac{\chi^*}{4}  \frac{d^2r}{dx^2} \bigl(\frac{\chi^*}{2}\bigr)  \right) \\
     &+  (\tau \eta_1^2 + (\varDelta t - \tau) \eta_2^2) \left((s^{tot} - 2\chi^*) \frac{1}{4} \frac{d^2r}{dx^2} \bigl(\frac{\chi^*}{2}\bigr) - \frac{1}{2} \frac{dr}{dx} \bigl(\frac{\chi^*}{2}\bigr) + (s^{tot} - \chi^*) \frac{\chi^*}{16} \frac{d^3r}{dx^3} \bigl(\frac{\chi^*}{2}\bigr) \right ) + \mathcal{O}(\xi^3),
      \\
           =& \varDelta t (s^{tot} - \chi^*) \frac{\chi^*}{2} \frac{dr}{dx} \Bigl(\frac{\chi^*}{2}\Bigr)  - \frac{1}{2} \xi^2 \Bigl(\frac{1}{\tau}+\frac{1}{\varDelta t - \tau}\Bigr) \Biggl[ \mathcal{R} \left( ( \frac{1}{2} s^{tot} - \chi^*)  \frac{dr}{dx} \bigl(\frac{\chi^*}{2}\bigr) + (s^{tot} - \chi^*) \frac{\chi^*}{4}  \frac{d^2r}{dx^2} \bigl(\frac{\chi^*}{2}\bigr)  \right)  \\
     &+  (\chi^* - \frac{1}{2} s^{tot}) \frac{d^2r}{dx^2} \bigl(\frac{\chi^*}{2}\bigr) + \frac{dr}{dx} \bigl(\frac{\chi^*}{2}\bigr) - (s^{tot} - \chi^*) \frac{\chi^*}{8} \frac{d^3r}{dx^3} \bigl(\frac{\chi^*}{2}\bigr) \Biggr] + \mathcal{O}(\xi^3),\\
\end{aligned}
\end{equation}
where the first term is the deviation under the CSTR with the dilution rate $\bar \phi$, $\log\left[\frac{\delta^{cstr}(\varDelta t)}{\chi^{cstr}(\varDelta t)}/\frac{\delta^{cstr}(0)}{\chi^{cstr}(0)}\right]$. 
When $\frac{dr}{dx} \bigl(\frac{\chi^*}{2}\bigr) = 0$,
\begin{equation}
    =  - \frac{1}{2} \xi^2 \Bigl(\frac{1}{\tau}+\frac{1}{\varDelta t - \tau}\Bigr) \left(  (\chi^* - \frac{1}{2} s^{tot}) \frac{d^2r}{dx^2} \bigl(\frac{\chi^*}{2}\bigr) - \frac{1}{16} (s^{tot} - \chi^*)^2 {\chi^*} \frac{\bigl(\frac{d^2r}{dx^2} \bigl(\frac{\chi^*}{2}\bigr) \bigr)^2}{r\bigl(\frac{\chi^*}{2}\bigr)} - \frac{1}{8} (s^{tot} - \chi^*) \chi^* \frac{d^3r}{dx^3} \bigl(\frac{\chi^*}{2}\bigr)  \right) + \mathcal{O}(\xi^3).
\end{equation}
Here, if we approximate that $\chi^*$ is nearly saturated, i.e., $\chi^* \sim s^{tot}$,
\begin{equation}
    \sim  - \frac{1}{4} \xi^2 \Bigl(\frac{1}{\tau}+\frac{1}{\varDelta t - \tau}\Bigr) s^{tot} \frac{d^2r}{dx^2} \bigl(\frac{\chi^*}{2}\bigr) + \mathcal{O}(\xi^3).
\end{equation}

Therefore, under the $\phi(t)$ with any choice of $\xi$ and $\tau$, $\frac{\delta(\varDelta t)}{\chi(\varDelta t)}/\frac{\delta(0)}{\chi(0)} < \frac{\delta^{cstr}(\varDelta t)}{\chi^{cstr}(\varDelta t)}/\frac{\delta^{cstr}(0)}{\chi^{cstr}(0)}$. Thus, if the stationary trajectory with $\delta = 0$ is stable under the CSTR, i.e., $\frac{\delta^{cstr}(\varDelta t)}{\chi^{cstr}(\varDelta t)}/\frac{\delta^{cstr}(0)}{\chi^{cstr}(0)} < 1$, it is also true for that under the protocol with the dilution rate $\phi(t)$, therefore, $\varDelta t_c \le \varDelta t_c^{cstr}$.

Further, we can divide the region for $\bar{\phi}_1$ or $\bar{\phi}_2$, and add more steps for the function $\phi(t)$. The same analysis as the above reveals that in each addition $\frac{\delta(\varDelta t)}{\chi(\varDelta t)}/\frac{\delta(0)}{\chi(0)}$ declines monotonically, thus the upper limit for heredity $\varDelta t_c$ also declines monotonically.

\paragraph{The perturbation from the serial dilution protocol:}
Secondly, we also consider the perturbation from the case with the serial dilution; here we consider $\bar{\phi}_1$ and $\bar{\phi}_2$ as $\bar{\phi}_1=0$ and $\bar{\phi}_2=\bar{\phi}\frac{\varDelta t}{\varDelta t-\tau}$ (see Fig.\,\ref{fig:for-proof}). Note that when $\varDelta t - \tau \rightarrow 0$, the dilution protocol becomes the same as the serial dilution. 

The deviation of $\frac{\delta}{\chi}$ in one cycle under the dilution protocol with $\phi(t)$,
\begin{equation}
    \begin{aligned}
             \log \left| \frac{\delta(\varDelta t)}{\chi(\varDelta t)} \right| - \log \left| \frac{\delta(0)}{\chi(0)} \right| =  \frac{1}{2}  
             \int_0^{\varDelta t}
             (s^{tot} - \chi) \chi \frac{dr}{dx} \Bigl(\frac{\chi}{2}\Bigr) dt, \\
     = \int_{\frac{\chi(0)}{2}}^{\frac{\chi(\tau)}{2}} \frac{\frac{dr}{dx}}{r} dx + \frac{1}{2} \int_\tau^{\varDelta t} (s^{tot} - \chi) \chi \frac{dr}{dx} \Bigl(\frac{\chi}{2}\Bigr) dt
    \end{aligned}
    \label{eq:delta_chi_phi}
\end{equation}
We assume $\chi^{sd}(\varDelta t) \sim \chi(\tau)$, because in the both cases $\chi(t)$ is saturated (i.e., S is exhausted) if $\varDelta t$ is enough large (see Fig.\,\ref{fig:for-proof}). Then, $\chi^{sd}(0) = \chi(\tau)\exp(-\bar{\phi}\varDelta t)$.
From the rate equation $\frac{d\chi}{dt} = \left( (s^{tot}-\chi) r\bigl(\frac{\chi}{2}\bigr) - \bar{\phi} \right) \chi$
\begin{equation}
    \begin{aligned}
        \chi(\varDelta t) = \chi(0) = \chi(\tau) \exp\left( -\varDelta t \bar{\phi} + \int_\tau^{\varDelta t} (s^{tot}-\chi) r\Bigl(\frac{\chi}{2}\Bigr) dt \right),\\
        = \chi^{sd}(0) \exp\left( \int_\tau^{\varDelta t} (s^{tot}-\chi) r\Bigl(\frac{\chi}{2}\Bigr) dt \right).
    \end{aligned}
\end{equation}

On the other hand, the deviation of $\frac{\delta^{sd}}{\chi^{sd}}$ in one cycle under SD is derived as
\begin{equation}
    \log \left| \frac{\delta^{sd}(\varDelta t)}{\chi^{sd}(\varDelta t)} \right| - \log \left| \frac{\delta^{sd}(0)}{\chi^{sd}(0)} \right| = 
    \int_{\frac{\chi^{sd}(0)}{2}}^{\frac{\chi(0)}{2}}\frac{\frac{dr}{dx}}{r} dx + \int_{\frac{\chi(0)}{2}}^{\frac{\chi(\tau)}{2}}\frac{\frac{dr}{dx}}{r} dx, 
        \label{eq:delta_chi_phi2}
\end{equation}
Here, the first term is,
\begin{equation}
    \int_{\frac{\chi^{sd}(0)}{2}}^{\frac{\chi(0)}{2}} \frac{\frac{dr}{dx}}{r} dx = \log \left| \frac{r\Bigl(\frac{\chi(0)}{2}\Bigr)}{r\Bigl(\frac{\chi^{sd}(0)}{2}\Bigr)} \right| = \frac{\chi^{sd}(0)}{2} \frac{\frac{dr}{dx}\Bigl(\frac{\chi^{sd}(0)}{2}\Bigr)}{r\Bigl(\frac{\chi^{sd}(0)}{2}\Bigr)} \int_\tau^{\varDelta t} (s^{tot}-\chi) r\Bigl(\frac{\chi}{2}\Bigr) dt + \mathcal{O}( (\varDelta t - \tau)^2 ),
\end{equation}
where in the approximation, we used the assumption that $\varDelta t - \tau$ is small.

From the comparison between Eq.\,\ref{eq:delta_chi_phi2} and Eq.\,\ref{eq:delta_chi_phi}, 
\begin{equation}
\begin{aligned}
    \log \left| \frac{\delta(\varDelta t)}{\chi(\varDelta t)} \right| - \log \left| \frac{\delta(0)}{\chi(0)} \right|  - \left(\log \left| \frac{\delta^{sd}(\varDelta t)}{\chi^{sd}(\varDelta t)} \right| - \log \left| \frac{\delta^{sd}(0)}{\chi^{sd}(0)} \right| \right) \\ = \frac{1}{2} \int_{\tau}^{\varDelta t} (s^{tot} - \chi(t')) \left[ \chi(t') \frac{dr}{dx} \Bigl(\frac{\chi(t')}{2}\Bigr) - \frac{r\Bigl(\frac{\chi(t')}{2}\Bigr) \chi^{sd}(0) \frac{dr}{dx}\Bigl(\frac{\chi^{sd}(0)}{2}\Bigr)}{r \Bigl(\frac{\chi^{sd}(0)}{2}\Bigr)} \right] dt'.
    \end{aligned}
\end{equation}
Thus, if $\frac{\chi(t) \frac{dr}{dx} \bigl(\frac{\chi(t)}{2}\bigr)}{r\bigl(\frac{\chi(t)}{2}\bigr)} > \frac{ \chi^{sd}(0) \frac{dr}{dx}\bigl(\frac{\chi^{sd}(0)}{2}\bigr)}{r \bigl(\frac{\chi^{sd}(0)}{2}\bigr)}$ is satisfied, the above is always positive. As $\chi(t) > \chi^{sd}(0)$, it 
reveals that $\frac{\delta(\varDelta t)}{\chi(\varDelta t)}/\frac{\delta(0)}{\chi(0)} > \frac{\delta^{sd}(\varDelta t)}{\chi^{sd}(\varDelta t)}/\frac{\delta^{sd}(0)}{\chi^{sd}(0)}$ $\frac{\delta(\varDelta t)}{\chi(\varDelta t)}$ for the serial dilution is always less than that for the protocol with the dilution rate function $\phi(t)$, if $\frac{x \frac{dr}{dx} (x)} {r (x)}$ is an increasing function of $x$. Therefore, the critical period of time for $\phi(t)$, $\varDelta t_c$ is always larger than that for the serial dilution, $\varDelta t^{sd}_c$, i.e., $\varDelta t^{sd}_c \le \varDelta t_c$.

\begin{figure}
    \centering
    \includegraphics[width=18cm]{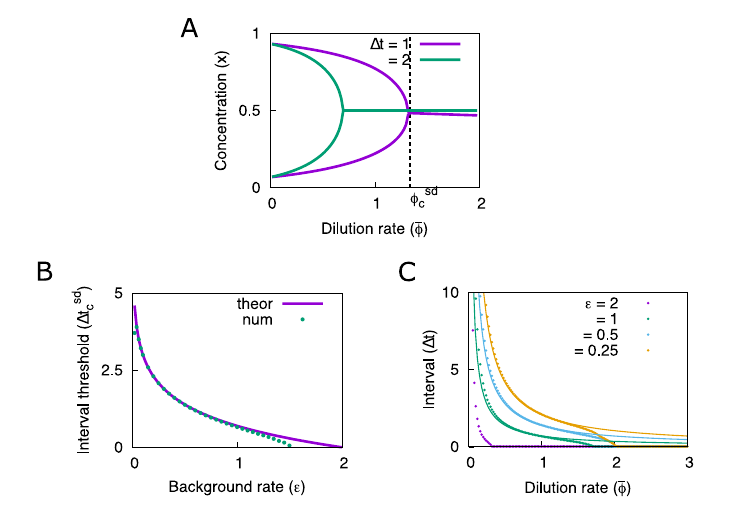}
    \caption{ (A) Bifurcation diagram with a varying long-term dilution rate $\bar{\phi}$.  
    (B) The dotted line represents the critical point $\varDelta t_c^{sd}$, which divides the regions where the system has bistability or not.
    The solid line represents the theoretical line for $\varDelta t_c^{sd}$ determined by the relation:
    \begin{math}
        \varDelta t_c^{sd} \sim \frac{1}{\bar{\phi}} \log \left( \frac{\kappa}{4\epsilon} (s^{tot})^2 \right).
    \end{math}
    (C) The dotted line represents the critical point $\varDelta t_c^{sd}$, which divides the regions where the system has bistability or not. Each colored dotted line represents the difference in the background reaction rate $\epsilon$. The theoretical lines (solid) are the same as in (B).
    We set the default parameters as $\epsilon = 0.5$, $\kappa = 8$ and $\bar\phi = 1$.}
    \label{fig:si_nullclines}
\end{figure}

\section{Robustness of results}
\label{sec:appdx-robust}

\begin{figure}
    \centering
    \includegraphics[width=14cm]{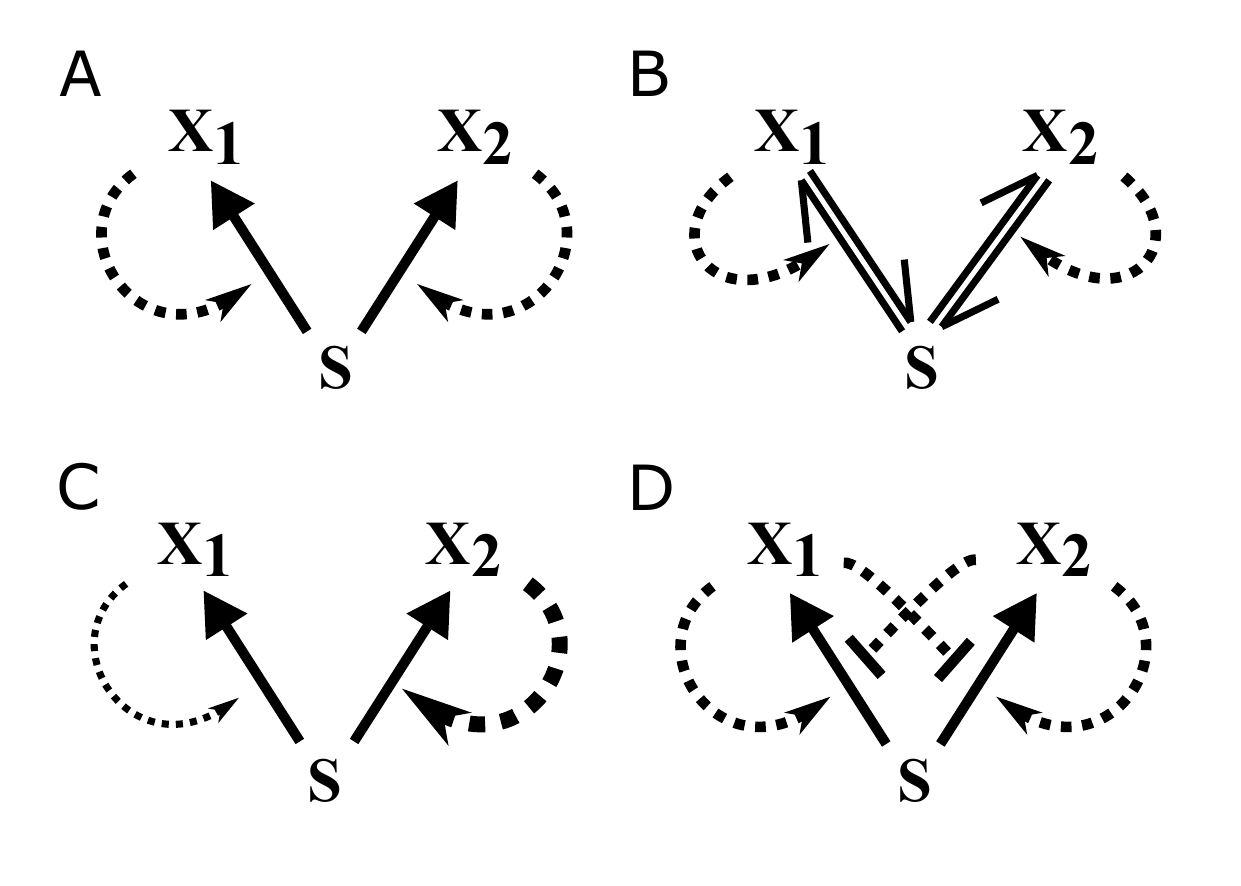}
    \caption{Schematics of the alternative models for the competing autocatalytic entities, $\rm{X}_1$ and $\rm{X}_2$; they are converted from a substrate S (solid arrows) catalyzed by itself (dashed arrows). (A) non-modified model (the same as Fig.\,\ref{fig:growth-division}A in the main text) (B) with reversible chemical reactions (Sec.\,\ref{sec:robust_reversible}) (C) with asymmetric kinetics rates (Sec.\,\ref{sec:asymmetric}) (D) with more general reproduction rate function (dashed arrow with bar head represents the inhibition; Sec.\,\ref{sec:robust_rate_function})}
    \label{fig:SImodel}
\end{figure}

\subsection{Autocatalytic sets with reversible chemical reactions} \label{sec:robust_reversible}

In the main text, we mainly considered chemical reaction systems with irreversible reactions. However, all reactions should be reversible to be chemically consistent and converge to thermal equilibrium in the absence of dilution protocols.
Here, we expand our previous analysis to examine how reversible reactions change the phase diagram of bistability. 

If the reactions in the system in Fig.\,\ref{fig:growth-division}A is reversible, the rate equations are
\begin{equation}
    \frac{d{x}_i}{dt} = (s - b x_i) r (x_i) x_i,
\end{equation}
where $b$ is the relative rate of the backward reaction (see Fig.\,\ref{fig:SImodel}B).

Then, $\frac{d\chi}{dt}$ and $\frac{d\delta}{dt}$ are modified as
\begin{equation} \label{eq:rev-model}
    \begin{aligned}
        \frac{d\chi}{dt} =& ( s - \frac{b}{2} \chi) r (\frac{\chi}{2}) \chi &+& \mathcal{O}(\delta^2),\\
        \frac{d\delta}{dt} =& s \left (r (\frac{\chi}{2}) +\frac{1}{2} \chi  \frac{dr}{dx} (\frac{\chi}{2}) \right) \delta - b \left( r (\frac{\chi}{2}) + \frac{1}{4} \chi \frac{dr}{dx} (\frac{\chi}{2})   \right) \chi \delta &+& \mathcal{O}(\delta^2).
    \end{aligned}
\end{equation}
Then,
\begin{equation}
    \frac{d}{dt} \bigl(\frac{\delta}{\chi}\Bigr) = \left[ ( s - \frac{b}{2} \chi) (  \frac{1}{2}  \frac{dr}{dx} \Bigl(\frac{\chi}{2}\Bigr) \chi )  - \frac{1}{2} b ( r (\frac{\chi}{2}) \chi ) \right] \frac{\delta}{\chi}.
\end{equation}

As a result,
\begin{equation}
        \log \left| \frac{\delta(t)}{\chi(t)} \right| = \left[ \log(r(x)) + \frac{b}{2+b} \log (s^{tot} - (2+ b) x) \right]_{x=\frac{\chi(0)}{2}}^{x=\frac{\chi(t)}{2}}
        + \log \left| \frac{\delta(0)}{\chi(0)} \right|,
\end{equation}
where we used $s = s^{tot} - \chi$. Thus, the reversible reactions alter Eq.\,2 in the main text as
\begin{equation}
    \frac{\delta(t)}{\chi(t)} =  \frac{ r(\frac{ \chi(t)}{2}) }{ r(\frac{\chi(0)}{2})} \left( \frac{s^{tot}-(1+\frac{b}{2})\chi( t)}{s^{tot}-(1+\frac{b}{2})\chi(0)} \right)^{\frac{b}{2+b}} \frac{\delta(0)}{\chi(0)}.
\end{equation}
Note that, from Eq.\,\ref{eq:rev-model}, $s^{tot}-(1+\frac{b}{2})\chi(\infty) = 0$ at the equilibrium, thus $\frac{\delta(t)}{\chi(t)}$ eventually converges to 0.

\begin{figure}
    \centering
    \includegraphics[width=16cm]{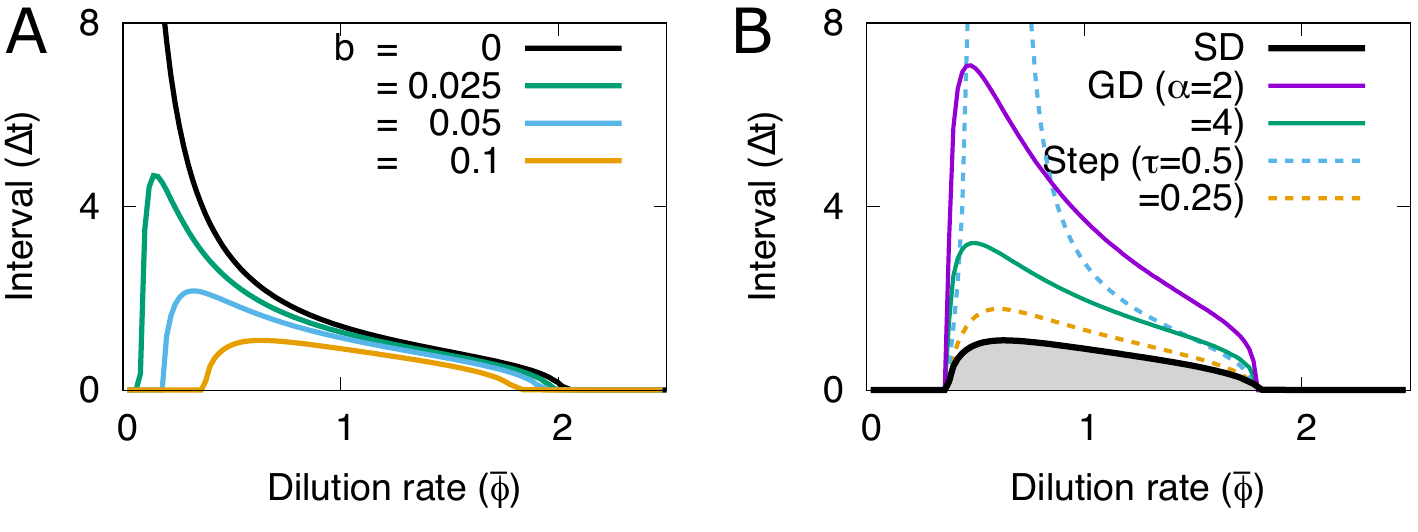} 
    \caption{Phase space of the bistability in the case with reversible reactions (Eq.\,\ref{eq:rev-model}). (A) The colored lines represent the boundary between with or without bistability in different backward reaction rates $b$. The other parameters are set as $\epsilon = 0.5$, $\kappa = 8$ and $\bar \phi =1$. (B) The boundary between regimes with and without bistability for different dilution protocols $\phi(t)$. 
GD denotes the growth--division protocol defined by Eq.~\ref{eq:phi_t} (Methods and Models) for different values of $\alpha$. 
Step denotes a pulsed dilution protocol given by $\phi(t)=\bar{\phi}\,\frac{\Delta t}{\tau}$ for $0 \le t \le \tau$ and $\phi(t)=0$ for $\tau < t \le \Delta t$. We set $b=0.1$.}
    \label{fig:rev-cdt}
\end{figure}

Similar to the case with irreversible reaction ($b=0$) in the main text, the system shows bifurcation that the bistability disappears when varying the cycle interval $\varDelta t$ or the dilution rate $\bar\phi$. 
Interestingly, unlike the irreversible reaction case ($b=0$), the region of bistability is bounded in the reversible case ($b>0$) for the parameter $\bar\phi$. Thus, for the SD protocol with  fixed $\varDelta t$, there is both an upper and a lower critical $\bar\phi$ (see Fig.\,\ref{fig:rev-cdt}). Note that the boundary of the region is close to the irreversible case when $\bar\phi$ is large and the critical value depends on $\epsilon$ similarly to the $b=0$ case. Importantly, even with reversible reactions, the parameter space for the general GD protocol contains that for SD protocols, as we found for irreversible reactions.

\subsection{ACSs with asymmetric kinetic rates}\label{sec:asymmetric}

In the previous sections, we considered symmetric competing ACSs. 
However, our results are similar even if the two competing ACSs have different kinetic rate constants (catalytic efficiency), although the bifurcation where the bistability disappears is discontinuous.

\begin{figure}
    \centering
    \includegraphics[width=16cm]{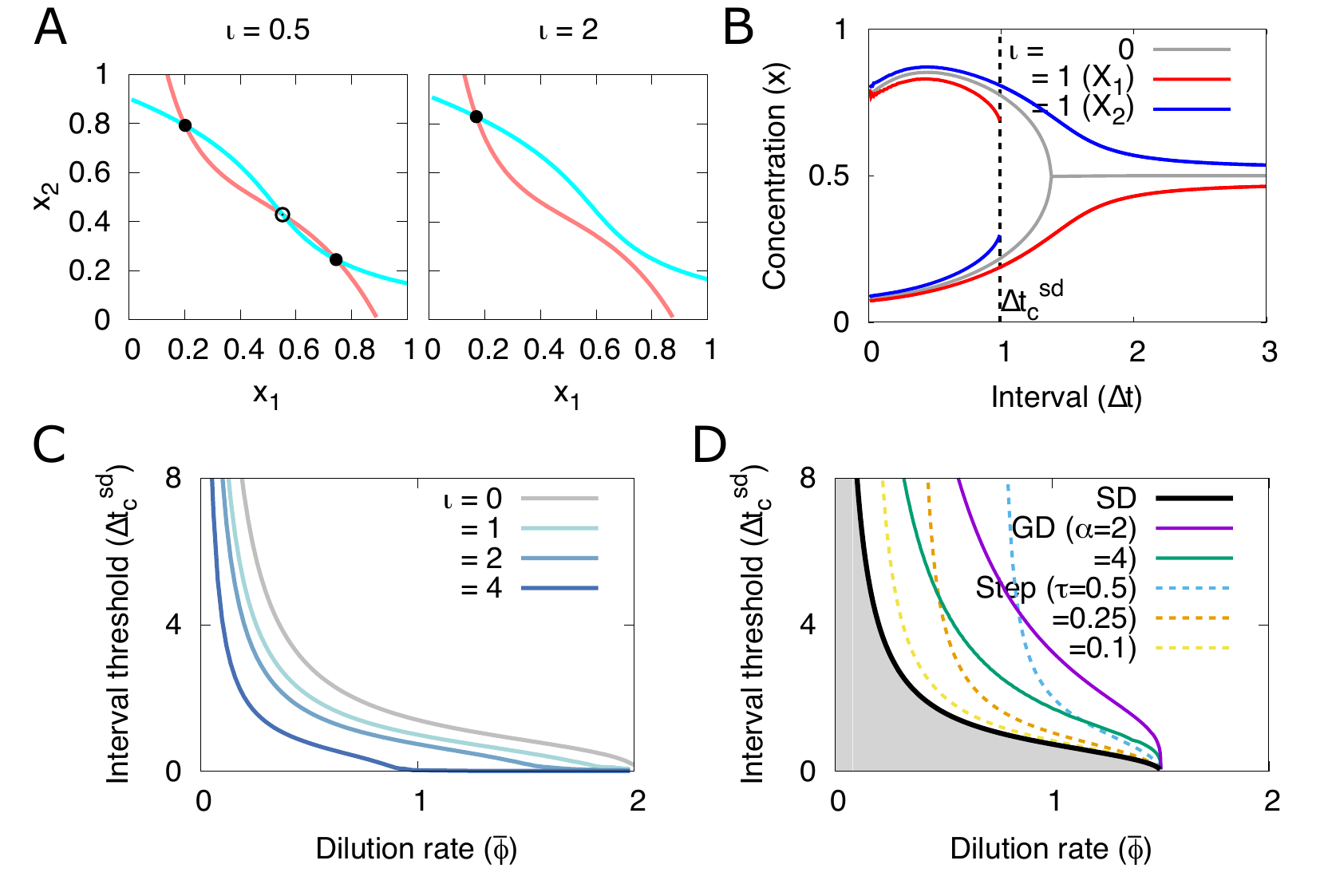}
    \caption{
    (A) The red and blue curves represent the nullclines for the concentrations of the entities $\rm{X}_1$ and $\rm{X}_2$ at just before the dilution, $x_1(-0)$ and $x_2(-0)$, respectively, for asymmetric reproduction-rate functions, $r_i(x)x = \epsilon + \kappa_i x^2$ ($i=1,2$) and $\kappa_1 < \kappa_2$. The intersection points represent the stable/unstable fixed points of concentrations of $\rm{X}_1$ and $\rm{X}_2$ at just before the dilution, $x_1^*(-0)$ and $x_2^*(-0)$. See Methods and Models for details of the drawing of the nullclines. We define  $\kappa_1=\kappa - \frac{\iota}{2}$ and $\kappa_2=\kappa + \frac{\iota}{2}$, and set $\varDelta t = 1$, $\kappa = 8$, $\epsilon = 0.5$, $\bar{\phi} = 1$, and $\kappa_2-\kappa_1=\iota=0.5$ (left) and $2$ (right). (B)  Bifurcation diagram of the concentrations of the entities $\rm{X}_1$ and $\rm{X}_2$ with a varying period of the dilution cycles $\varDelta t$. In contrast with the symmetric case (i.e., $\iota = 0$), the $\rm{X}_1$-dominant state (and thus the bistability) disappears discontinuously at around $\varDelta t_c^{sd} \sim 1.0$.   (C) Dependence of the critical period $\varDelta t_c$ of the cycle on $\bar\phi$ for the different $\iota$ values. (D) The boundary between regimes with and without bistability for different dilution protocols $\phi(t)$. 
GD denotes the growth--division protocol defined by Eq.~\ref{eq:phi_t} (Methods and Models) for different values of $\alpha$. 
Step denotes a pulsed dilution protocol given by $\phi(t)=\bar{\phi}\,\frac{\Delta t}{\tau}$ for $0 \le t \le \tau$ and $\phi(t)=0$ for $\tau < t \le \Delta t$. 
    }
    \label{fig:asymmetric}
\end{figure}

In Eq.\,\ref{eq:model} in the main text, we assumed the reproduction rate function, $r(x_i)$ is the same between the two self-reproducing entities. Here, we also discuss the rate functions of two entities, $r_1(x_1)$ and $r_2(x_2)$ are different, i.e., $r_1(x) \neq r_2(x)$. The simplest example is $r_i(x) x = \epsilon + \kappa_i x^2$, where $\kappa_1 \neq \kappa_2$. In this case, provided that the difference between the catalytic strength of two entities $|\kappa_1 - \kappa_2|$ is not too large, the system has bistability, as the nullclines show (Fig.\,\ref{fig:asymmetric}A). Further, as the bifurcation diagram shows (Fig.\,\ref{fig:asymmetric}B), even in this case, the critical point for the interval $\varDelta t$ and the dilution rate $\bar{\phi}$ exist, although the transition is discontinuous.

In such cases, the boundary in the parameter space for ACSs that have bistability in the general GD protocols is also bounded by the boundaries for the SD and CSTR protocols. 

\subsection{More general reproduction rate function} \label{sec:robust_rate_function}

Here, we consider the autocatalytic system is under the SD protocol, in which the reproduction rate function depends on both $x_1$ and $x_2$, including cases where $\rm{X}_1$ and $\rm{X}_2$ mutually inhibit their synthesis (see Fig.\,\ref{fig:SImodel}D). For example, in the case of the genetic toggle switch model~\cite{gardner2000construction}, $r(x_i, x_{\bar i}) = \frac{1}{x_i} ( \frac{1}{1 + x_{\bar i}^n })$, where $n$ is the Hill coefficient.

The rate equations are
\begin{equation}
    \frac{d{x}_i}{dt} = s r (x_i, x_{- i}) x_i.
\end{equation}
where $- i$ represents 2 or 1, if $i=1$ or 2, respectively.

Here, the same as before, we define
$\chi = x_1 + x_2, \delta = x_1 - x_2$, then
\begin{equation}
    \frac{d\chi}{dt} = s r (\frac{\chi}{2}, \frac{\chi}{2}) \chi + \mathcal{O}(\delta^2),  \quad \frac{d\delta}{dt}  =  s \left (r (\frac{\chi}{2},\frac{\chi}{2}) +\frac{1}{2} \chi  \frac{\partial r}{\partial x_{i}} (\frac{\chi}{2}, \frac{\chi}{2}) - \frac{1}{2} \chi  \frac{\partial r}{\partial x_{- i}} (\frac{\chi}{2}, \frac{\chi}{2}) \right) \delta + \mathcal{O}(\delta^2),
\end{equation}
where we used the expansion 
$$r \left( \frac{1}{2} (\chi \pm \delta), \frac{1}{2} (\chi \mp \delta) \right) = r (\frac{\chi}{2}, \frac{\chi}{2}) \pm \frac{\partial r}{\partial x_i} (\frac{\chi}{2}, \frac{\chi}{2}) \frac{\delta}{2} \mp \frac{\partial r}{\partial x_{- i}} (\frac{\chi}{2}, \frac{\chi}{2}) \frac{\delta}{2} + \mathcal{O}( \delta^2),$$
and assumed $\delta$ is small.

\begin{equation}
    \frac{d}{dt} \bigl(\frac{\delta}{\chi}\Bigr) = \frac{1}{\chi^2} (\frac{d\delta}{dt}\chi - \frac{d\chi}{dt}\delta) =  \frac{1}{2} s \chi \left(   \frac{\partial r}{\partial x_{i}} (\frac{\chi}{2}, \frac{\chi}{2})  -  \frac{\partial r}{\partial x_{- i}} (\frac{\chi}{2}, \frac{\chi}{2})   \right) \frac{\delta}{\chi}.
\end{equation}
The integration of $\frac{d}{dt} (\frac{\delta}{\chi})/(\frac{\delta}{\chi})$ lead to
\begin{equation}
    \begin{aligned}
        \log \left| \frac{\delta(t)}{\chi(t)} \right| &=  \int_0^t  \frac{1}{2} s \chi \left(   \frac{\partial r}{\partial x_{i}} (\frac{\chi}{2}, \frac{\chi}{2})  -  \frac{\partial r}{\partial x_{- i}} (\frac{\chi}{2}, \frac{\chi}{2})   \right) dt &+& \log \left| \frac{\delta(0)}{\chi(0)} \right| \\
        &=  \int_{\frac{\chi(0)}{2}}^{\frac{\chi(t)}{2}}    \frac{\frac{\partial r}{\partial x_{i}} (x, x)  -  \frac{\partial r}{\partial x_{- i}} (x, x)}{r(x,x)}  dx &+& \log \left| \frac{\delta(0)}{\chi(0)} \right|,
    \end{aligned}
\end{equation}
where we used $\frac{dt}{d\chi} = 1/ (  s r (\frac{\chi}{2}, \frac{\chi}{2}) \chi )$.
Therefore, 
\begin{equation}
    \frac{\delta(t)}{\chi(t)} =  \frac{ R_i(\frac{ \chi(t)}{2}) }{ R_i(\frac{\chi(0)}{2})} \frac{ R_{- i}(\frac{\chi(0)}{2})}{ R_{- i}(\frac{ \chi(t)}{2}) } \frac{\delta(0)}{\chi(0)},
\end{equation}
where $R_{i}(x)$ and $R_{- i}(x)$ are defined as $R_{i}(x) = \exp \left( \int^{x} \frac{\frac{\partial r}{\partial x_{i}} (x', x')}{r(x',x')} dx' \right)$ and $R_{\bar i}(x) = \exp \left( \int^{x} \frac{\frac{\partial r}{\partial x_{- i}} (x', x')}{r(x',x')} dx' \right)$. 
Therefore, the sufficient condition for the bistability of the system is
\begin{equation} \label{eq:gen-condition}
    \frac{ R_i(\frac{ \chi(t)}{2}) }{ R_{- i}(\frac{ \chi(t)}{2}) } > 
    \frac { R_i(\frac{\chi(0)}{2})}{ R_{- i}(\frac{\chi(0)}{2})}.
\end{equation}

Here, if we assume that the separation of variables is possible for the reproduction rate function $r(x_i, x_{- i})$, $r(x_i, x_{- i}) = f(x_i)g(x_{- i})$; then, $R_{i}(x) = f(x)$ and $R_{- i}(x) = g(x)$. Note that, if $g(x) = 1$, then the condition Eq.~\ref{eq:gen-condition} reduced to Eq.~\ref{eq:condition} in the main text. For example, in a case with the genetic toggle switch model, the above condition is $\frac{1 + (\frac{ \chi(\varDelta t)}{2})^n} { \frac{ \chi(\varDelta t)}{2}} > \frac{1 + (\frac{ \chi(0)}{2})^n} { \frac{ \chi(0)}{2}}$. Thus this model could show the bistability if $n>1$.

\section{Stochasticity in the reaction dynamics and transition between states (variation)} 
\label{sec:robust_stochasticity}

In contrast with the previously discussed deterministic model, Eq.\,\ref{eq:model}, if the volume of the compartment is small (i.e., the system size) the system, the stochastic fluctuation in the reaction dynamics is non-negligible~\cite{gardiner1985handbook}. 
The stochastic dynamics of the discrete number of entities $\rm{X}_i$ ($i=1,2$), $n_i$ $(= x_i V)$, under the chemical reaction system represented in Fig.\,\ref{fig:nullclines}A, is described by the chemical master equation: the probability of $n_i$ at a time $t$, $P(n_1, n_2, t)$ obeys 
\begin{equation} \label{eq:master_model}
\begin{split}
    \frac{d}{dt}P(n_1, n_2, t) =& P(n_1 - 1, n_2, t) \tau_1(n_1-1,n_2) \\
    &+ P(n_1, n_2-1, t) \tau_2(n_1,n_2-1) \\
    &- P(n_1 , n_2, t)(\tau_1(n_1,n_2) + \tau_2(n_1,n_2) ),
\end{split}
\end{equation}
where $\tau_i$ is the production reaction of $\rm{X}_i$, $\tau_i(n_1,n_2) = s r(x_i) n_i$.

Further, by the system size expansion \cite{gardiner1985handbook} and remaining only leading terms of $V^{-1}$, and using the concentration (i.e., continuous variable) $x_i$ instead of the  number of molecules $n_i$ ($x_i = n_i/V$), the above master equations is transformed into the chemical Langevin equations~\cite{gardiner1985handbook, gillespie2000chemical}, 

\begin{equation} \label{eq:langevin_model}
    \frac{d x_i}{dt} = s r_i(x_i) x_i + \sqrt{ s r_i(x_i) V^{-1} } \eta_i(t),
\end{equation}
where $\eta_i$ are i.i.d.\,Gaussian random variables with the correlation function $\langle \eta_i(t) \eta_i(t') \rangle = \delta (t-t')$. $V$ is the volume of the compartment, and $V^{-1}$ corresponds to the intensity of noise. Then, this equation becomes Eq.\,\ref{eq:model} when the volume $V$ is infinitely large (i.e., $V^{-1} \rightarrow 0$).

Here, we assume the serial dilution protocol, the same as the deterministic case, where the volume $V$ is fixed during one cycle. At each dilution, the probability of each entity remaining in the system is $m^{-1} = e^{-\bar\phi \varDelta t}$. Thus, the number of entities at the start of each cycle, $n_i(n\varDelta t + 0)$, follows the Binomial distribution $B(n,p)$ with the number of trials $n= n_i(n\varDelta t - 0)$ and the success probability at each trial $p = m^-1$, where $n_i(n\varDelta t - 0)$ is the number at the end of the previous cycle.

We numerically solved the dynamics of the master equation Eq.\,\ref{eq:master_model}  using the Gillespie method~\cite{gillespie1977exact}. Note that, for taking into account SD protocol, if time $t$ exceeds $n\varDelta t$, which is the time the $n$-th dilution took place, during one reaction step, then the reaction should be discarded, and the time is set to $t=n\varDelta t$.

In contrast with the deterministic case, the transition could occur from $\rm X_1$($\rm X_2$)-dominant state to the other one in the presence of the stochastic noise (Fig.\,\ref{fig:stochastic-reaction}A). We numerically calculated the averaged transition time (the first passage time until the numbers of two entities become equal, $n_1=n_2$) as in Fig.\,\ref{fig:stochastic-reaction}. 
The transition time depends on the parameters for the dilution protocols, $\varDelta t$ and $\bar \phi$, and also on the volume (system size) $V$. 
As Fig.\,\ref{fig:stochastic-reaction}B shows, the transition time depends on the volume $V$: the time depends on $V$ exponentially if the parameters of the protocol are below the critical value (i.e., $\bar \phi < \phi_c$), while the time depends on $V$ sublinearly and saturate as $V$ increases. 
if the system is in the region with heredity (in the deterministic case), the transition time is reasonably long even if the system size is small (see Fig.\,\ref{fig:stochastic-reaction}C). 

We also calculate the transition time in the case that the catalytic strengths of the entities are asymmetric (see Appendix Sec.\,\ref{sec:asymmetric}) in Fig.\,\ref{fig:appdx-stochastic-reaction}.
We calculated the transition time from $\rm{X}_1$- to $\rm{X}_2$-dominant states, $T_{1 \rightarrow 2}$, and the reverse direction $T_{2 \rightarrow 1}$. 
 If the catalytic strength for $\rm{X}_2$, $\kappa_2$, is larger than for $\rm{X}_1$, $\kappa_1$ ($\iota >0$), then the transition from $\rm{X}_2$ to $\rm{X}_1$ takes longer (i.e., $T_{2 \rightarrow 1}$ > $T_{1 \rightarrow 2}$) by orders of magnitude. The same as in the symmetric case, transition times $T_{1 \rightarrow 2}$ and $T_{2 \rightarrow 1}$ depend on $V$ (Fig.\,\ref{fig:appdx-stochastic-reaction}). The difference between $T_{1 \rightarrow 2}$ and $T_{2 \rightarrow 1}$ is relatively small if $V$ is small. 

\begin{figure}
    \centering
    \includegraphics[width=16cm]{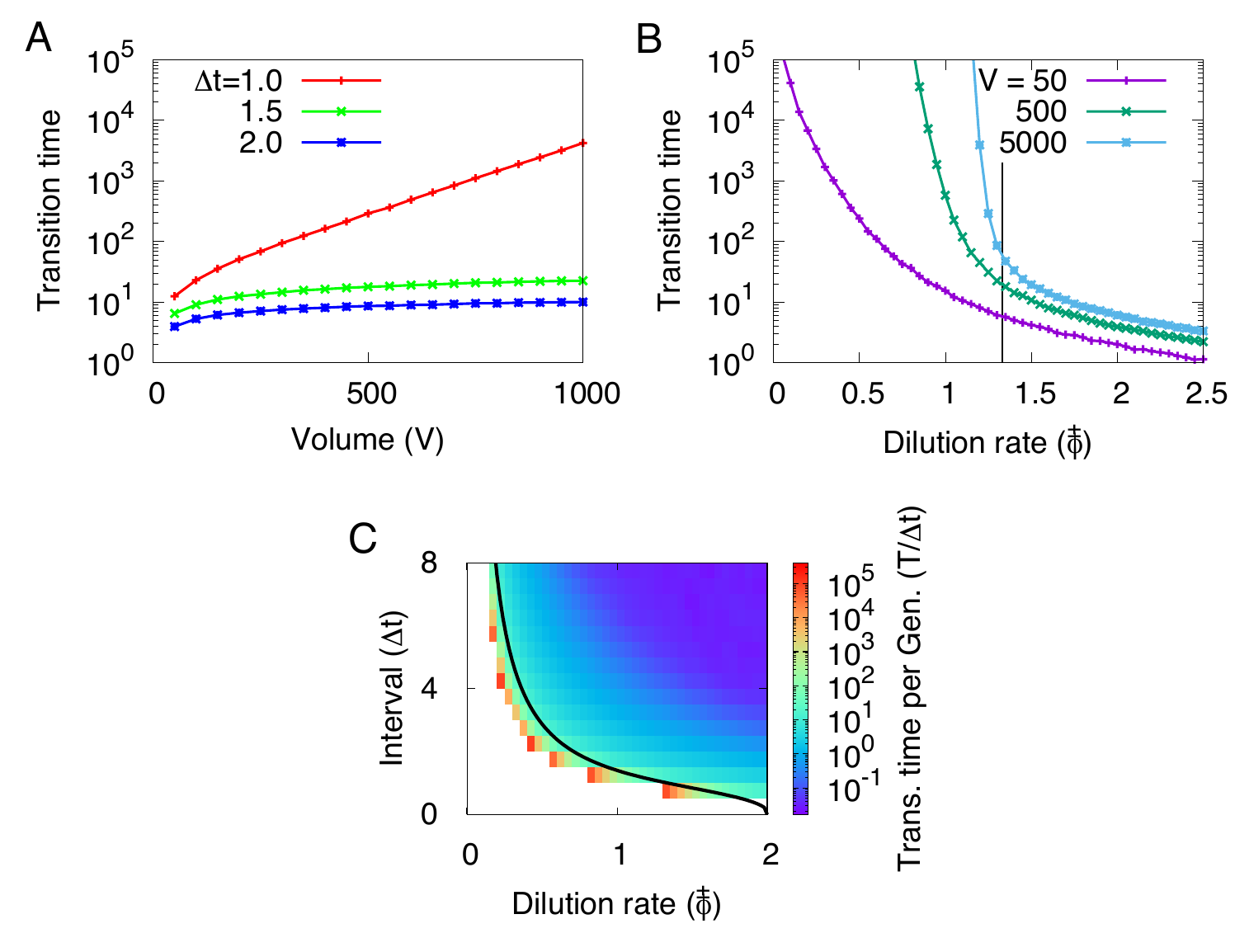}
    \caption{ (A) The transition time from $\rm X_1$ to $\rm X_2$, or the reverse direction, $T_{1 \rightarrow 2}$ or $T_{2 \rightarrow 1}$, varying the volume $V$, under SD protocol. The transition time $T_{1 \rightarrow 2}$ is calculated as the average of 10000 trials (the time until $x_2 > x_1$ starting from $(x_1, x_2) = (s^{tot},0)$). The lines with different colors represent the difference in the interval $\varDelta t$.
    We set $\epsilon = 0.5, \kappa = 8$, in $r(x)x = \epsilon + \kappa x^2$. (B) The same transition time between states, but varying the dilution rate $\bar \phi$. The lines with different colors represent the difference in the volume $V$. The vertical line represents the bifurcation point where the system loses bistability.
    (C) The color intensity represents the transition time divided by the interval ($T/\varDelta t$). The same figure as Fig.\,\ref{fig:stochastic-reaction}B in the main text, but with a larger volume. The solid curve is the boundary between with and without bistability in a deterministic case (the same as shown in Fig.\,\ref{fig:nullclines}C in the main text). 
    We set the parameters as $\kappa$ = 8, $\epsilon$ = 0.5, and $V=500$.}
    \label{fig:si_stochastic-reaction}
\end{figure}

\begin{figure}
    \centering
    \includegraphics[width=8cm]{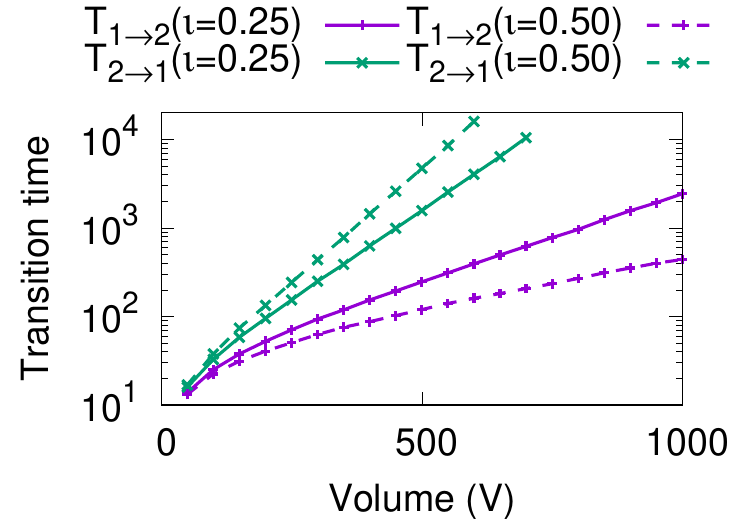}
    \caption{
    The transition time between states in the case with asymmetric catalytic strength (from $\rm X_1$-dominant to $\rm X_2$, $T_{1 \rightarrow 2}$ (blue), and the reverse direction (green)). We set $\iota = 0.25$, or $\iota = 0.5$ ($\kappa_1 = \kappa - \frac{\iota}{2}$ and $\kappa_2 = \kappa + \frac{\iota}{2}$) in the solid or dashed lines, respectively, and $\varDelta t = 1$. }
    \label{fig:appdx-stochastic-reaction}
\end{figure}

\section{Differential reproduction of different chemical compositional states}
\label{sec:robust_differential_rate}

\begin{figure}
    \centering
    \includegraphics[width=15cm]{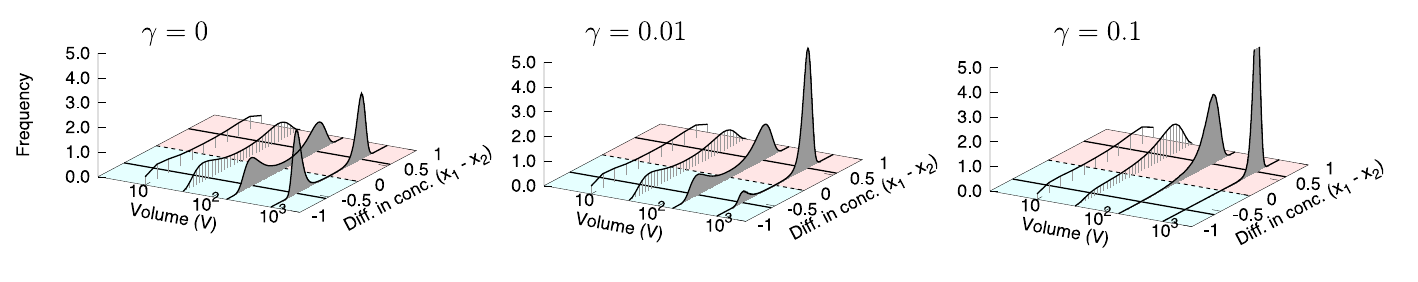}
    \caption{The probability density profile of $x_1-x_2$ at each volume $V$ value, with different differential reproduction $\gamma$. The solid lines represent the steady states of $x_1 - x_2$ in the deterministic case. We set $\varDelta t = 1$, $\bar \phi = 1$, $\epsilon = 0.5$, $\kappa = 8$, $\gamma = 0, 0.01$ and $0.1$ (from left to right). }
    \label{fig:si_differential-reproduction}
\end{figure}

\begin{figure*}
    \centering
    \includegraphics[width=16cm]{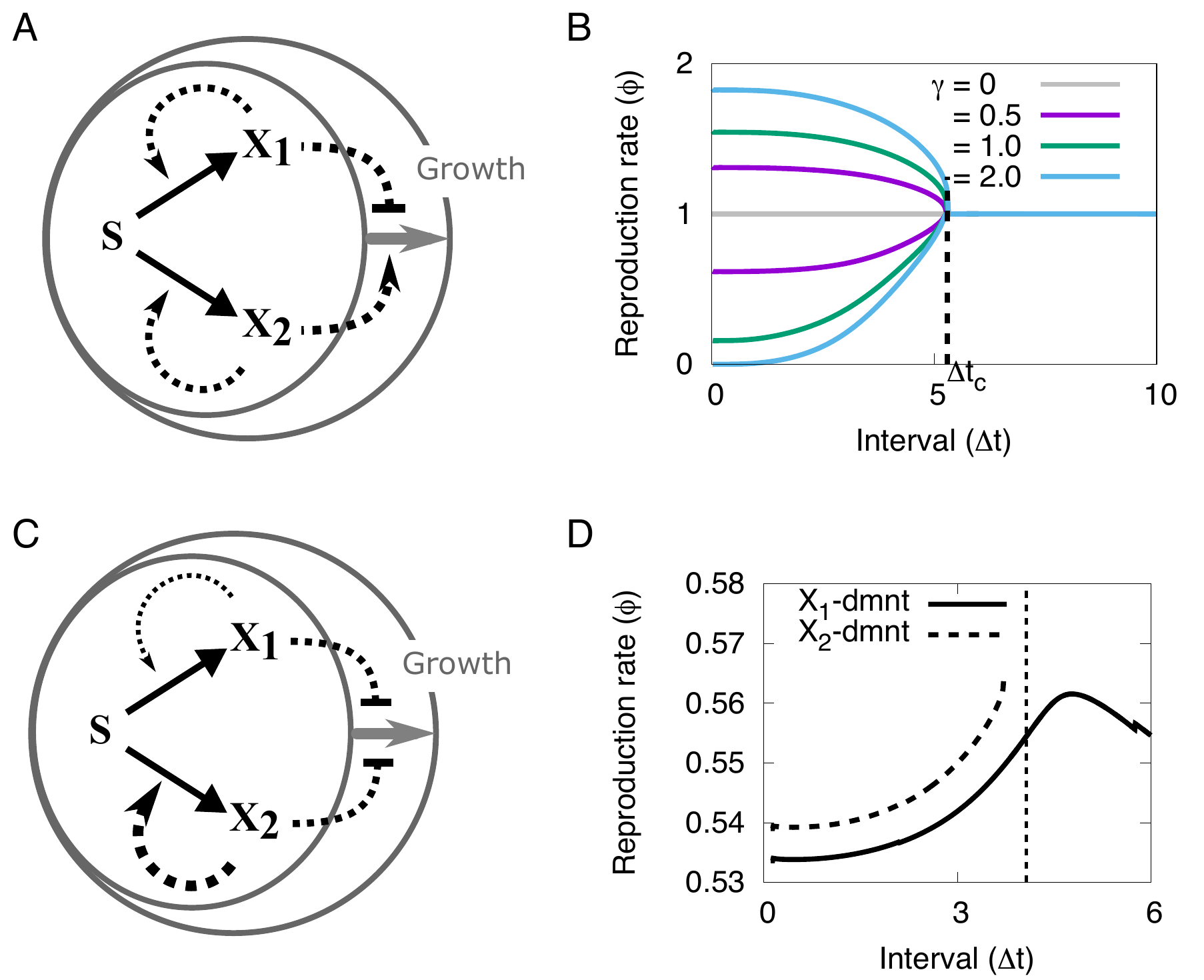}
    \caption{(A) Schematic of competing autocatalytic entities encapsulated by a growing compartment. The grey arrow represents the growth of the compartment, which is promoted (dotted arrow) or inhibited (bar-headed dotted arrow) by entities. The entities have symmetric catalytic strength (the same model as Fig.\,\ref{fig:SImodel}A), but affect the compartment growth asymmetrically (see Eq.\,\ref{eq:compo_dep_dilution}).
    (B) Reproduction rates of the compartment (i.e., the long-term cycle averaged dilution rate), $\frac{1}{\varDelta t}\int_T^{T + \varDelta t} \hat \phi(t, \bm x(t)) dt$ as $T\rightarrow \infty$, at each compositional state (stationary trajectory $\{\bm x(t)\}$), varying the interval $\varDelta t$. The dotted line represents the critical point $\varDelta t_c^{sd}$, which divides the regions where the system has bistability or not. All the colored curves, which represent the difference in $\gamma$, show the bifurcation from bistability to mono-stability at $\varDelta t_c^{sd}$. We set the parameters as $\kappa = 8, \epsilon=0.5$ and $\alpha=2$. 
    (C) The same schematic diagram as (A) in a case with $r_i(x)x = \epsilon + \kappa_i x^2$ and the symmetric dilution rate $\hat \phi (t, \bm x(t)) = \phi(t) + \gamma \chi$. 
    (D) The reproduction rates of the compartment, under the same setup as (B).  We set the parameters as $\iota = -1$, $\kappa = 8, \epsilon=0.5$ and $\alpha=2$.
    }
    \label{fig:si_differential_fitness}
\end{figure*}

We investigate situations where the dilution rates depend on the chemical composition, and the system has different growth rates between the two states. As noted in the introduction, this is, in fact, a crucial property necessary for a population of compartmentalized chemical reaction systems to undergo Darwinian evolution.

In such a case, the dilution rate $\hat\phi\big(t, \bm{x}(t)\big)$ is given by
\begin{math}
    \hat\phi\big(t, \bm{x}(t)\big) \equiv {\frac{dV}{dt}}/{V},
\end{math}
such that
\begin{equation}
    \int_0^{\varDelta t} \hat\phi\big(t, \bm{x}^*_{(i)}(t)\big) = \bar{\phi}_i \varDelta t,
\end{equation}
where $i=1,2$ and $\bm{x}^*_{(i)}(t)$ represents the steady-state trajectories of $\bm{x}$, where $x_i$ is dominant. Here, differential reproduction implies $\bar{\phi}_1 \neq \bar{\phi}_2$. Recall that $m=e^{\bar{\phi}\varDelta t}$ is the number of offspring at one generation with interval $\varDelta t$ (main text Sec.\,\ref{sec:growth-division}). 

As a simple example, we consider the dilution due to the growth of compartments which depends on their components,
\begin{equation}
\label{eq:compo_dep_dilution}
    \hat\phi\big(t, \bm{x}(t)\big) = \phi(t) - \gamma \delta,
\end{equation}
where $\gamma$ is a constant (control parameter), $\delta = x_1 - x_2$, and ${\phi}(t)$ is the dilution rate independent with the component; here, we choose Eq.\,\ref{eq:phi_t} in Methods and Models as ${\phi}(t)$, such that $\int_0^{\varDelta t} \phi(t) dt = \bar\phi \varDelta t$. Naturally, in this case, the growth rate of the two states are different, $\bar{\phi}_1 < \bar{\phi}_2$, if $\gamma > 0$. Notably, this dependency of $\hat\phi$ on the composition does not change the critical long-term dilution rate at which the system loses bistability, $\phi_c$ (Fig.\,\ref{fig:differential_fitness}A), i.e., $\phi_c$ does not depend on $\gamma$. Generally, the bistability of the system does not seem to be affected by the asymmetric factor of $\hat\phi$. 
We can see this by considering the autocatalytic system is under arbitrary protocols with the dilution rate depending on the chemical composition, 
\begin{equation}
    \frac{d x_i}{dt} = s r (x_i) x_i - \phi(t, \chi, \delta ) x_i.
\end{equation}
where $\chi = x_1 + x_2, \delta = x_1 - x_2$. The dependency of $\phi$ on $\delta$ is interpreted as differential reproduction between the states.
The same calculation as in Appendix Sec.\,\ref{sec:appdx-model} leads to 
\begin{equation}
    \frac{d\chi}{dt} = s r (\frac{\chi}{2}) \chi  - \phi(t, \chi, 0) \chi + \mathcal{O}(\delta),  \quad \frac{d\delta}{dt}  =  s \left (r (\frac{\chi}{2}) +\frac{1}{2} \chi  \frac{dr}{dx} (\frac{\chi}{2}) \right) \delta - \phi(t, \chi, 0) \delta + \mathcal{O}(\delta^2),
\end{equation}
where $\chi \gg \delta$ is assumed.
Then, the deviation of $\delta$ in one cycle is calculated as the same as in Appendix Sec.\,\ref{sec:appdx-model},
\begin{equation}
    \begin{aligned}
        \log \left| \frac{\delta(t)}{\chi(t)} \right| &=  \int_0^t  \frac{1}{2}  s \frac{dr}{dx} \Bigl(\frac{\chi}{2}\Bigr) \chi  dt &+& \log \left| \frac{\delta(0)}{\chi(0)} \right|, 
    \end{aligned}
\end{equation}
which, notably, does not depend on the asymmetric part of $\phi(t, \chi, \delta)$.

Next, we consider a case where $\phi$ depends on symmetrically $x_1$ and $x_2$, i.e., $\hat\phi\big(t, x_1, x_2\big) = \hat\phi(t, x_2, x_1)$:
\begin{equation}
    \hat\phi\big(t, \bm{x}(t)\big) = \phi(t) - \gamma \chi,
\end{equation}
where $\chi = x_1 + x_2$, while the catalytic strength of two entities, $\kappa_1$ and $\kappa_2$, are asymmetrical, i.e., $\kappa_1 \neq \kappa_2$, as discussed in Appendix Sec.\,\ref{sec:asymmetric}. In this case, also the growth rate is different for different states (Fig.\,\ref{fig:differential_fitness}).

We also consider a situation where the division interval $\varDelta t$ depends on the composition $\bm{x}$. For example, we decide whether $\varDelta t$ is equal to $\varDelta t_1$ if $x_1 > x_2$ at the beginning of cycle, or $\varDelta t_2$ otherwise. Trivially, in this case also, the system has the bistability if $\varDelta t_1$ and $\varDelta t_2$ are below the critical value discussed in the main text Sec.\,C, $\varDelta t_1, \varDelta t_2 < \varDelta t_c$.

Overall, even in the case that growth rate $\phi$ or/and the division interval $\varDelta t$ depend on the chemical composition of the system, we can guarantee the bistability if $\phi$ at each stationary trajectory, $\bm x_{(1)}^*(t)$ and $\bm x_{(2)}^*(t)$ satisfies $\bar\phi_1, \bar\phi_2 < \phi_c^{sd}$ and $\varDelta t_1, \varDelta t_2 < \varDelta t_c^{sd}$.

\section{Darwinian population of growing and dividing protocells}
\label{sec:appdx-darwinian}

\begin{figure}
    \centering
    \includegraphics[width=0.5\linewidth]{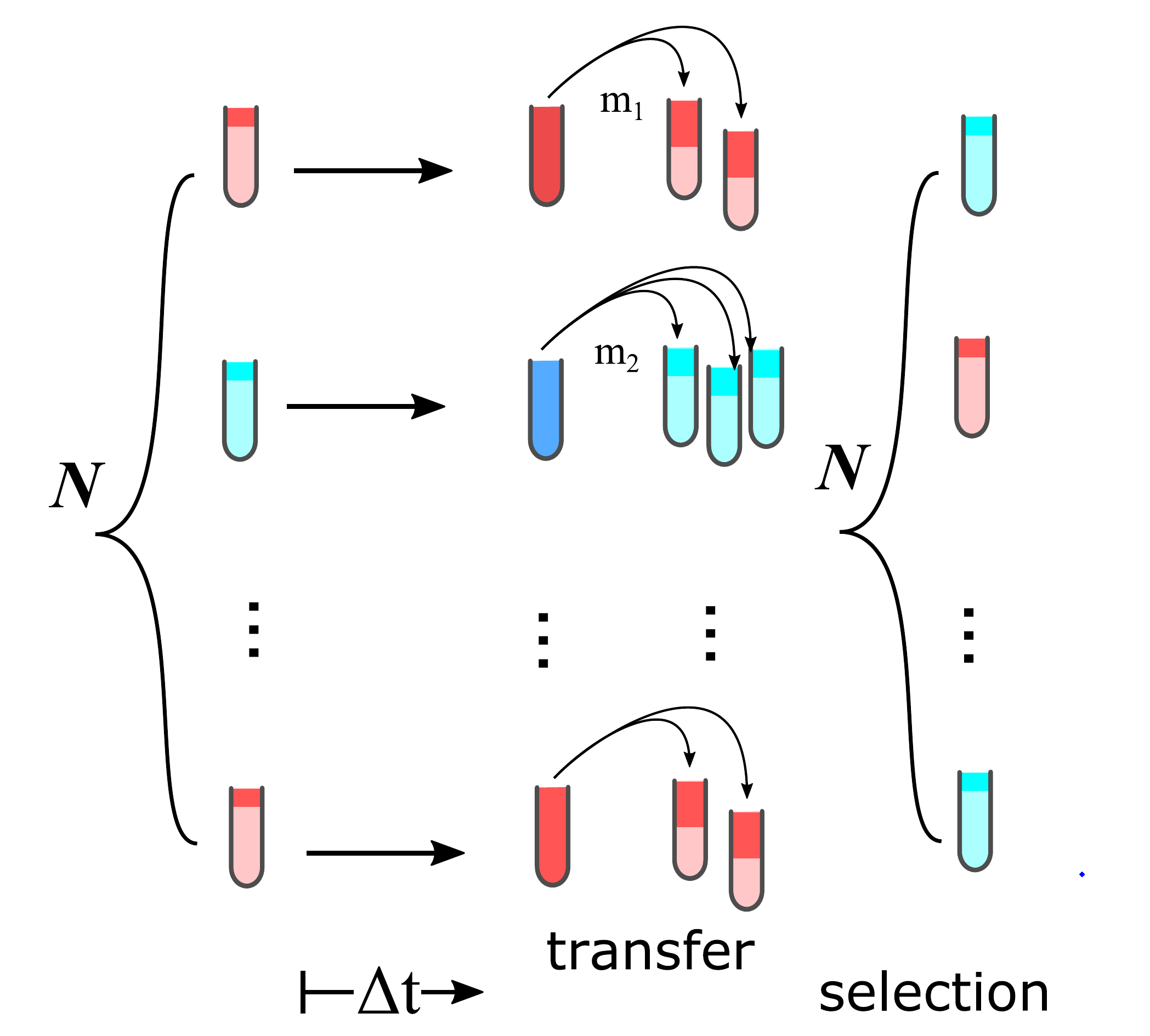}
    \caption{Schematic of the population dynamics of parallelized serial dilution lineages. The red and blue circles represent $\rm X_1$- and $\rm X_2$-dominant tubes, respectively. Schematics of the corresponding scenario of the compartment population, which grows and divides. }
    \label{fig:si_WF_schematics}
\end{figure}

We consider a population of $N$ protocells, each containing a copy of the autocatalytic system discussed above, described by Eq.\,\ref{eq:model}. We choose $r(x)x=\epsilon+\kappa x^2$ with parameters such that the system is bistable in CSTR. Initially, all protocells are given random chemical compositions and a volume $V_0$. The population undergoes a Moran process \cite{ewens2004mathematical}: whenever any protocell divides, it is replaced by its two daughter cells, and additionally, one random protocell is removed to maintain the population size of $N$. On shorter timescales, the chemical reactions in each protocell occur stochastically as in Sec.~\ref{sec:variation}. Depending on the selection pressure present each compartment grows in volume at a specified rate based on its chemical composition at that time. Cells divide when their volume reaches $2V_0$. We subject the system to three regimes of selection pressure:\\
1. Initially, no selection pressure is imposed; neither growth state is favored. The growth rate of the volume of a protocell is given by $\frac{dV}{dt} = \bar \phi V$. We run the Moran process under these conditions until the population stabilizes.\\
2. We then switch to a selection pressure that favors state 1 by making $\frac{dV}{dt} = (\bar \phi + \gamma \delta)V$.\\
3. After the population stabilizes, we again switch conditions such that now state 2 is favored, by making $\frac{dV}{dt} = (\bar \phi - \gamma \delta)V$.\\

\begin{figure*}
    \centering
    \includegraphics[width=17.8cm]{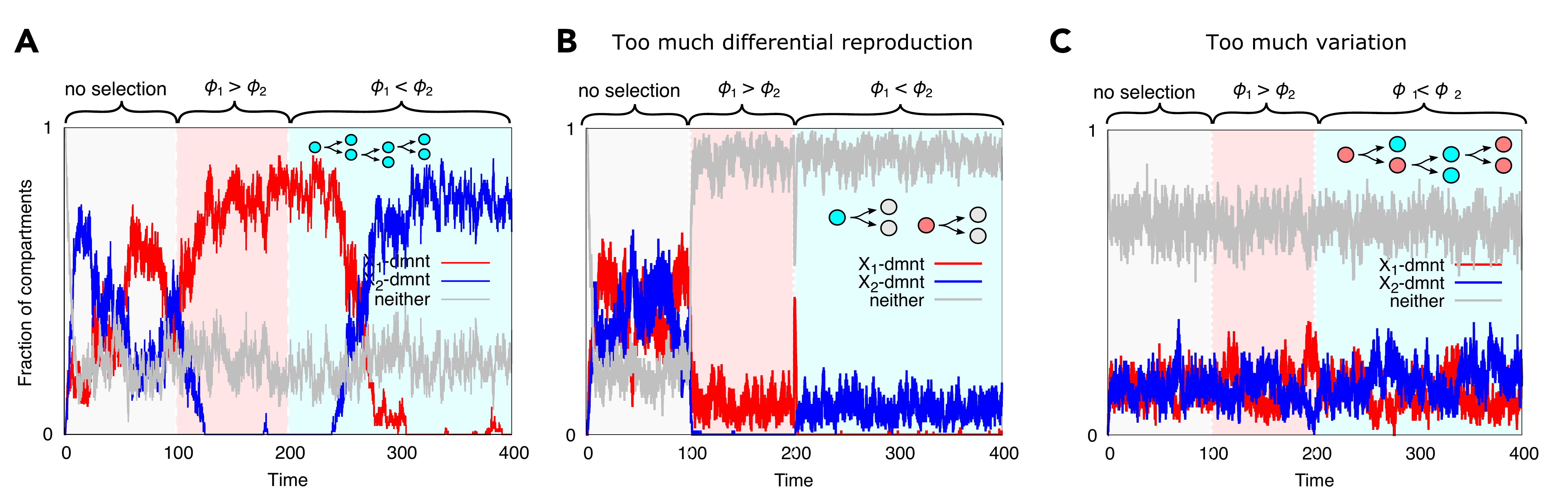}
\caption{
Population dynamics of compartmentalized autocatalytic sets under a Moran-like process with heredity, variation, and differential reproduction.
(A) Frequency of $\mathrm{X}_1$-dominant, $\mathrm{X}_2$-dominant, and neither compartments through generations. The population size is fixed ($N=100$), and compartments grow and divide when their volume reaches twice the initial value. The volume dynamics follow $dV/dt = (\phi + \gamma)V$, where $\gamma = \gamma_0$ if $x_1 > x_2$, and $\gamma = -\gamma_0$ otherwise (selection favoring $\mathrm{X}_1$-dominant compartments). Initially (from time 0 to 100), no selection is applied ($\gamma = 0$), while from the time 100 to 200, and from the 200 onward, selection favors $\mathrm{X}_2$- and $\mathrm{X}_1$-dominant compartments, respectively, by reversing the sign of $\gamma$. When variation (mutation) is large, selection becomes ineffective compared with the small-variation case. Compartments are classified as $\mathrm{X}_1$- or $\mathrm{X}_2$-dominant if $x_1 - x_2 > 0.5$ or $< -0.5$, respectively. Parameters: $\epsilon = 0.5$, $\kappa = 8$, initial volume $V=50$, $\gamma_0 = 0.05$.
(B) Case of excessive differential reproduction, where $\gamma_0 = 0.9$. Strong selection suppresses stable dominance patterns, leading to reduced observable selection at the population level.
(C) Case of excessive variation, implemented by reducing compartment volume to $V=10$, which enhances stochastic fluctuations and weakens selection.
}
  \label{fig:si_moran}
\end{figure*}

\section{ACS based on \emph{Azoarcus} ribozyme coupled with metabolism}
\label{sec:azo_result}

In this section, we apply the framework we have developed to an experimentally realized ACS based on the \emph{Azoarcus} ribozyme~\cite{vaidya2012spontaneous, yeates2016dynamics, ameta2021darwinian}.
We examine a few simplifications and variants of this ribozyme system, along with systems consisting of two \emph{Azoarcus} ribozymes competing for the same food set. We show that some of these variants can exhibit two (exponential) growth states, and some are not. In particular, we find a modified version of the \emph{Azoarcus} ribozyme, which incorporates additional catabolic and anabolic steps~\cite{arsene2018coupled}, can exhibit bistability under competition for shared resources. Applying our critical-threshold results to this system, we propose a serial dilution protocol to test whether the modified \emph{Azoarcus} system inherits its phenotypic state.

The \textit{Azoarcus} ribozyme $\mathbf{WXYZ}$ can be assembled from two fragments by the reaction,
\begin{equation}
    {}_{\mathrm{{M}}}\mathbf{WXY}_{\mathrm{{N}}} + \mathbf{Z} \rightarrow \ {}_{\mathrm{{M}}}\mathbf{WXY}_{\mathrm{{N}}} \mathbf{Z}.
\end{equation}
${}_{\mathrm{{M}}}\mathbf{WXY}_{\mathrm{{N}}}$ and $\mathbf{Z}$ are the fragments, and ${\mathrm{{M}}}$ and ${\mathrm{{N}}}$ are bases at the ends of $\mathbf{WXY}$~\footnote{The bases M and N are the middle nucleotide of the 3nt recognition element in WXY, called the internal guide sequence (IGS) and the tag sequence~\cite{yeates2016dynamics}.}. $\mathrm{{M}}$ and $\mathrm{{N}}$ can be arbitrary bases $\mathrm{  \{A, U, C, G \}}$, thus there are 16 different types of this engineered \textit{Azoarcus} ribozyme. This reaction is catalyzed specifically by a ribozyme if $\mathrm{{M}}$ in the fragment is a complementary base to $\mathrm{{N}}$ in the ribozyme.
Certain types of the ribozyme can catalyze the formation reaction of themselves: e.g., ${}_{\mathrm{{M}}}\mathbf{WXY}_{\mathrm{{N}}}\mathbf{Z}$ such that $\mathrm{{M}} = \mathrm{C}$ and $\mathrm{{N}} = \mathrm{G}$ or $\mathrm{{M}} = \mathrm{U}$ and $\mathrm{{N}} = \mathrm{A}$. Besides being catalyzed by the corresponding ribozymes, the reaction is also (weakly) catalyzed by a non-covalent complex between the corresponding fragments, or non-specifically by non-corresponding ribozymes; we call these `background reactions'~\cite{yeates2016dynamics}.
 
\subsubsection{Absence of bistability in competing ACSs based on the original engineered　{Azoarcus} ribozyme}
\label{sec:azoarcus-noheredity}

We imagine a particular case of the system in Fig.\,\ref{fig:azoarcus}A, where $\rm{X_1}$ and $\rm{X_2}$ are two distinct types of self-catalyzing \textit{Azoarcus} ribozyme, made from two distinct $\mathbf{WXY}$ fragments and a common $\mathbf{Z}$ fragment. For simplicity, we assume the two sets of reactions occur with symmetric kinetic rates. 
We assume the $\mathbf{WXY}$ fragments are abundant, whereas $\mathbf{Z}$ is not and limits the reaction rates. Thus, $\mathbf{Z}$ acts as the common substrate S. 

Then, the rate equations for the concentrations of $\rm{X_1}$ and $\rm{X_2}$, $x_1$ and $x_2$ are described as Eq.\,\ref{eq:model} with the linear reproduction function $r(x)x = \epsilon + \kappa x$, where $\epsilon$ is the rate constant of the background reaction, and $\kappa$ is the catalytic strength of $\rm{X}_1$ and $\rm{X}_2$. \footnote{Here, the background reaction rate is approximated as a constant, although this reaction is due to catalyzed reaction by the non-covalent ribozymes or non-corresponding ribozymes. Assuming that the WXY fragments are abundant, the concentration of non-covalent ribozymes WXY:Z is $\sim s$ approximately. We assume their catalytic activity is non-specific, whose reaction rate is $\epsilon$. Next, we assume corresponding and non-corresponding ribozymes catalyze with efficiencies $\tilde \kappa$ and $\tilde \kappa'$, respectively. Then, the rate of the reaction $\rm{S} \rightarrow \rm{X_1}$ is $s \epsilon + x_2 \tilde \kappa' + x_1 \tilde \kappa$. If we assume that $\epsilon = \tilde \kappa'$, the rate is given as $\epsilon + \kappa x_1$, where $\kappa = \tilde \kappa - \epsilon$.} As mentioned earlier, for simplicity, we assume the catalytic efficiency of $\rm{X_1}$ and $\rm{X_2}$ are equal.

As already discussed in Sec.\,\ref{sec:acs-model}, competing ACSs with such linear reproduction rate functions cannot be bistable -- for all initial conditions, the system eventually reaches the state with an equal amount of the two ribozymes. More precisely, in Sec.\,\ref{sec:acs-model}, only the local stability of the symmetric state is shown.
In this case, however, we can further show the \emph{global} stability of this symmetric state~\footnote{Here, the time derivative of difference of concentration between $\rm{X_1}$ and $\rm{X_2}$ is 
\begin{math}
\frac{d}{dt} \left(\frac{\delta}{\chi}\right) = \frac{1}{\chi^2} ( \frac{d\delta}{dt}\chi - \delta\frac{d\chi}{dt} ) = - 2\epsilon s \frac{\delta}{\chi},
\end{math}
where $\delta = x_1 - x_2$ and $\chi = x_1 + x_2$.
Therefore, $\delta/\chi$ decays exponentially into zero in the characteristic relaxation time scale $\tau = \frac{1}{2 s \epsilon}$. For example, if $ s \epsilon \sim 0.01 \min^{-1}$, the half time $\tau$ is estimated as $\sim 50 \min$.}. Therefore, the system has no heredity,  which is consistent with previous experiments~\cite{ameta2021darwinian}.

\subsubsection{{Azoarcus} system coupled with metabolism exhibits bistability}
\label{sec:coupledazoarcus}

As discussed in Sec.\,\ref{sec:acs-model}, a reproduction rate function $r(x)$ with a higher order of catalysis is necessary for bistability. In the \textit{Azoarcus} system, this has previously been realized by engineering a variant where the system is coupled to catabolism and anabolism reactions: 
\begin{equation}
    \begin{split}
        {}_{\mathrm{{M}}}\mathbf{WXY}_{\mathrm{{N}}}\mathbf{\mathchar`- mod} \rightarrow {}_{\mathrm{{M}}}\mathbf{WXY}_{\mathrm{{N}}} + \mathbf{\mathchar`- mod}, \\
        {}_{\mathrm{{M}}}\mathbf{WXY}_{\mathrm{{N}}} + \mathbf{Z} \rightarrow \ {}_{\mathrm{{M}}}\mathbf{WXY}_{\mathrm{{N}}} \mathbf{Z},
    \end{split}
\end{equation}
where $\mathbf{\mathchar`-mod}$ represents an extra sequence joined to fragments $\mathbf{WXY}$.
Here, the first reaction represents a catabolism reaction that processes the modified fragment to a substrate that can participate in ribozyme synthesis. The second one represents an anabolic reaction that joins the fragments to form the ribozyme.
We denote $\mathbf{Z}$ as $\rm{S}$, $\mathbf{WXYZ}$ as $\rm{X}_1$ and $\rm{X}_2$ and $\mathbf{WXY}$ as $\rm{X'_1}$ and $\rm{X'_2}$.
Two ribozymes $\rm{X_1}$ and $\rm{X_2}$ synthesize themselves from the shared substrate $\rm{S}$, and there is the intermediate state $\rm{X'_1}$ and $\rm{X'_2}$ during the synthesis (see Fig.\,\ref{fig:azoarcus}B).

The concentrations of chemical species obey the rate equations
\begin{equation}\label{eq:azoarcus}
    \begin{aligned}
        \frac{d x'_i}{dt} &= ( \epsilon + \kappa x_i ) ( 1 - (s+b) x'_i + b x_i ), \\
        \frac{d x_i}{dt} &= ( \epsilon + \kappa x_i ) (s x'_i - b x_i),
    \end{aligned}
\end{equation}
where $i=1$ or $2$, $\epsilon$ is the spontaneous reaction rate, $\kappa$ is the catalytic efficiency of the ribozymes, and $b$ ($ \ll 1$) is the relative rate of backward reaction compared with the forward one \footnote{The ACSs based on the \emph{Azoarcus} system show slow backward reactions since they are based on the recombination reactions of nucleotides~\cite{arsene2018coupled}.}.
We again consider the ACS system under the SD protocol with the interval $\varDelta t$ and an $m$-fold dilution factor.
Also, the total concentration of S, ${\rm X_1}$ and ${\rm X_2}$ is kept as a constant $s^{tot} = s+x_1+x_2$.

In contrast with the original system (Fig.\,\ref{fig:azoarcus}A), this system indeed exhibits bistability for appropriate values of the parameters $\varDelta t$ and $\bar\phi$ (Fig.\,\ref{fig:az_delta_t}), i.e.,
it reaches either the $\rm{X}_1$-dominated state or the $\rm{X}_2$-dominated state, depending on the initial composition of the species (Fig.\,\ref{fig:azoarcus}C). 

\subsubsection{Bounds on the critical $\varDelta t$ and $\bar\phi$ for observing inheritance of compositional state} \label{sec:azo_bounds}

\begin{figure*}
    \centering
    \includegraphics[width=16cm]{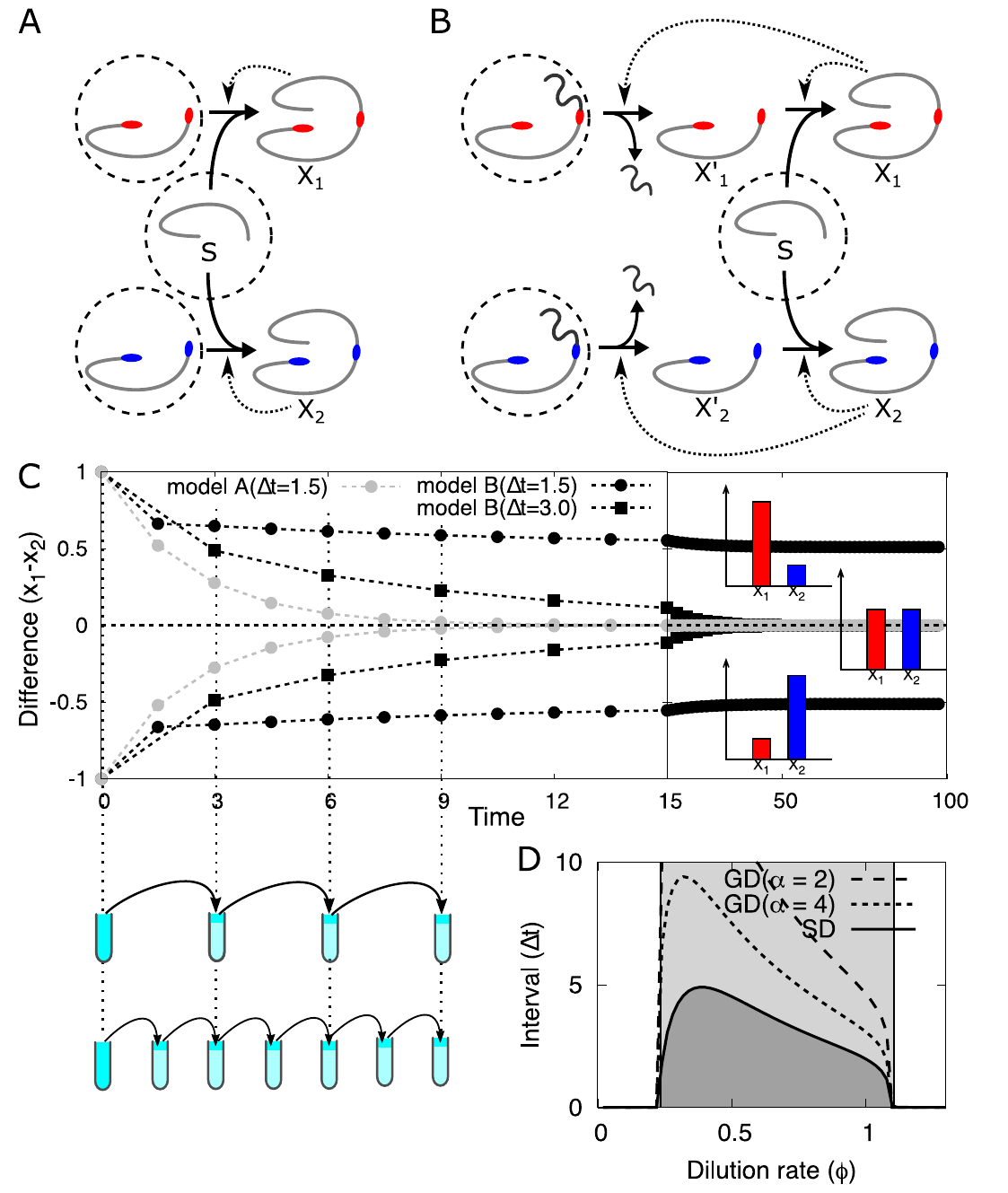}
    \caption{Schematic diagram of the model chemical reaction systems:  (A) non-modified two competing autocatalytic ribozymes and (B) the ones coupled with anabolism and catabolism reactions. Solid arrows represent reactions and dashed represent catalysis. Species enclosed by dashed circles are supplied from outside.  (C) Time course of the difference of the concentration of the ribozyme species, $\delta = x_1 - x_2$, starting from the initial condition where $\delta = 1$ or $=-1$. The gray circle represents the time course for the model (A) under the SD protocol with the interval $\varDelta t = 1.5$ (Eq.\,1 with $r(x)x = \epsilon + \kappa x$; we considered the reversible reaction). The black circle and square represent time courses for the  model (B) under the SD with $\varDelta t = 1.5$ and $=3.0$, respectively. We set $\epsilon = 0.25, \kappa = 2.5$, and $b = 0.1$. (D) The parameter region for the system with the bistability. In the dark-grey area, the system has bistability under SD protocol. The thin vertical lines (the boundaries for the light-grey area) represent the lower and upper boundary for $\bar \phi$ for the bistability under the CSTR. We set $\epsilon = 0.25, \kappa = 2.5$, and $b=0.1$.}
    \label{fig:azoarcus}
\end{figure*}

As in the previous models, there is the region for the kinetic parameters, $\varDelta t$ and $\bar \phi$, where the system exhibits bistability (see Fig.\,\ref{fig:azoarcus}D). 
For this modified \emph{Azoarcus} system, 
the concrete value that would exhibit heredity under SD is predicted that the dilution interval lies within 50-125 min, and the dilution factor per cycle lies between 2.5-11-fold
(We assumed $\kappa = 0.1 \rm{\mu M}^{-2} \min^{-1}$, $\epsilon = 0.01 \rm{\mu M}^{-1} \min^{-1}$, $\bar \phi = 0.04 \min^{-1}$ and $b = 0.1$; see Fig.\,\ref{fig:az_dt-m})

Further, we checked numerically (see Fig.\,\ref{fig:azoarcus}D and Appendix Sec.\,\ref{sec:azo_result}) that under the alternative dilution protocols (various functions of $\phi(t)$) upper bound of $\varDelta t$ for the bistability is bounded by $\varDelta t_c^{sd}$.

\begin{figure}
    \centering
    \includegraphics[width=15cm]{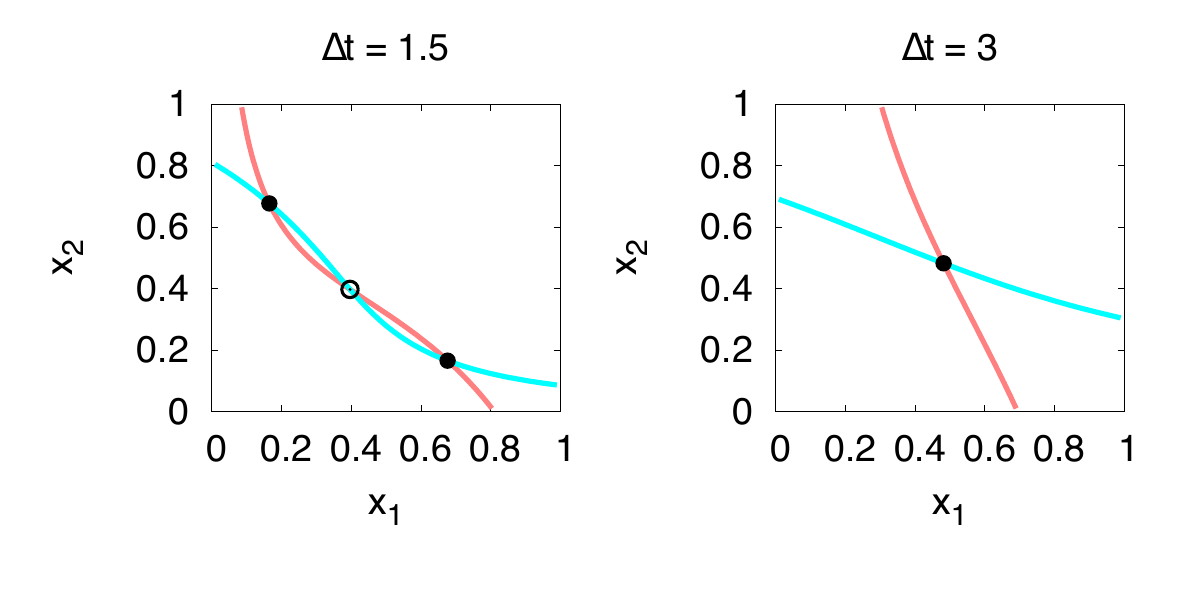}
    \caption{The red and blue curves represent the nullclines for $\rm{X}_1$ and $\rm{X}_2$, respectively, just before dilution. 
    We set $\varDelta t = 1.5$, $3$, $\kappa = 2.5$, $\epsilon = 0.25$, $\bar{\phi}=1$, and $b=0.1$. }
    \label{fig:az_nullclines}
\end{figure}

\begin{figure}
    \centering
        \includegraphics[width=16cm]{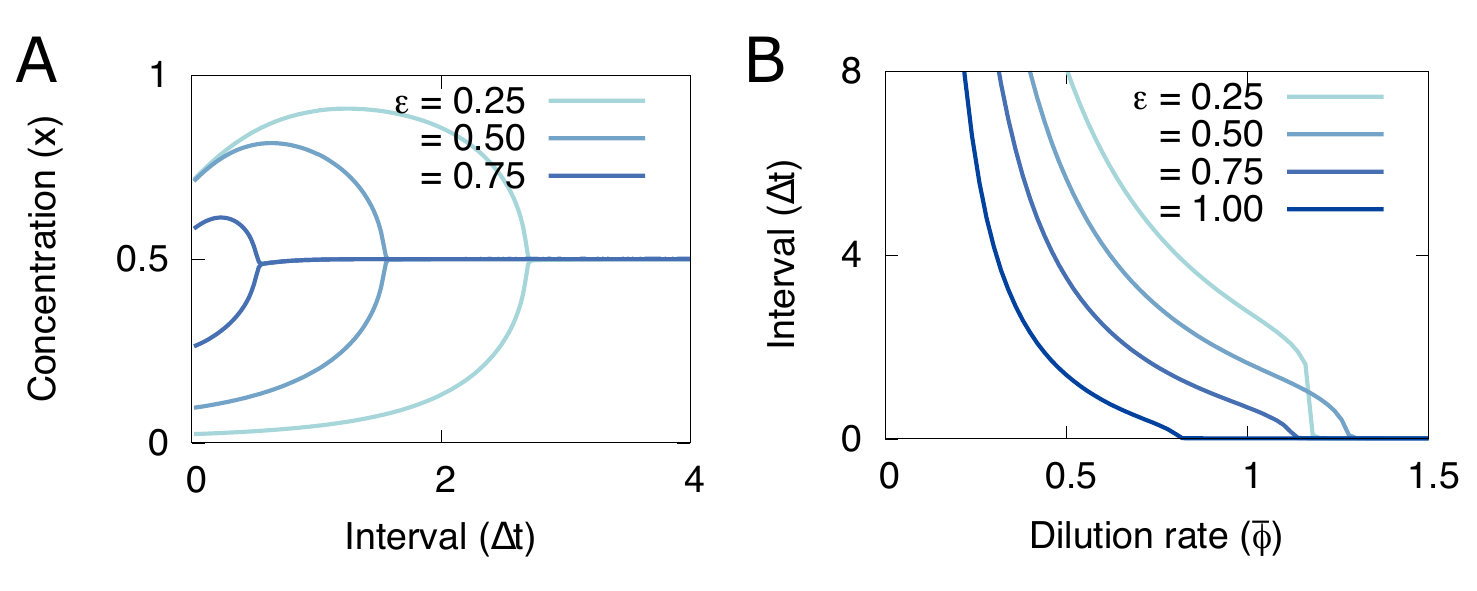}
    \caption{(A) The bifurcation diagram for the concentration at just before dilution varying the cycle interval $\varDelta t$. The lines with different colors represent the difference in the background reaction rate $\epsilon$. (B) The lines represent the critical point $\varDelta t_c$, which divides the regions where the system has bistability or not. We set the parameters as $\kappa=2.5$ and $\bar{\phi}=1$.}
    \label{fig:az_delta_t}
\end{figure}

\begin{figure}
    \centering
        \includegraphics[width=16cm]{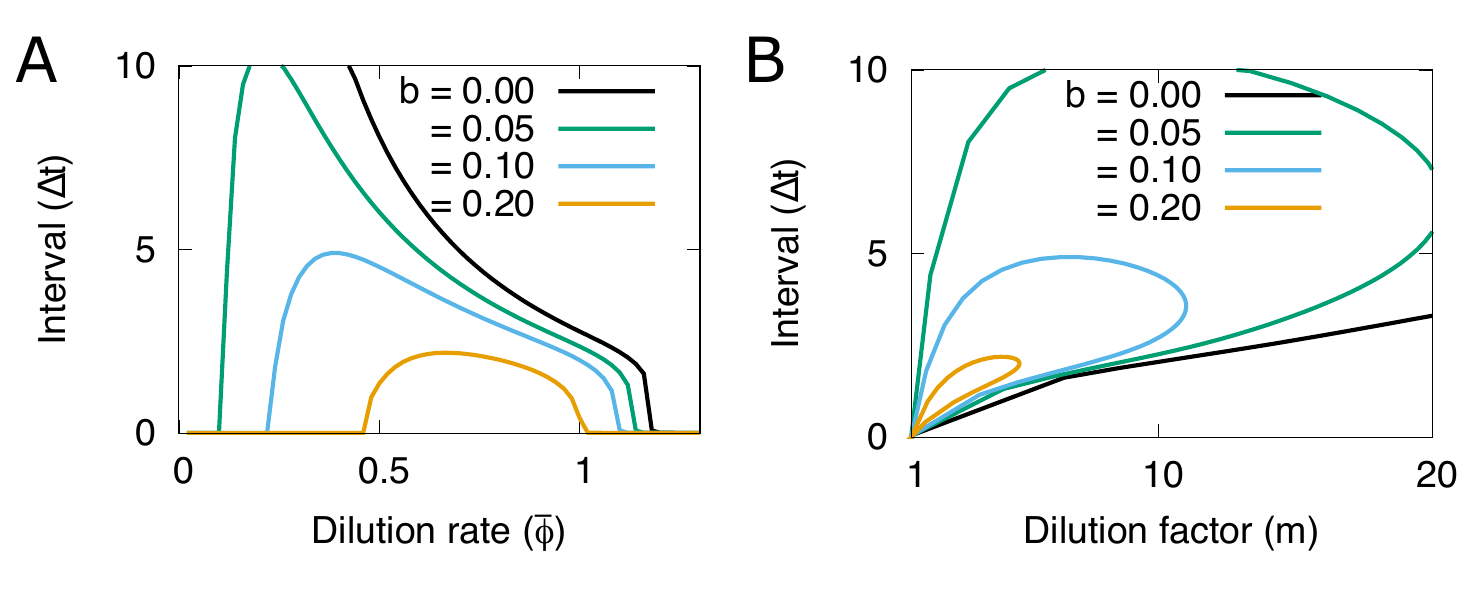}
    \caption{The parameter region for the system based on \emph{Azoarcus} ribozyme (Eq.~\ref{eq:azoarcus}) with the bistability under SD protocol. The same plot as Fig.\,\ref{fig:azoarcus}D, but its horizontal axis is the dilution factor $m$, $m=\exp(\bar \phi \varDelta t)$, instead of $\bar \phi$. The lines with different colors represent the boundary between with/without bistability under different backward reaction rates $b$. We set $\epsilon = 0.25, \kappa = 2.5$. }
    \label{fig:az_dt-m}
\end{figure}

In the case of the system based on \emph{Azoarcus} ribozyme coupled with catabolism/anabolism reaction (Eq.\,\ref{eq:azoarcus}), we drew the same figures as the model (Eq.\,\ref{eq:model}) in Sec.\,II B, C and D. These results are qualitatively similar as shown in Fig.\,\ref{fig:az_nullclines} and Fig.\,\ref{fig:az_delta_t}.

\section{Robustness of the results for the system based on \emph{Azoarcus} ribozyme}\label{sec:azo_robust}

\subsection{In the case of asymmetric catalytic efficiency}

\begin{figure}
    \centering
    \includegraphics[width=16cm]{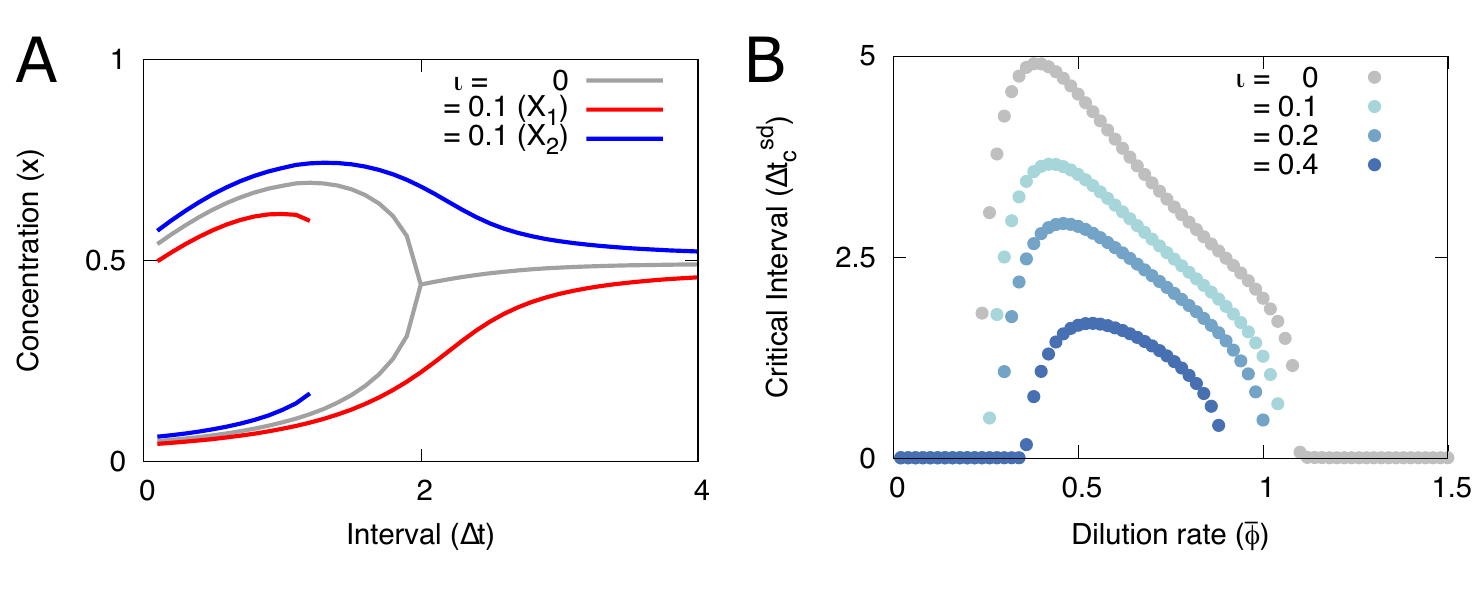}
    \caption{(A) The concentrations of the species $\rm{X}_1$ and $\rm{X}_2$ with a varying the period of the dilution cycles $\varDelta t$, when the catalytic strengths are asymmetric, $\kappa_1 \neq \kappa_2$. We set $\kappa_1=\kappa-\frac{\iota}{2}$ and $\kappa_2=\kappa+\frac{\iota}{2}$, and $\kappa_2-\kappa_1=\iota=0.1$. In contrast with the symmetric case (i.e., $\iota =0$), the $\rm{X}_2$-dominant state (and thus the bistability) disappears discontinuously at around $\varDelta t \sim 1.71$. We set the other parameters as $\kappa=2.5, \epsilon=0.25$, $b=0.1$ and $\bar{\phi}=1$. (B) The dependence of the critical interval $\varDelta t_c^{sd}$ of the cycle on the dilution rate $\bar \phi$ when the catalytic strengths are asymmetric: $\kappa_2-\kappa_1=\iota=0,0.05, 0.1$, and $0.2$. We set the other parameters as $\kappa=2.5, \epsilon = 0.25$. }
    \label{fig:k1-k2}
\end{figure}

A similar relation also appears even when the catalytic activities of two species are different, i.e., $\kappa_1 \neq \kappa_2$ (Fig.\,\ref{fig:k1-k2}A). Although the bifurcation at which the bistability disappears is discontinuous, it is at the similar $\varDelta t_c$ provided that the difference between $\kappa_1$ and $\kappa_2$ is not too large, as shown in Fig.\,\ref{fig:k1-k2}B.

\subsection{The variation of two-step ACSs with or without heredity}
\begin{figure}
    \centering
    \includegraphics[width=16cm]{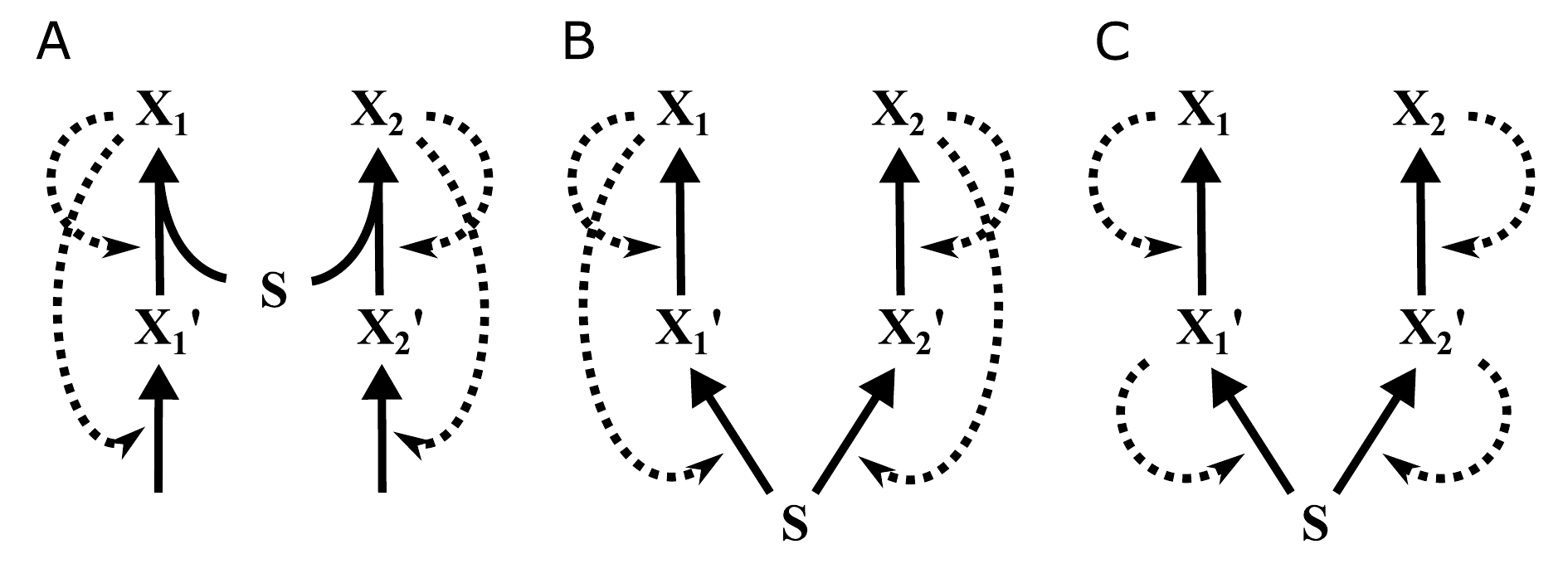}
    \caption{Schematics of the alternative models for the symmetric competing autocatalytic entities with two reaction steps. Solid arrows represent the reactions, and dotted arrows represent the catalysis. (A) The autocatalytic ribozymes coupled with anabolism and catabolism reactions in the main text Fig.\,\ref{fig:azoarcus}B. (B) A similar reaction system with two catalyzed reaction steps, but the substrate S is consumed to produce $\rm X'_1$ and $\rm X'_2$ instead of $\rm X_1$ and $\rm X_2$. (C) Two competing Lotka-Volterra models.}
    \label{fig:SI2steps}
\end{figure}

Here, assuming the mass action kinetics, we investigate alternative models with two-step reactions, similar to the model discussed in the main text (Fig.\,\ref{fig:SI2steps}A). 

Firstly, if we assume the substrate $\rm{S}$ is consumed in another reaction (see Fig.\,\ref{fig:SI2steps}B), the result does not change qualitatively; Here, assuming the CSTR condition, we consider the modified rate equations,
\begin{equation}
       \frac{d x'_i}{dt} = ( \epsilon + \kappa x_i ) ( s - x'_i ) - \bar\phi x'_i, \quad
       \frac{d x_i}{dt} = ( \epsilon + \kappa x_i ) x'_i - \bar\phi x_i,
\end{equation}
where $s^{tot} = s + x'_1 + x'_2+ x_1 + x_2$. $\rm{X}'_1$ and $\rm{X}'_2$ are converted from the shared substrate S, and which are further converted into $\rm{X}_1$ and $\rm{X}_2$, respectively. 

Then, we consider the dynamics near the steady state, and if we assume $x'_1$ and $x'_2$ can be adiabatically eliminated from $\frac{d x'_i}{dt}=0$,
\begin{equation}
    \frac{d x_i}{dt} = \frac {s (\epsilon + \kappa x_i)^2 } { \epsilon + \kappa x_i + \bar\phi} - \bar\phi x_i.
\end{equation}
Thus, this rate equation corresponds to Eq.~\ref{eq:model} with the reproduction rate function $x r(x) = \frac {(\epsilon +\kappa x)^2} {\epsilon + \kappa x + \bar\phi}$. 
Further, in this model also $\rm{X}_1$ and $\rm{X}_2$ compete for the same substrate $\rm{S}$; therefore, this model also shows the bistability, if the condition $\frac{d}{dt}r(\chi^*/2) > 0$ is satisfied, where $s^*r(\chi^*/2) = \phi$. (Later, the condition for this reaction system to be bistable under the serial dilution protocol is derived.)

Secondly, for the system to exhibit bistability, both of the reactions, from $\rm{S}$ to $X'_1$ and from $X'_1$ to $\rm{X}_1$ have to be catalyzed by $\rm{X}_1$; for example, we modify the model as the reaction $\rm{S}$ to $X'_1$ is catalyzed by $X'_1$ instead of $\rm{X}_1$ (Fig.\,\ref{fig:SI2steps}C). Under the CSTR condition, the rate equations of the model are
\begin{equation}
        \frac{dx'_i}{dt} = ( \epsilon + \kappa x'_i ) ( s - x'_i ) - \bar\phi x'_i, \quad
        \frac{dx_i}{dt} = ( \epsilon + \kappa x_i ) x'_i  - \bar\phi x_i,
\end{equation}
where $s^{tot} = s + x'_1 + x'_2 + x_1 + x_2$. This system has only one stable fixed point and does not show bistability. 
This is because $\rm{X}_1$ and $\rm{X}_2$ do not compete for the same resource for their replications, but $\rm{X}'_1$ and $\rm{X}'_2$ do. Then, the effective reproduction rate functions for 
$X'_1$ and $X'_2$ do not satisfy the condition for the symmetric state to be unstable. 

In conclusion, if the mass action kinetics is assumed, for the competing autocatalytic chemical reaction networks sharing the same substrate to have bistability (i.e., the nonlinear reproduction rate function), it is required at least two reaction steps to produce the autocatalytic entities (catalysts), which is catalyzed by the entities themselves.

\subsection{derivation of $\varDelta t_c^{sd}$ in a case with two-step catalyzed reactions}
We further consider the two competing chemical reaction networks with two catalyzed reaction steps in Fig.\,\ref{fig:SI2steps}B, under the SD protocol:
\begin{equation}
        \frac{dx'_i}{dt} = (s(\{x'_j\}, \{x_j\}, t) - x'_i) r(x_i),\quad
        \frac{dx_i}{dt} = x'_i r(x_i),
\end{equation}
where $r(x) = \epsilon + \kappa x$.
Here, we define
\begin{equation}
    \begin{split}
        \chi = x_1 +x_2,  \quad \chi' = x_1'  + x_2', \quad
        \delta  = x_1 -x_2, \quad \delta' = x_1'  - x_2'.
    \end{split}
\end{equation}
The time derivative of the above is derived as 
\begin{equation}
    \begin{aligned}
        \frac{d\delta}{dt} &= \frac{1}{2} \delta \chi'\frac{dr}{dx} \Bigl(\frac{\chi}{2}\Bigr) + \delta' r \Bigl(\frac{\chi}{2}\Bigr) &+& \mathcal{O}(\delta^2, \delta \delta', \delta'^2),\\
        \frac{d\chi}{dt} &= \chi' r\Bigl(\frac{\chi}{2}\Bigr) &+& \mathcal{O}(\delta^2, \delta \delta', \delta'^2),\\
        \frac{d\delta'}{dt} &= s \delta \frac{dr}{dx} \Bigl(\frac{\chi}{2}\Bigr) - \delta' r \Bigl(\frac{\chi}{2}\Bigr) - \frac{1}{2} \chi' \delta \frac{dr}{dx} \Bigl(\frac{\chi}{2}\Bigr) &+& \mathcal{O}(\delta^2, \delta \delta', \delta'^2),\\
        \frac{d \chi'}{dt} &= 2sr \Bigl(\frac{\chi}{2}\Bigr) - \chi'r \Bigl(\frac{\chi}{2}\Bigr) &+& \mathcal{O}(\delta^2, \delta \delta', \delta'^2),
    \end{aligned}
\end{equation}
where we used $r(x_i) = r(\frac{\chi}{2}) \pm \frac{\delta}{2} \frac{dr}{dx}(\frac{\chi}{2}) + \mathcal{O} (\delta^2)$.

To determine the deviation of $\delta$ in one cycle, we integrate $\frac{d\delta}{dt}/\delta$, 
\begin{equation}
    \begin{aligned}
        \int_{0}^{t} \frac{\frac{d \delta}{dt}}{\delta} dt &= \frac{1}{2} \int_{0}^{t} \chi' \frac{dr}{dx} \Bigl(\frac{\chi}{2}\Bigr) dt &+& \int_{0}^{t} r \Bigl(\frac{\chi}{2}\Bigr) \frac{\delta'}{\delta} dt,\\
            &= \int_{\frac{\chi(0)}{2}}^{\frac{\chi(t)}{2}} \frac{\frac{dr}{dx}}{r} dx &+& \int_{\frac{\chi(0)}{2}}^{\frac{\chi(t)}{2}} \frac{2 \delta'}{\chi' \delta} dx,
    \end{aligned}
\end{equation}
where we used the change of the variable $\frac{dt}{d\chi} = 1 / (\chi' r (\frac{\chi}{2}))$.

\begin{figure}
    \centering
    \includegraphics[width=16cm]{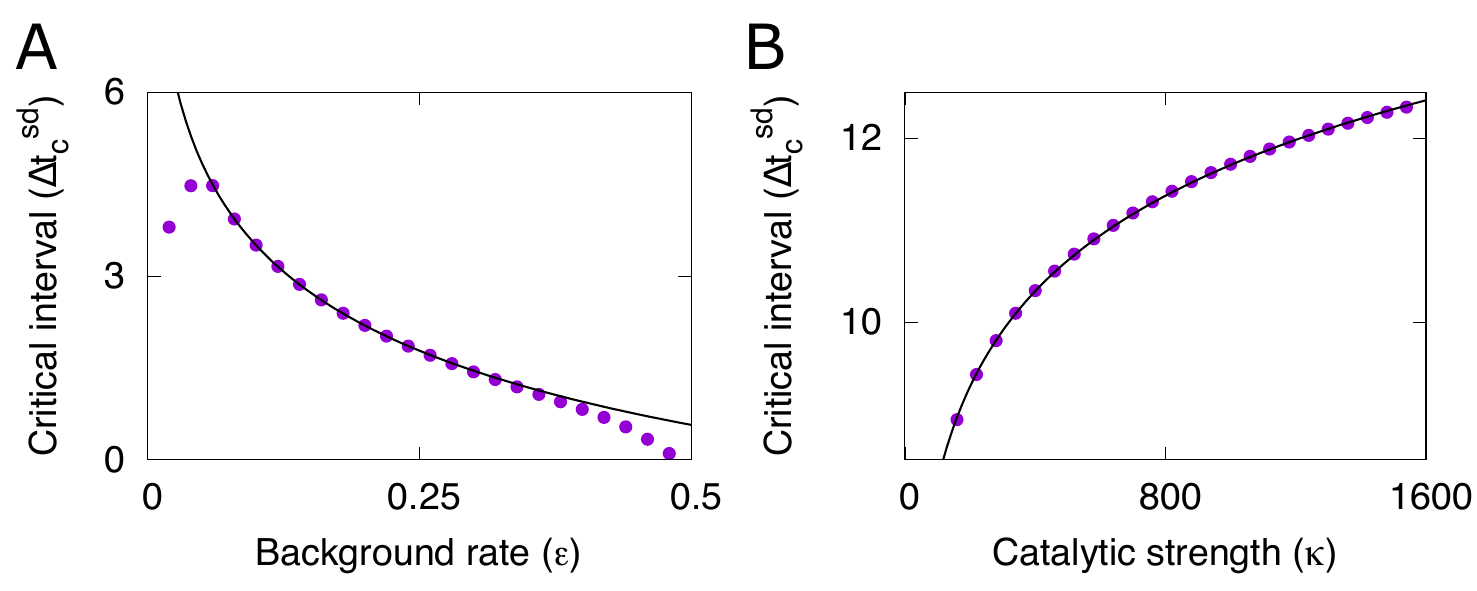}
    \caption{(A) The dotted line represents the critical point $\varDelta t_c$, which divides the regions where the system has bistability or not.
    The solid line represents the theoretical line for $\varDelta t_c$ determined by the relation in Eq.~\ref{eq:alt-delta_t}. (B) The interval threshold $\varDelta t_c$ vs the catalytic efficiency $\kappa$, fitted by $\frac{3}{2}\log(1+\kappa/2/\epsilon) - C_1$. We set $\bar{\phi}=\bar{\sigma}=1$.}
    \label{fig:alt-dleta_t}
\end{figure}

Here, to calculate Eq.\,S7, we have to estimate $\delta' / \chi'$ in the second integral in Eq.\,S7. 
If we assume in the second term, in the dominant part of the integrate, $\delta' / \chi'$ has the scaling relation $\delta' / \chi' \sim \mathcal{O}(1)$,
\begin{equation}
	\frac{\delta'}{\chi'} \approx \frac{\frac{d\delta'}{dt}}{\frac{d\chi'}{dt}} = \frac{1}{2} \frac{\frac{dr}{dx}}{r} \delta - \frac{\delta'}{2s-\chi'} \approx \frac{1}{2} \frac{\frac{dr}{dx}}{r} \delta,
\end{equation}
where we used $\frac{d}{dt} ( \frac{\delta'}{\chi'}) = 0$ in the first approximation, and in the second approximation, we assumed $s\gg\chi'$.
We substitute this to the above equation,
\begin{equation}
        \log \left| \frac{\delta(t)}{\delta(0)} \right| \approx 2 \log \left| \frac{ r (\frac{\chi(t)}{2}) }{ r (\frac{\chi(0)}{2}) } \right|.
\end{equation}
The threshold of $\varDelta t$ for the symmetric (i.e., $\delta =0$) trajectory to be unstable is,
\begin{equation} \label{eq:alt-delta_t}
    \varDelta t_c^{sd} = \frac{2}{\bar{\phi}} \log \left(1 + \frac{\kappa s^{tot}}{2 \epsilon}\right) - C_0,
\end{equation}
where $C_0$ is a constant value, which is determined numerically as $C_0 \sim 3.6 $ (see Fig.\,\ref{fig:alt-dleta_t}A).

While 
we use the asymptotic relation $\frac{\delta'}{\chi'^2} \sim \mathcal{O}(1)$ as $t$ becomes large. Then,
\begin{equation}
	\frac{\delta'}{\chi'} \approx \frac{1}{2} \frac{\frac{d\delta'}{dt}}{\frac{d\chi'}{dt}} \approx \frac{1}{4} \frac{\frac{dr}{dx}}{r} \delta,
\end{equation}
where $\frac{d}{dt} ( \frac{\delta'}{\chi'^2}) = 0$ in the first approximation, and $s \gg \chi'$ in the second approximation, as above. Thus,
\begin{equation}
        \log \left| \frac{\delta(t)}{\delta(0)} \right| \approx \frac{3}{2} \log \left| \frac{ r (\frac{\chi(t)}{2}) }{ r (\frac{\chi(0)}{2}) } \right|,
\end{equation}
and then
\begin{equation}
    \varDelta t_c^{sd} =  \frac{3}{2\phi} \log \left(1 + \frac{\kappa s^{tot}}{2 \epsilon}\right) - C_1,
\end{equation}
where $C_1$ is numerically determined as $C_1 \sim 0.477$ (see Fig.\,\ref{fig:alt-dleta_t}B).

\end{document}